%% file: article2_arXiv_3.tex
\documentclass[article, noheadings, nofooter, nojss]{jss}\usepackage[]{graphicx}\usepackage[]{color}
\makeatletter
\def\maxwidth{ %
  \ifdim\Gin@nat@width>\linewidth
    \linewidth
  \else
    \Gin@nat@width
  \fi
}
\makeatother

\definecolor{fgcolor}{rgb}{0, 0, 0}
\newcommand{\hlnum}[1]{\textcolor[rgb]{0,0,0}{#1}}
\newcommand{\hlstr}[1]{\textcolor[rgb]{0,0,0}{#1}}
\newcommand{\hlcom}[1]{\textcolor[rgb]{0,0,0}{#1}}
\newcommand{\hlopt}[1]{\textcolor[rgb]{0,0,0}{#1}}
\newcommand{\hlstd}[1]{\textcolor[rgb]{0,0,0}{#1}}
\newcommand{\hlkwa}[1]{\textcolor[rgb]{0,0,0}{#1}}
\newcommand{\hlkwb}[1]{\textcolor[rgb]{0,0,0}{#1}}
\newcommand{\hlkwc}[1]{\textcolor[rgb]{0,0,0}{#1}}
\newcommand{\hlkwd}[1]{\textcolor[rgb]{0,0,0}{#1}}

\usepackage{framed}
\makeatletter
\newenvironment{kframe}{%
 \def\at@end@of@kframe{}%
 \ifinner\ifhmode%
  \def\at@end@of@kframe{\end{minipage}}%
  \begin{minipage}{\columnwidth}%
 \fi\fi%
 \def\FrameCommand##1{\hskip\@totalleftmargin \hskip-\fboxsep
 \colorbox{shadecolor}{##1}\hskip-\fboxsep
     \hskip-\linewidth \hskip-\@totalleftmargin \hskip\columnwidth}%
 \MakeFramed {\advance\hsize-\width
   \@totalleftmargin\z@ \linewidth\hsize
   \@setminipage}}%
 {\par\unskip\endMakeFramed%
 \at@end@of@kframe}
\makeatother

\definecolor{shadecolor}{rgb}{1, 1, 1}
\definecolor{messagecolor}{rgb}{0, 0, 0}
\definecolor{warningcolor}{rgb}{1, 0, 1}
\definecolor{errorcolor}{rgb}{1, 0, 0}
\newenvironment{knitrout}{}{} 

\usepackage{alltt}

\include{packages}

\include{header}

\author{Sarah Brockhaus\\ LMU Munich \And 
		David R\"ugamer\\LMU Munich \And
		Sonja Greven\\LMU Munich}
		
\title{Boosting Functional Regression Models with \pkg{FDboost}}

\Abstract{
\noindent 
The \proglang{R} add-on package \pkg{FDboost} is a flexible toolbox for the estimation of functional regression models by model-based boosting. It provides the possibility to fit regression models for scalar and functional response with effects of scalar as well as functional covariates, i.e., scalar-on-function, function-on-scalar and function-on-function regression models.
In addition to mean regression, quantile regression models as well as generalized additive models for location scale and shape can be fitted with \pkg{FDboost}. Furthermore, boosting can be used in high-dimensional data settings with more covariates than observations. 
We provide a hands-on tutorial on model fitting and tuning, including the visualization of results. 
The methods for scalar-on-function regression are illustrated with spectrometric data of fossil fuels and those for functional response regression with a data set including bioelectrical signals for emotional episodes.
}
\Plainkeywords{functional data analysis, function-on-function regression, function-on-scalar regression, gradient boosting, model-based boosting, scalar-on-function regression} 

\Address{
Sarah Brockhaus\\
  Ludwig-Maximilans-Universit\"at M\"unchen\\
  E-mail: \email{sarah.brockhaus@stat.uni-muenchen.de}\\[.3cm]
  David R\"ugamer\\
  Ludwig-Maximilans-Universit\"at M\"unchen\\
  E-mail: \email{david.ruegamer@stat.uni-muenchen.de}\\[.3cm]
  Sonja Greven\\
  Ludwig-Maximilans-Universit\"at M\"unchen\\
  E-mail: \email{sonja.greven@stat.uni-muenchen.de}\\
}

\definecolor{myc}{rgb}{0,0,0}
\IfFileExists{upquote.sty}{\usepackage{upquote}}{}
\begin{document}



\section{Introduction} 

With the progress of technology today, we have the ability to observe more and more data of a functional nature, such as curves, trajectories or images \citep{ramsay2005}. 
Functional data can be found in many scientific fields like demography, biology, medicine, meteorology and economics \citep[see, e.g.,][]{ullah2013}. 
In practice, the functions are observed on finite grids. 
In this paper, we deal with one-dimensional functional data that are observed over a real valued interval. 
Examples for such data are growth curves over time, acoustic signals, temperature curves and spectrometric measurements in a certain range of wavelengths.  
Regression models are a versatile tool for data analysis and various models have been proposed for regression with functional variables; see \citet{morris2015} and \citet{greven2017discussion} for recent reviews of functional regression models. One can distinguish between three different types of functional regression models: 
scalar-on-function regression, a regression with scalar response and functional covariates, function-on-scalar regression referring to models with functional response and scalar covariates and function-on-function regression, which is used when both response and covariates are functional. 
Models for scalar-on-function regression are sometimes also called signal regression. 

\citet{greven2017discussion} lay out a generic framework for functional regression models including the three mentioned model types. Many types of covariate effects are discussed including linear and non-linear effects of scalar covariates as well as linear effects of functional covariates and interaction terms. They describe that estimation can be based on a mixed models framework \citep{scheipl2015,scheipl2016gfamm} or on component-wise gradient boosting \citep{brockhaus2015,brockhaus2015hist}. 
In this paper, we describe the latter approach and provide a hands-on tutorial for its implementation in \proglang{R} \citep{R2017} in the comprehensive \proglang{R} package \pkg{FDboost} \citep{FDboost_gen}.  

Boosting estimates the model by iteratively combining simple models and can be seen as a method that conducts gradient descent \citep{buhlmann2007}. Boosting is capable of estimating models in high-dimensional data settings and implicitly does variable selection. 
The modeled features of the conditional response distribution can be chosen quite flexibly by minimizing different loss functions. The framework includes linear models (LMs), generalized linear models (GLMs) as well as quantile and expectile regression. 
Furthermore, generalized additive models for location, scale and shape \citep[GAMLSS,][]{rigby2005} can be fitted \citep{mayr2012}. GAMLSS model all distribution parameters of the conditional response distribution simultaneously depending on potentially different covariates. \citet{brockhaus2016} discuss GAMLSS with scalar response and functional covariates. \citet{stoecker2017} introduce GAMLSS for functional response.  
Due to variable selection and shrinkage of the coefficient estimates, no classical inference concepts are available for the boosted models. 
However, it is possible to quantify uncertainty by bootstrap \citep{efron1979} and stability selection \citep{meinshausen2010}. 
The main advantages of the boosting approach are the possibility to fit models in high dimensional data settings with variable selection and to estimate not only mean regression models but also GAMLSS and quantile regression models. The main disadvantage is the lack of formal inference.  
 
Other frameworks for flexible regression models with functional response exist. \cite{morris2006wavelet} and \cite{meyer2015} use a basis transformations approach  and Bayesian inference to model functional variables. Usually, loss-less transformations like a wavelet transformation are used. See \citet{morris2017comparison} for a detailed comparison of the two frameworks. 


In this tutorial, we present the \proglang{R}~package \pkg{FDboost} \citep{FDboost_gen}, which is designed to fit a great variety of functional regression models by boosting. \pkg{FDboost} builds on the \proglang{R}~package \pkg{mboost} \citep{mboost_gen} for statistical model-based boosting. Thus, in the back-end we rely on a well-tested implementation. 
\pkg{FDboost} provides a comprehensive implementation of the most important methods for boosting functional regression models. In particular, the package can be used to conveniently fit models with functional response. For effects of scalar covariates on functional responses, we provide base-learners with suitable identifiability constraints. 
In addition, base-learners that model effects of functional covariates are implemented. 
The package also contains functions for model tuning and for visualizing results. 

As a case study for scalar-on-function regression, we use a dataset on fossil fuels, which was analyzed in \cite{fuchs2015} and \cite{brockhaus2015} and is part of the \pkg{FDboost} package. In this application, the heat value of fossil fuels should be predicted based on spectral data. 
As a case study for function-on-scalar and function-on-function regression, we use the emotion components data set, which is analyzed in \citet{ruegamer2018} in the context of factor-specific historical effect estimation and which is provided in an aggregated version in \pkg{FDboost}. \textcolor{myc}{Note that we use both data sets as a running example to illustrate the capabilities of the package. We give a more complex example with a stronger focus on answering the underlying research question in Appendix~E.}
  
The remainder of the paper is structured as follows. 
We shortly review the generic functional regression model (Section~\ref{sec.fun_reg}) for scalar and for functional response. Then the boosting algorithm used for model fitting is introduced in Section~\ref{sec.boosting}. 
In Section~\ref{sec.fdboost}, we give details on the infrastructure of the package \pkg{FDboost}. 
Scalar-on-function regression with \pkg{FDboost} is described in Section~\ref{sec.fdboost_sof}. Regression models for functional response with scalar and/or functional covariates are described in Section~\ref{sec.fdboost_fof}. 
We present possible covariate effects as well as discuss model tuning and show how to extract and display results. 
In Section~\ref{sec.fdboost_family}, we discuss regression models that model other characteristics of the response distribution than the mean, in particular median regression and GAMLSS. 
In Section~\ref{sec.fdboost_stabsel}, we shortly comment on stability selection in combination with boosting. 
\textcolor{myc}{In Section~\ref{sec.computational_cost} we comment on the computational burden of fitting models with \pkg{FDboost}.} 
We conclude with a discussion in Section~\ref{sec.discussion}. 
The paper is structured such that the subsections on functional response can be skipped if one is only interested in scalar-on-function regression.


\section{Functional regression models}
\label{sec.fun_reg}

%
In Section~\ref{sec.sof} we first introduce a generic model for scalar response with functional and scalar covariates. Afterwards, we deal with models with functional response in Section~\ref{sec.fof}. 
%
\subsection{Scalar response and functional covariates}
\label{sec.sof} 
Let the random variable $Y$ be the scalar response with realization $y \in \real$.    
The covariate set $\mX$ can include both scalar and functional variables.  
We denote a generic scalar covariate by $Z$ and a generic functional covariate by $X(s)$, with $s \in \mathcal{S} = [S_1, S_2]$ and $S_1 < S_2$, $S_1, S_2 \in \real$.   
We assume that we observe $i=1,\ldots,N$ data pairs $(y_i, \mx_i)$, where $\mx_i$ comprises the realizations $z_i$ of scalar covariates as well as the realizations $x_i(s)$ of $X_i(s)$. In practice, $x_i(s)$ is observed on a grid of evaluation points $s_1,\ldots, s_R$, such that each curve is observed as a vector $(x_i(s_1), \ldots, x_i(s_R))^\top$. While different functional covariates may be observed on different grid points over different intervals, \textcolor{myc}{which is supported by \pkg{FDboost} as also the following example will show,} we do no introduce additional indices here for ease of notation. 

We model the expectation of the response by an additive regression model     
\begin{equation}
\label{eq.sof}
\EV( Y_i  | \mX_i = \mx_i ) = h(\mx_i) = \sum_{j=1}^{J} h_j(\mx_i),
\end{equation} 
where $h(\mx_i)$ is the additive predictor containing the additive effects $h_j(\mx_i)$.  
Each effect $h_j(\mx_i)$ can depend on one or more covariates in $\mx_i$. Possible effects include linear, non-linear and interaction effects of scalar covariates as well as linear effects of functional covariates. Moreover, group-specific effects and interaction effects between scalar and functional variables are possible. 
To give an idea of possible effects $h_j(\mx)$, Table~\ref{tab.sof_effects} lists effects of functional covariates that are currently implemented in \pkg{FDboost}.  
\renewcommand\arraystretch{1.4}
\begin{table}[htb]\centering
\begin{small}
\begin{tabular}{p{.45\textwidth}| p{.31\textwidth}| p{.19\textwidth} }
	covariate(s) & type of effect & $h_j(x)$  \\
\hline 
	functional covariate $x(s)$ & linear functional effect & $\int_{\mathcal{S}} x(s) \beta(s)\,ds $  \\
	scalar and functional covariate, $z$ and  $x(s)$ & linear interaction & $z \int_{\mathcal{S}} x(s) \beta(s)\,ds $  \\
	& smooth interaction & $\int_{\mathcal{S}} x(s) \beta(z,s)\,ds $  \\ 
\end{tabular}
\end{small}
\caption[Possible covariate effects]{Overview of possible covariate effects of functional covariates, including interaction effects with scalar covariates.}
\label{tab.sof_effects}
\end{table}
\renewcommand\arraystretch{1}
\textcolor{myc}{A scalar-on-function model with only one functional covariate would be  
$\EV( Y_i  | \mX_i = \mx_i ) = \beta_0 + \int_{\mathcal{S}} x_i(s) \beta(s)\,ds$, see 
Section~\ref{sec.fdboost_sof} for concrete examples of scalar-on-function models for the fossil fuel data set.}

The effects $h_j(\mx_i)$ are linearized using a basis representation:  
\begin{equation}
\label{eq.sof_eff}
h_j(\mx_i) = \mb_{j}(\mx_i)^{\top} \mtheta_j, \ \ j=1, \ldots, J, 
\end{equation}
with basis vector $\mb_{j}(\mx_i) \in \real^{K_{j}}$ and coefficient vector $\mtheta_j \in \real^{K_{j}}$ that has to be estimated. 
The $N \times K_j$ design matrix for the $j$th effect consists of rows $\mb_{j}(\mx_i)^{\top}$ for all observations $i=1,\ldots,N$.  
A ridge-type penalty term $\lambda_j \mtheta_{j}^\top \mP_{j} \mtheta_{j}$ is used for regularization, where $\mP_{j}$ is a suitable penalty matrix for $\mb_j$ and $\lambda_j$ is a non-negative smoothing parameter. 
The smoothing parameter controls the degrees of freedom of the effect. 

Consider, for example, a linear effect of a functional covariate $\int_{\mathcal{S}} x_i(s) \beta(s)\,ds$. 
Using $\mtheta_{j} = (\theta_{j1}, \ldots, \theta_{jK_j})^\top$, this effect is computed as 
\begin{equation*}
\begin{aligned}
\int_{\mathcal{S}} x_i(s) \beta(s)\,ds
&= \int_{\mathcal{S}}  x_i(s) 
\underbrace{ \sum_{k=1}^{K_j} \ba_k(s) \theta_{jk}}_{\approx \beta(s)} \,ds \\
&\approx \sum_{r=1}^R \Big( \Delta(s_r) x_i(s_r) 
 \sum_{k=1}^{K_j} \ba_k(s_r) \theta_{jk} \Big) \\
&= \sum_{k=1}^{K_j} \Big( \underbrace{ \sum_{r=1}^R \Delta(s_r) x_i(s_r)  \ba_k(s_r)}_{\text{entries in } \mb_{j}(\mx_i)} \theta_{jk} \Big) \\
&= \mb_j(\mx_i)^\top \mtheta_{j}, \\
\end{aligned}
\end{equation*}
where first, the smooth effect $\beta(s)$ is expanded in basis functions, second, the integration is approximated by a weighted sum and, third, the terms are rearranged such that they fit into the scheme $\mb_j(\mx_i)^\top \mtheta_{j}$.  
The basis $\mb_{j}(\mx_i)$ is thus computed as 
\begin{equation}
\label{eq.sof_example_fun_effect}
\begin{aligned}
 \mb_j(\mx_i)^\top 
 &= \left[\sum_{r=1}^R \Delta(s_r) x_i(s_r) \ba_1(s_r)\ \cdots\ \sum_{r=1}^R \Delta(s_r) x_i(s_r) \ba_{K_j}(s_r)\right] \\
&\approx \left[ \int_{\mathcal{S}}  x_i(s) \ba_1(s)\, ds \ \cdots\ \int_{\mathcal{S}} x_i(s) \ba_{K_j}(s)\, ds \right],
\end{aligned}
\end{equation}
with spline functions $\ba_k$, $k=1,\ldots, K_j$, for the expansion of the smooth effect $\beta(s)$ in $s$ direction and integration weights $\Delta(s_r)$ for numerical computation of the integral.   
The penalty matrix~$\mP_j$ is chosen such that it is suitable to regularize the splines $\ba_k$. 
\textcolor{myc}{In the current implementation only P-splines are readily available to estimate smooth effects.}
To set up a P-spline basis \citep{eilers1996} for the smooth effect, $\ba_k$ in Equation~\ref{eq.sof_example_fun_effect} are B-splines and the penalty~$\mP_j$ is a squared difference matrix. 


\subsubsection{Case study: Heat value of fossil fuels}
The aim of this application is to predict the heat value $y$ of fossil fuels using spectral data \citep[Siemens AG]{fuchs2015}. 
For $N=129$ samples, the dataset contains the heat value, the percentage of humidity $z_\ho$ and two spectral measurements, which can be thought of as functional variables $x_\NIR (s_\NIR)$ observed over $\mathcal{S}_{\NIR}$\textcolor{myc}{$= [250.4, 876.8]$} and $x_\UVVIS (s_\UVVIS)$ observed over $\mathcal{S}_{\UVVIS}$\textcolor{myc}{$= [800.4, 2761.0]$}. 
One spectrum is ultraviolet-visible (UVVIS), the other a near infrared spectrum (NIR). For both spectra, the observation points are not equidistant. 
The dataset is contained in the \proglang{R} package \pkg{FDboost}. 

\begin{knitrout}
\definecolor{shadecolor}{rgb}{1, 1, 1}\color{fgcolor}\begin{kframe}
\begin{alltt}
\hlstd{R> }\hlkwd{library}\hlstd{(FDboost)}
\hlstd{R> }\hlkwd{data}\hlstd{(}\hlstr{"fuelSubset"}\hlstd{,} \hlkwc{package} \hlstd{=} \hlstr{"FDboost"}\hlstd{)}
\hlstd{R> }\hlkwd{str}\hlstd{(fuelSubset)}
\end{alltt}
\begin{verbatim}
List of 7
 $ heatan      : num [1:129] 26.8 27.5 23.8 18.2 17.5 ...
 $ h2o         : num [1:129] 2.3 3 2 1.85 2.39 ...
 $ nir.lambda  : num [1:231] 800 803 805 808 810 ...
 $ NIR         : num [1:129, 1:231] 0.2818 0.2916 -0.0042 -0.034 -0.1804 ...
 $ uvvis.lambda: num [1:134] 250 256 261 267 273 ...
 $ UVVIS       : num [1:129, 1:134] 0.145 -1.584 -0.814 -1.311 -1.373 ...
 $ h2o.fit     : num [1:129] 2.58 3.43 1.83 2.03 3.07 ...
\end{verbatim}
\end{kframe}
\end{knitrout}

Figure~\ref{fig:fuel01} shows the two spectral measurements colored according to the heat value. Predictive models for the heat values, discussed in the next sections, will include scalar-on-function terms to accommodate the spectral covariates. 

\begin{knitrout}
\definecolor{shadecolor}{rgb}{1, 1, 1}\color{fgcolor}\begin{figure}[ht]

{\centering \includegraphics[width=\maxwidth]{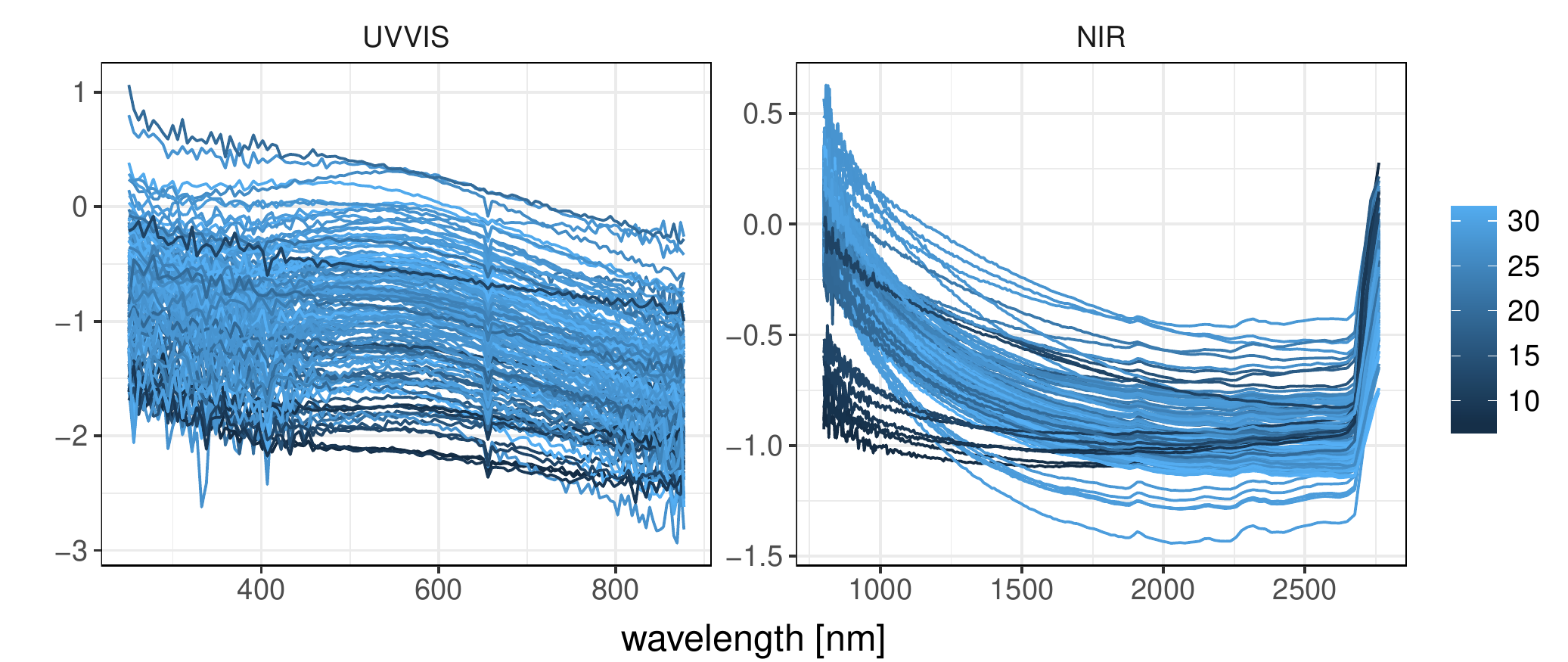} 

}

\caption{Spectral data of fossil fuels. Coloring of the spectral data depicts the corresponding heat value.}\label{fig:fuel01}
\end{figure}

\end{knitrout}

\hfill $\blacklozenge$


%
\subsection{Functional response}
\label{sec.fof}
We denote the functional response by $Y(t)$, where $t$ is the evaluation point at which the function is observed. We assume that $t \in \mathcal{T}$, where $\mathcal{T}$ is a real-valued interval $[T_1, T_2]$, for example a time-interval.   
All response curves can be observed on one common grid or on curve-specific grids. For responses observed on one common grid, we write $y_i(t_g)$ for the observations, with $t_g \in \{t_1, \ldots, t_G\}$ denoting the grid of evaluation points.  For curve-specific evaluation points, the observations are denoted by $y_i(t_{ig})$, with $t_{ig} \in \{t_{i1}, \ldots, t_{iG_i}\}$.  
As above, the covariate set $\mX$ can contain both scalar and functional variables. 

As in model (\ref{eq.sof}), we model the conditional expectation of the response. In this case, the expectation is modeled for each point $t \in \mathcal{T}$:      
\begin{equation}
\label{eq.fof}
\EV( Y_i(t) | \mX_i = \mx_i ) = h(\mx_i,t) = \sum_{j=1}^{J} h_j(\mx_i,t). 
\end{equation} 
As the response $Y_i(t)$ is a function of~$t$, the linear predictor $h(\mx_i,t)$ as well as the additive effects $h_j(\mx_i,t)$ are functions of~$t$.   
Each effect $h_j(\mx_i,t)$ can depend on one or more covariates in $\mx_i$ as well as on $t$. 
To give an idea of possible effects $h_j(\mx_i,t)$, Table~\ref{tab.fun_effects} lists some effects that are currently implemented. 
\renewcommand\arraystretch{1.4}
\begin{table}[htb]\centering
\begin{small}
\begin{tabular}{p{.34\textwidth}| p{.33\textwidth}| p{.19\textwidth}}
	covariate(s) & type of effect & $h_j(x,t)$ \\
\hline
	(none) & smooth intercept & $\beta_0(t)$\\
	scalar covariate $z$ & linear effect & $z \beta(t)$\\
	& smooth effect & $f(z, t)$\\
	two scalars  $z_1$, $z_2$ & linear interaction & $z_1 z_2 \beta(t)$\\
	& functional varying coefficient & $z_1 f(z_2,t)$\\
	& smooth interaction & $f(z_1, z_2, t)$\\
\hline 
	functional covariate $x(s)$ & linear functional effect & $\int_{\mathcal{S}} x(s) \beta(s,t)\,ds $\\
	scalar $z$ and functional  $x(s)$ & linear interaction & $z \int_{\mathcal{S}} x(s) \beta(s,t)\,ds $\\
	& smooth interaction & $\int_{\mathcal{S}} x(s) \beta(z,s,t)\,ds $\\
\hline
	functional covariate $x(s)$,      & concurrent effect & $ x(t) \beta(t)$\\
	with $\mathcal{S}=\mathcal{T}=[T_1,T_2]$    & historical  effect & $\int_{T_1}^{t} x(s) \beta(s,t)\,ds $\\
	& lag effect, with lag $\delta>0$ & $\int_{t-\delta}^{t} x(s) \beta(s,t)\,ds $\\
	& lead effect, with lead $\delta>0$ & $\int_{T_1}^{t-\delta} x(s) \beta(s,t)\,ds $\\
	& effect with $t$-specific integration limits $[l(t), u(t)]$ 
	                   & $\int_{l(t)}^{u(t)} x(s) \beta(s,t)\,ds $\\
\hline
	grouping variable $g$ & group-specific smooth intercepts & $\beta_g(t)$\\
	grouping variable $g$ and scalar $z$ & group-specific linear effects & $z \beta_g(t)$\\
	curve indicator $i$ & curve-specific smooth residuals & $e_i(t)$\\
\end{tabular}
\end{small}
\caption[Possible covariate effects]{Overview of some possible covariate effects that can be represented within the framework of functional regression. 
}
\label{tab.fun_effects}
\end{table}
\renewcommand\arraystretch{1}
\textcolor{myc}{A function-on-function model with only one functional covariate would be  
$\EV( Y_i  | \mX_i = \mx_i ) = \beta_0(t) + \int_{\mathcal{S}} x_i(s) \beta(s,t)\,ds$. In Section~\ref{sec.fdboost_fof}, we give several examples for concrete models with functional response.}

All effects mentioned in Table~\ref{tab.fun_effects} are varying over~$t$ but can also be modeled as constant in~$t$. 
The upper part of the table contains linear, smooth and interaction effects for scalar covariates. 
The middle part of the table gives possible effects of functional covariates and interaction effects between scalar and functional covariates. 
The lower part of the table in addition shows some group-specific effects. 
 
In practice, all effects $h_j(\mx_i,t_{ig})$ are linearized using a basis representation \citep{brockhaus2015hist}:  
\begin{equation}
\label{eq.b_jY}
h_j(\mx_i,t_{ig}) = \mb_{jY}(\mx_i,t_{ig})^{\top} \mtheta_j, \ \ j=1, \ldots, J, 
\end{equation}
where the basis vector $\mb_{jY}(\mx_i,t_{ig}) \in \real^{K_{jY}}$ depends on covariates $\mx_i$ and the observation-point of the response~$t_{ig}$. The corresponding coefficient vector $\mtheta_j \in \real^{K_{jY}}$ has to be estimated. 
The design matrix for the $j$th effect consists of rows $\mb_{jY}(\mx_i,t_{ig})^{\top}$ for all observations $i=1,\ldots,N$ and all time-points $t_{ig}$, $g=1,\ldots,G_i$.    

In the following, we will use a modularization of the basis into a first part depending on covariates 
and a second part that only depends on~$t$. This modular structure reduces the problem of specifying the basis $\mb_{jY}(\mx_i,t_{ig})$ to that of creating two suitable marginal bases. For many effects, the marginal bases are easy to define as they are known from regression with scalar response. 
%

First, we focus on responses observed on one common grid $(t_1, \ldots, t_G)^\top$ which does not depend on~$i$.  
In this case, we represent the effects using the Kronecker product~$\otimes$ of two marginal bases \citep{brockhaus2015}     
\begin{equation}
\label{eq.effect_o}
h_j(\mx_i,t_{g}) = \big( \mb_j(\mx_i)^\top \otimes \mb_Y(t_{g})^\top \big) \mtheta_{j},   
\end{equation}
where the marginal basis vector $\mb_{j}(\mx_i) \in \real^{K_j}$, $i=1,\ldots,N$, depends on covariates in $\mx_i$ and the marginal basis vector $\mb_Y(t_{g}) \in \real^{K_Y}$, $g=1,\ldots,G$, depends on the grid point~$t_g$. 
The $NG \times K_j K_Y$ design matrix is computed as the Kronecker product of the two marginal design matrices, which have dimensions $N \times K_j$ and $G \times K_Y$. 
If the effect can be represented as in Equation~\ref{eq.effect_o} it fits into the framework of linear array models \citep{currie2006}. 
The representation as array model has computational advantages, saving time and memory. \cite{brockhaus2015} discuss array models in the context of functional regression.  

Note that the representation in Equation~\ref{eq.effect_o} is only possible for responses observed on one common grid, as otherwise $\mb_Y(t_{ig})$ depends on the curve-specific grid points $t_{ig}$. In this case, the marginal bases are combined by the row-wise tensor product \citep{scheipl2015,brockhaus2015hist}. This is a rather technical detail and is thoroughly explained in \cite{brockhaus2015hist}, also for the case where the basis for the covariates depends on $t_{ig}$ such as for historical effects.  

We regularize the effects by a ridge-type penalty term $\mtheta_{j}^\top \mP_{jY} \mtheta_{j}$. 
The penalty matrix for the composed basis can be constructed as \citep[Sec.\ 4.1.8]{wood2006}  
\begin{equation}
\label{eq.penalty_ani}
\mP_{jY} = \lambda_{j} (\mP_{j} \otimes \mI_{K_Y} ) + \lambda_Y ( \mI_{K_{j}} \otimes \mP_{Y}),
\end{equation} 
where $\mP_{j}$\textcolor{myc}{ $=[p_{j,\varkappa,\varsigma}]_{\varkappa,\varsigma\in \{1,\ldots,K_s\}}$} is a suitable penalty for $\mb_j$ and $\mP_{Y}$ is a suitable penalty for $\mb_Y$. The non-negative smoothing parameters $\lambda_j$ and $\lambda_Y$ determine the degree of smoothing in each direction. 
\textcolor{myc}{
To illustrate the resulting penalty matrix, we explicitly compute the Kronecker products in  Equation 7: 
\begin{equation*}
  \mP_{jY} = \lambda_j 
  \begin{bmatrix}
  p_{j,1,1} \cdot \bm{I}_{K_y}   & \cdots &  p_{j,1,K_s} \cdot \bm{I}_{K_y} \\
  \vdots &  \ddots  & \vdots & \\
  p_{j,K_s,1} \cdot \bm{I}_{K_y}  &    \cdots &  p_{j,K_s,K_s} \cdot \bm{I}_{K_y} \\
  \end{bmatrix}
  +
  \lambda_Y
  \begin{bmatrix}
  \bm{P}_{Y}&  &  \bm{0}  \\
    &  \ddots &    \\
  \bm{0} & &  \bm{P}_{Y}\\
    \end{bmatrix}
\end{equation*}
This shows the block structure of the penalty matrix and how the two marginal penalty matrices are combined. 
}
The anisotropic penalty in Equation~\ref{eq.penalty_ani} can be simplified in the case of an isotropic penalty depending on only one smoothing parameter $\lambda_{j} \geq 0$: 
\begin{equation}
\label{eq.penalty_iso}
\mP_{jY} = \lambda_{j} (\mP_{j} \otimes \mI_{K_Y} + \mI_{K_{j}} \otimes \mP_{Y}). 
\end{equation} 
In this simplified case only one instead of two smoothing parameters has to be estimated. 
If $\mP_{j} = \boldsymbol{0}$ in Equation~\ref{eq.penalty_iso}, this results in a penalty that only penalizes the marginal basis in $t$ direction:
\begin{equation}
\label{eq.penalty_0}
\mP_{jY} = \lambda_{j} ( \mI_{K_{j}} \otimes \mP_{Y}). 
\end{equation} 


Consider, for example, a linear effect of a functional covariate $\int_{\mathcal{S}} x_i(s) \beta(s,t)\,ds $. 
The basis vector $\mb_j(\mx_i)$ and the penalty~$\mP_j$ are the same as in Equation~\ref{eq.sof_example_fun_effect}. For the basis in~$t$ direction, we use a spline representation    
\begin{equation}
\label{eq.fof_example_fun_effect}
\mb_Y(t_{g})^\top = [ \ba_1(t_{g}) \; \cdots \; \ba_{K_Y}(t_{g}) ]
\end{equation}
with spline functions $\ba_k$, $k=1,\ldots, K_Y$ and the penalty matrix $\mP_Y$ has to be chosen such that it is suitable for the chosen spline basis. Using P-splines again, $\ba_k$ are B-splines and $\mP_Y$ is a squared difference matrix \citep{eilers1996}.  
The complete basis is
\begin{equation*}
\mb_j(\mx_i)^\top  \otimes \mb_Y(t_{g})^\top = 
\left[ \int_{\mathcal{S}}  x_i(s) \ba_1(s)\, ds \; \cdots\; \int_{\mathcal{S}} x_i(s) \ba_{K_j}(s)\, ds \right] 
 \otimes
\big[ \ba_1(t_{g}) \; \cdots \; \ba_{K_Y}(t_{g}) \big].
\end{equation*}
This choice expands $\beta(s,t)$ in a tensor-product spline basis and approximates the integral using numerical integration. 
\textcolor{myc}{
For this effect, the penalty matrix from Equation~\ref{eq.penalty_ani} ensures smoothness of $\beta(s,t)$ in s- and in t-direction. 
}



\subsubsection{Case study: Emotion components data with EEG and EMG}

The emotion components data set is based on a study of \citet{Gentsch.2014}, in which brain activity (EEG) as well as facial muscle activity (EMG) was simultaneously recorded during a computerised game. 
As the facial muscle activity should be traceable to the brain activity for a certain game situation, \citet{ruegamer2018} analyzed the synchronization of EEG and EMG signal using function-on-function regression models with factor-specific historical effects.
During the gambling rounds, three binary game conditions were varied, resulting in a total of $8$ different study settings:
\begin{itemize}
\item the goal conduciveness (\code{game\_outcome}) corresponding to the monetary outcome (\code{gain} or \code{loss}) at the end of each game round, 
\item the \code{power} setting, which determined whether the player was able or not able to change the final outcome in her favor (\code{high} or \code{low}, respectively) and,
\item the \code{control} setting, which was manipulated to change the participant's subjective feeling about her ability to cope with the game outcome. The player was told to frequently have high power in rounds with \code{high} control and have frequently low power in \code{low} control situations.
\end{itemize}
We focus on the EMG of the frontalis muscle, which is used to raise the eyebrow. The EMG signal is a functional response $Y(t)$, with $t \in \mathcal{T} = [0, 1560]$ ms, which is measured at a frequency of $256Hz$ resulting in $384$ equidistant observed time points given by the vector \texttt{t}. 
The experimental conditions are scalar covariates. The EEG signal $x_\EEG(s)$ is observed over the same time interval as the EMG signal. We use the EEG signal from the Fz electrode, which is in the center front of the head. 

In the following, we consider an aggregated version of the data, in which the EEG and EMG signals are aggregated per subject and game condition. One participant is excluded, yielding $N =  23$ subjects.

\begin{knitrout}
\definecolor{shadecolor}{rgb}{1, 1, 1}\color{fgcolor}\begin{kframe}
\begin{alltt}
\hlstd{R> }\hlkwd{data}\hlstd{(}\hlstr{"emotion"}\hlstd{,} \hlkwc{package} \hlstd{=} \hlstr{"FDboost"}\hlstd{)}
\hlstd{R> }\hlkwd{str}\hlstd{(emotion)}
\end{alltt}
\begin{verbatim}
List of 8
 $ power       : Factor w/ 2 levels "high","low": 1 1 2 2 1 1 2 2 1 1 ...
 $ game_outcome: Factor w/ 2 levels "gain","loss": 1 2 1 2 1 2 1 2 1 2 ...
 $ control     : Factor w/ 2 levels "high","low": 1 1 1 1 2 2 2 2 1 1 ...
 $ subject     : Factor w/ 23 levels "1","2","3","4",..: 1 1 1 1 1 1 1 1 2 2 ...
 $ EEG         : num [1:184, 1:384] -0.14 0.303 -0.715 0.7 0.11 ...
 $ EMG         : num [1:184, 1:384] -2.56 -4.06 -1.15 4.11 8.09 ...
 $ s           : int [1:384] 1 2 3 4 5 6 7 8 9 10 ...
 $ t           : int [1:384] 1 2 3 4 5 6 7 8 9 10 ...
\end{verbatim}
\end{kframe}
\end{knitrout}

%
In order to fit simple and meaningful models for function-on-function regression, we define a subset of the data that contains only the observations for a certain game condition. We use the game condition with high control, gain and low power: 

\begin{knitrout}
\definecolor{shadecolor}{rgb}{1, 1, 1}\color{fgcolor}\begin{kframe}
\begin{alltt}
\hlstd{R> }\hlstd{subset} \hlkwb{<-} \hlstd{emotion}\hlopt{$}\hlstd{control} \hlopt{==} \hlstr{"high"} \hlopt{&}
\hlstd{+ }  \hlstd{emotion}\hlopt{$}\hlstd{game_outcome} \hlopt{==} \hlstr{"gain"} \hlopt{&}
\hlstd{+ }  \hlstd{emotion}\hlopt{$}\hlstd{power} \hlopt{==} \hlstr{"low"}
\hlstd{R> }\hlstd{emotionHGL} \hlkwb{<-} \hlkwd{list}\hlstd{()}
\hlstd{R> }\hlstd{emotionHGL}\hlopt{$}\hlstd{subject} \hlkwb{<-} \hlstd{emotion}\hlopt{$}\hlstd{subject[subset]}
\hlstd{R> }\hlstd{emotionHGL}\hlopt{$}\hlstd{EMG} \hlkwb{<-} \hlstd{emotion}\hlopt{$}\hlstd{EMG[subset,]}
\hlstd{R> }\hlstd{emotionHGL}\hlopt{$}\hlstd{EEG} \hlkwb{<-} \hlstd{emotion}\hlopt{$}\hlstd{EEG[subset,]}
\hlstd{R> }\hlstd{emotionHGL}\hlopt{$}\hlstd{s} \hlkwb{<-} \hlstd{emotionHGL}\hlopt{$}\hlstd{t} \hlkwb{<-} \hlstd{emotion}\hlopt{$}\hlstd{t}
\end{alltt}
\end{kframe}
\end{knitrout}

In Figure~\ref{fig:explData} the EEG and EMG signal is depicted for each of the 23 participants and the 384 observation points.

\begin{knitrout}
\definecolor{shadecolor}{rgb}{1, 1, 1}\color{fgcolor}\begin{figure}[ht]

{\centering \includegraphics[width=\maxwidth]{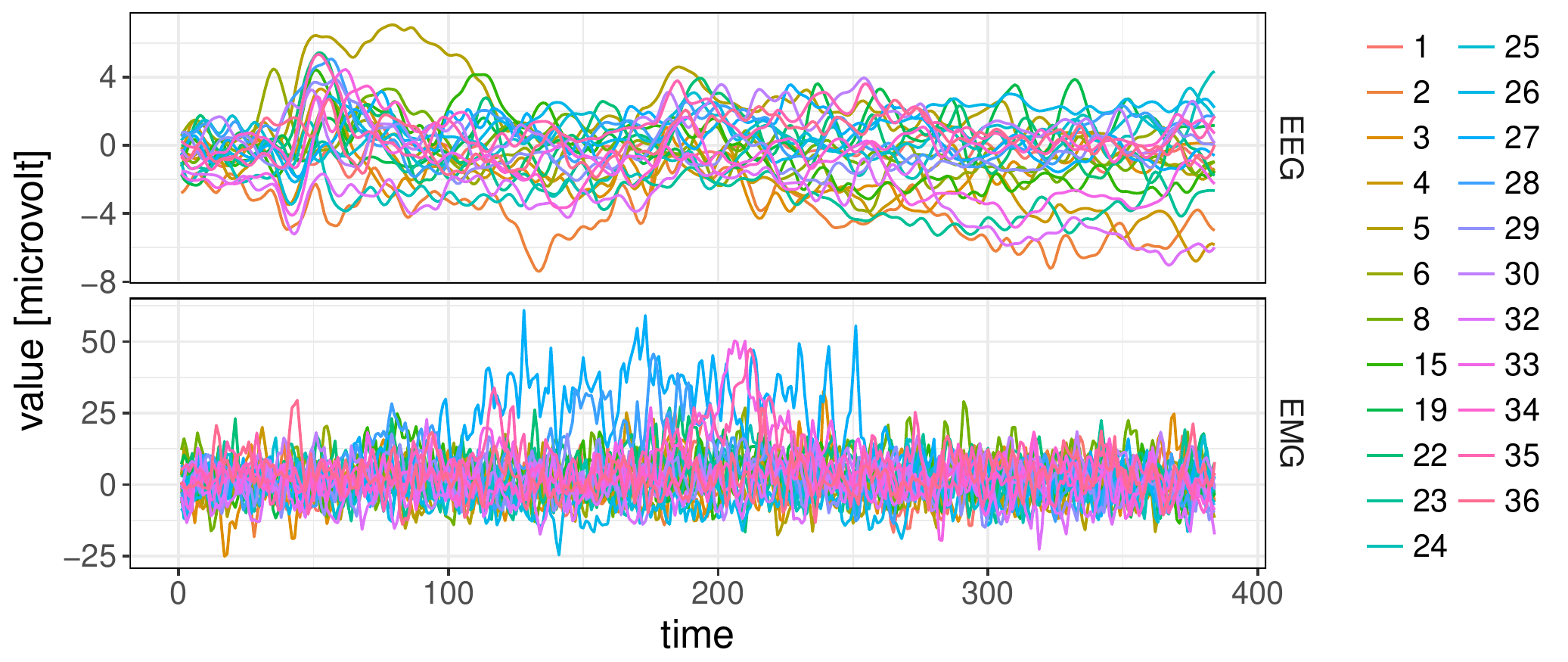} 

}

\caption{EEG signal (Fz electrode) and EMG signal (frontalis muscle) for each of the 23 participants (line colours) and the chosen game condition.}\label{fig:explData}
\end{figure}

\end{knitrout}

\hfill $\blacklozenge$

\section{Estimation by gradient boosting}
\label{sec.boosting}

Initially, boosting was proposed as a technique to 
iteratively improve the predictive performance of simple models or 
\textit{base-learners} \citep{ridgeway1999}. 
Boosting was soon recognized as a model fitting technique for statistical applications. 
Based on the 
idea of \citet{friedman2001}, \citet{buhlmann2007} proposed the model-based 
boosting framework, which allows for a component-wise fitting of 
additive terms in the linear predictor and can handle complex additive effects. 
Many boosting algorithms, which are purely used for 
prediction, fit a rather simple model using all covariates. 
In contrast, in model-based boosting it is possible 
to define the effects of each covariate separately in different base-learners. 
By iteratively selecting 
only one base-learner at a time, model-based boosting performs variable selection 
as base-learners that are never selected for the model update are excluded from the model. 
This framework is implemented in the \texttt{mboost} package. 
In contrast to other implementations of gradient boosting, 
such as \texttt{gbm} \citep{gbm2017}, the focus of  
model-based boosting lies in estimating an interpretable additive structure rather
than aiming at optimal predictive performance. 
 
Component-wise gradient boosting minimizes the expected loss (risk) 
via gradient descent in a step-wise procedure.
In each boosting step, each base-learner is fitted separately to the 
negative gradient and only the best fitting base-learner is selected for 
the model update; hence the term 'component-wise'.  
To fit a model for the expectation, like the models in 
Equation~\ref{eq.sof} and~\ref{eq.fof}, the squared error 
loss ($L_2$ loss) is minimized. In this case, the negative 
gradient corresponds to the residuals.  

\textcolor{myc}{Resulting estimation and prediction performance of boosting
depend on different tuning parameters, 
namely the \textit{number of boosting iterations} $m_{\text{stop}}$, 
the \textit{step-length} $\nu$, and the specification of the base-learners,  
e.g., whether a continuous covariate has a linear or smooth effect and 
the set-up of spline functions and penalties for smooth effects. 
We will give guidance on the choice of these 
parameters in the following by briefly describing the functionality of the algorithm.}

\textcolor{myc}{
The most important tuning parameter of boosting is the number of boosting iterations, 
as the algorithm is usually stopped before convergence. This so-called early stopping }
leads to regularized effect estimates and therefore yields more stable predictions. 
Since some of the base-learners are never selected in the course of all iterations,
 boosting also performs variable selection. The optimal stopping iteration can be 
determined by methods like cross-validation, sub-sampling or bootstrap. For each fold, 
the empirical out-of-bag risk is computed and the stopping iteration that yields the 
lowest empirical risk is chosen. 
As resampling must be conducted on the level of independent observations, this is 
done on the level of curves for functional response.

\textcolor{myc}{In order to avoid overshooting the minimum of 
the loss function in each iteration, only a small step in the chosen 
direction is made. The length of the update is determined by the step-length $\nu$. 
Some boosting frameworks adapt the choice of the step-length in each iteration. 
\cite{buhlmann2007} show that the estimation performance is barely affected by 
setting $\nu$ to a fixed and sufficiently small value for all iterations. 
They there propose to use a fixed step-length in the range $0.01$ to $0.1$. 
The appropriate size of the step-length depends on the loss that is minimized.  
In practice, the default value $\nu = 0.1$ works well for 
most applications when the model is specified using the $L_2$-loss. 
A smaller step-length than $0.01$ 
is sometimes needed for loss functions, which result in discontinuous gradients, 
such as the check-function for quantile regression \citep{fenske2011} or for loss functions, 
which can result in infinite pseudo-residuals, such as the Poisson likelihood loss. 
Since 
base-learner-specific tuning parameter are fixed for all iterations, the 
model fit is determined by the number of iterations for a given 
step-length.}


By representing all base-learners as linear effects of covariates (if necessary, by using a basis representation for non-linear effects), base-learners also define the covariate effects in the sense of additive regression models and can be associated with a specific hat matrix as well as a certain number of degrees of freedom. 
\textcolor{myc}{The degrees of freedom for each base-learner and other base-learner-specific 
tuning parameters have an influence on the prediction and estimation performance. 
The degrees of freedom $df_j$ for each base-learner $j=1,\ldots,J$ -- not to be confused 
with the \textit{effective degrees of freedom} for each model term in the final model -- 
determine the flexibility of each base-learner prior to the model fit. In the model-based
boosting framework each base-learner is fitted to the pseudo-residuals using a 
(penalized) least squares fit with fixed smoothing parameter 
$\lambda_j$, which is determined via the pre-specified degrees of freedom.  
Whereas defining a fixed smoothness for each model term prior to the model fit 
might seem restrictive at first sight, the final smoothness of each model term is in fact 
determined through the number of iterations in which the respective base-learner is chosen. 
The \textit{effective degrees of freedom} 
for each smooth component after the model fit are cumulated over the iterations where 
the model term is selected and typically differ from 
the initially specified $df_j$. The model fit can thus adapt even to relatively complex 
functions by repeatedly selecting and updating a particular model term \citep[cf.][]{brockhaus2015}. 
Determining the smoothness through the number of 
iterations works well in practice and allows for a closed-form solution of 
the penalized least squares fit in each update. As boosting chooses base-learners 
in a greedy manner, selection in each step is biased towards more flexible base-learners 
with higher degrees of freedom, if base-learners exhibit different degrees of freedom. 
This is due to the fact that these base-learner more likely yield larger improvements 
of the fit in each iteration \citep[see][for details]{hofner2011}.
For parameter estimation quality, it is essential 
to facilitate a fair base-learner selection in each step \citep{hofner2011}. 
It is recommended to set $df_j$ to an equal and rather small number 
for all base-learners $j=1,\ldots,J$ \citep{kneib2009,hofner2011}. In the case of 
scalar-on-function regression, fulfilling this constraint is not straightforward  
as functional covariates must usually be incorporated with more than one degree 
of freedom whereas scalar linear effects are restricted to have one degree of 
freedom. In order to maintain a fair base-learner selection, more complex effects 
can be orthogonalized such that they represent deviations from 
less complex effects. For example, a smooth effect can be centered around its
linear effect, thereby allowing both terms to have one degree of freedom. In Section~4.3 
as well as in Appendix~E different examples demonstrate how to facilitate a fair selection 
in this respect.}

\textcolor{myc}{Due to the nature of the algorithm, other base-learner-specific tuning parameters 
are also defined prior to the model fit and kept fixed over the iterations. 
The number of knots is of primary interest for functional or smooth predictors 
and should be chosen considering as a trade-off between 
computing time and flexibility of each base-learner. Per default, 10 knots are used, which  
can be rather large for some applications, but allows for a large flexibility of the estimated effects. 
The number of knots can be decreased if computing time is a concern. Moreover, due to the 
smoothness penalty, with the default penalizing deviations from linearity for smooth functions, 
users need not to be concerned about overfitting when increasing the number of knots.}

\subsubsection{Functional Response}
To adapt boosting for a functional response, we compute the loss at each point $t$ and integrate it over the domain of the response $\mathcal{T}$ \citep{brockhaus2015}.

For the $L_2$ loss the optimization problem for functional response aims at minimizing  
\begin{equation}
\label{eq.boosting_risk}
\sum_{i=1}^N \int \big[ y_i(t) - h(\mx_i, t) \big]^2 \,dt,
\end{equation}
which is approximated by numerical integration. 
To obtain identifiable models, suitable identifiability constraints for the base-learners are necessary and implemented. \pkg{FDboost} also contains base-learners that model the effects of functional covariates. For a discussion of both points, please see \citet{brockhaus2015}.


\section{The package FDboost} 
\label{sec.fdboost}


Fitting functional regression models via boosting is implemented in the \proglang{R}~package \pkg{FDboost}. The package uses the fitting algorithm and other infrastructure from the \proglang{R}~package \pkg{mboost} \citep{mboost_gen}. All base-learners and distribution families that are implemented in \pkg{mboost} can be used within \pkg{FDboost}. Many naming conventions and methods in \pkg{FDboost} are implemented in analogy to \pkg{mboost}.   
A tutorial for \pkg{mboost} can be found in \cite{hofner2014mboost}. 
\textcolor{myc}{We will mention all features of \pkg{mboost} that are important when working with \pkg{FDboost} in the following.}
 
The main fitting function to estimate functional regression models, like the models in Equation~\ref{eq.sof} and~\ref{eq.fof}, is called \code{FDboost()}.    
The interface of \code{FDboost()} is as follows:\footnote{Note that for the presentation of functions we restrict ourselves to the most important function arguments. For the full list of arguments, we refer to the corresponding manuals. } 

\begin{knitrout}
\definecolor{shadecolor}{rgb}{1, 1, 1}\color{fgcolor}\begin{kframe}
\begin{alltt}
\hlstd{R> }\hlkwd{FDboost}\hlstd{(formula, timeformula,} \hlkwc{id} \hlstd{=} \hlkwa{NULL}\hlstd{,} \hlkwc{numInt} \hlstd{=} \hlstr{"equal"}\hlstd{,}
\hlstd{+ }  \hlstd{data,} \hlkwc{offset} \hlstd{=} \hlkwa{NULL}\hlstd{, ...)}
\end{alltt}
\end{kframe}
\end{knitrout}
First, we focus on the arguments that are necessary for regression models both with scalar and with functional response. 
\code{formula} specifies the base-learners for the covariate effects $\mb_j$ and \code{timeformula} specifies $\mb_Y$, which is the basis along~$t$. Per default, this basis $\mb_Y$ is the same for all effects $j=1, \ldots, J$. 
\textcolor{myc}{To specify different base-learners along $t$, it is necessary to set up the Kronecker product of two base-learners explicitly in \code{formula}. For a detailed explanation, we refer to  Appendix~\ref{app.operatorsOX}.}
The data is provided in the \code{data} argument as a \code{data.frame} or a named \code{list}. 
The \code{data}-object has to contain the response, all covariates and the evaluation points of functional variables. 
Prior to the model fit, an offset is subtracted from the response to center it. This corresponds to initializing the fit with this offset, e.g., an overall average, and leads to faster convergence and better stability of the boosting algorithm.
For mean regression, by default the offset is the smoothed point-wise mean of the response over time without taking into account covariates. This offset is part of the intercept and corresponds to an initial estimate that is then updated. 
In the dots-argument, '\code{...}', further arguments passed to \code{mboost()} and \code{mboost\_fit()} can be specified. 
The most important argument is \code{family} determining the loss- and link-function for the model fit. The default is \code{family = Gaussian()}, which minimizes the squared error loss and uses the identity as link function. Thus, per default a mean regression model for continuous response is fitted. For the duality of loss-function and the \code{family} argument, we refer to Section~\ref{sec.fdboost_family}. 
Further important arguments are \code{control}, which determines the number of boosting iterations and the step-length~$\nu$ of the boosting algorithm specified by \code{nu}. 
\textcolor{myc}{The argument \code{control} must be supplied as a call to the function 
\code{boost\_control()}. 
For example, \code{control = boost\_control(mstop = 100, nu = 0.1)} 
implies 100 boosting iterations and step-length $\nu = 0.1$, which also corresponds to the default settings. 
Note that while 100 iterations are the default chosen to avoid a computationally expensive default, 
this might not be sufficient and should be chosen appropriately for the given application.}

\textcolor{myc}{\pkg{FDboost} allows for (tensor product) spline or functional principle component bases, but user-specified base-learner allow for possible extensions \citep[see, e.g.][]{hofner2014mboost}. Although the package only provides base-learners with ridge- or $L_2$-type penalization, model selection as facilitated by an $L_1$-penalty is achieved by early stopping of the algorithm. The covariance of final effects results from the additive fit with Kronecker separable penalty structure. Dependent functions can be modelled by including regularized cluster-specific functional intercepts or smooth temporal / spatial effects.}

\subsubsection{Specification for scalar response}
For scalar response, we set \code{timeformula = NULL} as no expansion of the effects in $t$ direction is necessary. \code{formula} specifies the base-learners for the covariates effects $\mb_j$ as in Equation~\ref{eq.sof_eff}. 
The arguments \code{id} and \code{numInt} are only needed for functional responses. 
For scalar response, \code{offset = NULL} results in a default offset, as, for example, the overall mean for mean regression.

\subsubsection{Arguments needed for functional response}
For functional response, the set-up of the covariate effects generally follows Equation~\ref{eq.effect_o} by separating the effects into two marginal parts.  
The marginal effects $\mb_j$, $j=1,\ldots, J$, are represented in the \code{formula} as \code{y ~ b\_1 + b\_2 + \ldots + b\_J}. The marginal effect $\mb_Y$ is represented in the \code{timeformula}, which has the form \code{~ b\_Y}. The base-learners for the marginal effects also contain suitable penalty matrices.   
Internally, the base-learners specified in \code{formula} are combined with the base-learner specified in \code{timeformula} as in Equation~\ref{eq.effect_o} and a suitable penalty matrix is constructed according to Equation~\ref{eq.penalty_iso}. 
Per default, the response is expected to be a matrix. In this case \code{id = NULL}. 
The matrix representation is not possible for a response which is observed on curve specific grids. In this case the response is provided as vector in long format and \code{id} specifies which position in the vector is attributed to which curve; see section~\ref{sec.fdboost_fof} for details.  
The argument \code{numInt} provides the numerical integration scheme for computing the integral of the loss over $\mathcal{T}$ in Equation~\ref{eq.boosting_risk}. Per default, \code{numInt = "equal"}, and thus all integration weights are set to one; for \code{numInt = "Riemann"} Riemann sums are used. 
For functional response, \code{offset = NULL} induces a smooth offset varying over~$t$. For \code{offset = "scalar"}, a scalar offset is computed. This corresponds to an offset that is constant along $t$.   
For more details and the full list of arguments, see the manual of \code{FDboost()}. 


\subsection{Scalar response and functional covariates}
\label{sec.fdboost_sof}
In this subsection, we give details on models with scalar response and functional covariates like the model in Equation~\ref{eq.sof}. Such models are called scalar-on-function regression models. As case study the data on fossil fuels is used. 

\subsubsection{Potential covariate effects: base-learners}
\label{subsec.sof_bl} 
In order to fit a scalar-on-function model as in Equation~\ref{eq.sof}, the \code{timeformula} is set to \code{NULL} and potential covariate effects $h_j(\mx_i)$ are specified in the \code{formula} argument. 
The effects of scalar covariates can be linear or non-linear. 
A linear effect \textcolor{myc}{$z \beta$ for the covariate $z$} is obtained using the base-learner \code{bols(z)}, which is also suitable for factor variables, in which case dummy variables are constructed for each factor level \citep{hofner2014mboost}. Per default, \code{bols()} contains an intercept. If the specified degrees of freedom are less than the number of columns in the design matrix, \code{bols()} penalizes the linear effect by a ridge penalty with the identity matrix as penalty matrix. 
The base-learner \code{brandom()} for factor variables sets up an effect, which is centered around zero and is penalized by a ridge penalty, having similar properties to a random effect, 
\textcolor{myc}{but no underlying distributional assumption. It is not possible to estimate random effects in the classical sense that they are estimated using variance parameters.}  
See the web appendix of \citet{kneib2009} for a discussion on \code{brandom()}. \textcolor{myc}{The ridge penalized effects, however, have a similar interpretation as random effects as a quadratic penalty is mathematically equivalent to a Gaussian prior. Note that this also allows for other types of random effects such as cluster-specific random effect functions.}  
A non-linear effect expanded by P-splines is obtained by the base-learner \code{bbs()}. 
\textcolor{myc}{Within \code{bbs()}, the argument \code{knots} determines the number of knots of the P-spline basis, 
\code{degree} specifies the degree of the spline basis and 
\code{differences} the order of the differences in the penalty matrix. 
Per default, cubic B-splines on 20 knots with a second order difference penalty are used. }
For more details on base-learners with scalar covariates, we refer to \cite{hofner2014mboost}. 

Potential base-learners for functional covariates can be seen in Table~\ref{tab.sof_FDboost_effects}. In this table exemplary linear predictors are listed in the left column. In the right column, the corresponding call to \code{formula} is given. Because of the scalar response, the call to \code{timeformula} is set to \code{NULL}.
For simplicity, only one possible parameterization which leads to simple interpretations and one corresponding model call are shown, although \pkg{FDboost} allows to specify several parameterizations.   
\renewcommand\arraystretch{1.6}
\begin{table}[ht]\centering
\begin{small}
\begin{tabular}{p{.35\textwidth}| p{.6\textwidth} }
additive predictor $h(\mx) = \sum_j h_j(\mx)$ & call  \\ 
\hline  
 $\beta_0 + \int_{\mathcal{S}} x(s) \beta_1(s)\,ds $ & 	\code{y ~ 1 + bsignal(x, s = s)} \\ 
 & \code{y ~ 1 + bfpc(x, s = s)} \\ 
 $\beta_0 + z \beta_1 + \int_{\mathcal{S}} x(s) \beta_2(s)\,ds$ &  \code{y ~ 1 + bolsc(z) + bsignal(x, s = s)} \\
 \qquad $ + z \int_{\mathcal{S}} x(s) \beta_3(s)\,ds $ &  \qquad \code{ + bsignal(x, s = s) \%X\% bolsc(z)}\\ 
\end{tabular}
\end{small}
\caption{Additive predictors for scalar-on-function regression models.}
\label{tab.sof_FDboost_effects} 
\end{table}
\renewcommand\arraystretch{1}

For a linear effect of a functional covariate $\int_{\mathcal{S}} x(s) \beta_1(s)\,ds $, two base-learners exist that use different basis expansions. 
Assuming $\beta_1(s)$ to be smooth, \code{bsignal()} uses a P-spline representation for the expansion of $\beta_1(s)$. \textcolor{myc}{In this case, the observations $x(s)$ are used directly without any basis representation.}  
Assuming that the main modes of variation in the functional covariate are the important directions for the coefficient function $\beta_1(s)$, a representation with functional principal components is suitable \citep{ramsay2005}. 
In the base-learner \code{bfpc()}, the coefficient function $\beta_1(s)$ and the functional covariate $x(s)$ are both represented by an expansion in the estimated functional principal components of $x(s)$. As penalty matrix, the identity matrix is used.  
In Appendix~\ref{app.funbl}, technical details on the representation of functional effects are given. 


The specification of a model with an interaction term between a scalar and a functional covariate is given at the end of Table~\ref{tab.sof_FDboost_effects}. The interaction term is centered around the main effect of the functional covariate using \code{bolsc} for the scalar covariate (as is the linear effect of the scalar covariate around the intercept). Thus, the main effect of the functional covariate has to be included in the model. For more details on interaction effects, we refer to \cite{brockhaus2015} and \cite{ruegamer2018}. 
The interaction is formed using the operator \code{\%X\%} that builds the row-wise tensor product of the two marginal bases, see Appendix~\ref{app.operatorsOX}.  


As explained in Section~\ref{sec.boosting}, all base-learners in a model should have equal and rather low degrees of freedom. 
The number of degrees of freedom that can be given to a base-learner is restricted. On the one hand, the maximum number is bounded by the number of columns of the design matrix (more precisely by the rank of the design matrix). On the other hand, for rank-deficient penalties, the minimum number of degrees of freedom is given by the rank of the null space of the penalty matrix.

The interface of \code{bsignal()} is as follows:
\begin{knitrout}
\definecolor{shadecolor}{rgb}{1, 1, 1}\color{fgcolor}\begin{kframe}
\begin{alltt}
\hlstd{R> }\hlkwd{bsignal}\hlstd{(x, s,} \hlkwc{knots} \hlstd{=} \hlnum{10}\hlstd{,} \hlkwc{degree} \hlstd{=} \hlnum{3}\hlstd{,} \hlkwc{differences} \hlstd{=} \hlnum{1}\hlstd{,}
\hlstd{+ }  \hlkwc{df} \hlstd{=} \hlnum{4}\hlstd{,} \hlkwc{lambda} \hlstd{=} \hlkwa{NULL}\hlstd{,} \hlkwc{check.ident} \hlstd{=} \hlnum{FALSE}\hlstd{)}
\end{alltt}
\end{kframe}
\end{knitrout}
The arguments \code{x} and \code{s} specify the name of the functional covariate and the name of its argument. \code{knots} gives the number of inner knots for the P-spline basis, \code{degree} the degree of the B-splines and \code{differences} the order of the differences that are used for the penalty. Thus, per default, 14 cubic P-splines with first order difference penalty are used. The argument \code{df} specifies the number of degrees of freedom for the effect and \code{lambda} the smoothing parameter. Only one of those two arguments can be supplied. If \code{check.ident = TRUE} identifiability checks proposed by \cite{scheipl2016} for functional linear effects are additionally performed. 

The interface of \code{bfpc()} is:
\begin{knitrout}
\definecolor{shadecolor}{rgb}{1, 1, 1}\color{fgcolor}\begin{kframe}
\begin{alltt}
\hlstd{R> }\hlkwd{bfpc}\hlstd{(x, s,} \hlkwc{df} \hlstd{=} \hlnum{4}\hlstd{,} \hlkwc{lambda} \hlstd{=} \hlkwa{NULL}\hlstd{,} \hlkwc{pve} \hlstd{=} \hlnum{0.99}\hlstd{,} \hlkwc{npc} \hlstd{=} \hlkwa{NULL}\hlstd{)}
\end{alltt}
\end{kframe}
\end{knitrout}
The arguments \code{x, s, df} and \code{lambda} have the same meaning as in \code{bsignal()}. The two other arguments allow to control how many functional principal components are used as basis. Per default the number of functional principal components is chosen such that the proportion of the explained variance is 99\%. This proportion can be changed using the argument \code{pve} (proportion variance explained). Alternatively, the number of components can be set to a specific value using \code{npc} (number principal components). 

The interface of \code{bolsc()} is very similar to that of \code{bols()}, which is laid out in detail in \cite{hofner2014mboost}. In contrast to \code{bols()}, \code{bolsc()} centers the design matrix such that the resulting linear effect is centered around zero. More details on \code{bolsc()} are given in Section~\ref{sec.fdboost_fof}. 
\begin{knitrout}
\definecolor{shadecolor}{rgb}{1, 1, 1}\color{fgcolor}\begin{kframe}
\begin{alltt}
\hlstd{R> }\hlkwd{bolsc}\hlstd{(...,} \hlkwc{df} \hlstd{=} \hlkwa{NULL}\hlstd{,} \hlkwc{lambda} \hlstd{=} \hlnum{0}\hlstd{,} \hlkwc{K} \hlstd{=} \hlkwa{NULL}\hlstd{)}
\end{alltt}
\end{kframe}
\end{knitrout}
In the dots argument, \code{...}, one or more covariates can be specified. For factor variables \code{bolsc()} sets up a design matrix in dummy-coding. 
The arguments \code{df} and \code{lambda} have the same meaning as above. If \code{lambda} > 0 or \code{df} < the number of columns of the design matrix a ridge-penalty is applied. Per default, \code{K = NULL}, the penalty matrix is the identity matrix. Setting the argument \code{K} to another matrix allows for customized penalty matrices.

\subsubsection*{Case study (ctd.): Fossil fuel data}
For the heat values $Y_i$, $i=1,\ldots,129$, we fit the model 
\begin{equation}
\label{eq.modSOF}
 \EV(Y| \mx)  
 = \beta_0 + f(z_\ho) + \int_{\mathcal{S}_\NIR} x_\NIR (s_\NIR)  \beta_\NIR(s_\NIR) \,ds_\NIR + \int_{\mathcal{S}_\UVVIS} x_\UVVIS (s_\UVVIS)  \beta_\UVVIS(s_\UVVIS) \,ds_\UVVIS, 
\end{equation}
with water content $z_\ho$ and centered spectral curves $x_\NIR$ and $x_\UVVIS$, which are observed over the wavelengths $s_\NIR \in \mathcal{S}_\NIR$ and $s_\UVVIS \in \mathcal{S}_\UVVIS$. 
We center the NIR and the UVVIS measurement per wavelength such that  \textcolor{myc}{$\sum_{i=1}^N  x_{\NIR,i} (s_\NIR) = 0 \, \forall s_\NIR$} and analogously for UVVIS. 
Thus, the functional effects have mean zero, $\sum_{i=1}^N \int_{\mathcal{S}_\NIR} x_{\NIR,i} (s_\NIR) \beta(s_\NIR) \,d s_\NIR= 0$ and analogously for UVVIS. 
This does not affect the interpretation of $\beta_\NIR(s_\NIR)$ and $\beta_\UVVIS(s_\UVVIS)$, it only changes the interpretation of the intercept of the regression model.  
If all effects are centered, the intercept can be interpreted as overall mean and the other effects as deviations from the overall mean. 

Note that the functional covariates have to be supplied as $<$number of curves$>$ by $<$number of evaluation points$>$ matrices.  
The non-linear effect of the scalar variable H2O is specified using the \code{bbs()} base-learner. 
For the linear functional effect of NIR and UVVIS, we use the base-learner \code{bsignal()}. 
The degrees of freedom are set to 4 for each base-learner. For the functional effects, we use a P-spline basis with 20 inner knots. 
Because of the scalar response \code{timeformula = NULL}.  

\begin{knitrout}
\definecolor{shadecolor}{rgb}{1, 1, 1}\color{fgcolor}\begin{kframe}
\begin{alltt}
\hlstd{R> }\hlstd{fuelSubset}\hlopt{$}\hlstd{UVVIS} \hlkwb{<-} \hlkwd{scale}\hlstd{(fuelSubset}\hlopt{$}\hlstd{UVVIS,} \hlkwc{scale} \hlstd{=} \hlnum{FALSE}\hlstd{)}
\hlstd{R> }\hlstd{fuelSubset}\hlopt{$}\hlstd{NIR} \hlkwb{<-} \hlkwd{scale}\hlstd{(fuelSubset}\hlopt{$}\hlstd{NIR,} \hlkwc{scale} \hlstd{=} \hlnum{FALSE}\hlstd{)}
\hlstd{R> }\hlstd{sof} \hlkwb{<-} \hlkwd{FDboost}\hlstd{(heatan} \hlopt{~} \hlkwd{bbs}\hlstd{(h2o,} \hlkwc{df} \hlstd{=} \hlnum{4}\hlstd{)}
\hlstd{+ }  \hlopt{+} \hlkwd{bsignal}\hlstd{(UVVIS,} \hlkwc{s} \hlstd{= uvvis.lambda,} \hlkwc{knots} \hlstd{=} \hlnum{20}\hlstd{,} \hlkwc{df} \hlstd{=} \hlnum{4}\hlstd{)}
\hlstd{+ }  \hlopt{+} \hlkwd{bsignal}\hlstd{(NIR,} \hlkwc{s} \hlstd{= nir.lambda,} \hlkwc{knots} \hlstd{=} \hlnum{20}\hlstd{,} \hlkwc{df} \hlstd{=} \hlnum{4}\hlstd{),}
\hlstd{+ }  \hlkwc{timeformula} \hlstd{=} \hlkwa{NULL}\hlstd{,} \hlkwc{data} \hlstd{= fuelSubset)}
\end{alltt}
\end{kframe}
\end{knitrout}

\hfill $\blacklozenge$

\subsubsection{Model tuning and early stopping}
\label{subsec.sof_tune}
Boosting iteratively selects base-learners to update the additive predictor. 
Fixing the base-learners and the step-length, the model complexity is controlled by the number of boosting iterations. With more boosting iterations the model becomes more complex \citep{buhlmann2003}.  
The step-length $\nu$ is chosen sufficiently small in the interval $(0,1]$, usually as $\nu = 0.1$, which is also the default. For smaller step-length, more boosting iterations are required and vice versa \citep{friedman2001}. 
Note that the default number of boosting iterations is 100. This is arbitrary and in most cases not adequate. 
The number of boosting iterations and the step-length of the algorithm can be specified in the argument \code{control}. This argument must be supplied as a call to \code{boost\_control()}. For example, \code{control = boost\_control(mstop = 50, nu = 0.2)} implies 50 boosting iterations and step-length $\nu = 0.2$.  

The most important tuning parameter is the number of boosting iterations. 
For regression with scalar response, the function \code{cvrisk.FDboost()} can be used to determine the optimal stopping iteration. This function directly calls \code{cvrisk.mboost()} from the \pkg{mboost} package, which performs an empirical risk estimation using a specified resampling method. The interface of \code{cvrisk.FDboost()} is: 

\begin{knitrout}
\definecolor{shadecolor}{rgb}{1, 1, 1}\color{fgcolor}\begin{kframe}
\begin{alltt}
\hlstd{R> }\hlkwd{cvrisk.FDboost}\hlstd{(object,}
\hlstd{+ }  \hlkwc{folds} \hlstd{=} \hlkwd{cvLong}\hlstd{(}\hlkwc{id} \hlstd{= object}\hlopt{$}\hlstd{id,} \hlkwc{weights} \hlstd{=} \hlkwd{model.weights}\hlstd{(object)),}
\hlstd{+ }  \hlkwc{grid} \hlstd{=} \hlnum{1}\hlopt{:}\hlkwd{mstop}\hlstd{(object))}
\end{alltt}
\end{kframe}
\end{knitrout}

In the argument \code{object}, the fitted model object is specified. 
\code{grid} defines the grid on which the optimal stopping iteration is searched. Per default the grid from 1 to the current stopping iteration of the model object is used as search grid. 
But it is also possible to specify a larger grid, e.g., \code{1:5000}.  
The argument \code{folds} expects an integer weight matrix with dimension $N \times \kappa$ (<number of observations> times <number of folds>). Depending on the range of values in the weight matrix, different types of resampling are performed. For example, if the weights sum to $N$ for each column but also have values larger than one, the resampling scheme corresponds to bootstrap while a $\kappa$-fold cross-validation is employed by using an incidence matrix, for which the rows sum to $\kappa-1$. If not manually specified, \pkg{mboost} and \pkg{FDboost} provide convenience functions -- \code{cv()} and \code{cvLong()} -- that construct such matrices on the basis of the given model object. The function \code{cvLong()} is suited for functional response and treats scalar response as the special case with one observation per curve. 
For scalar response, the function \code{cv()} from package \pkg{mboost} can be used, which has a simpler interface. 

\begin{knitrout}
\definecolor{shadecolor}{rgb}{1, 1, 1}\color{fgcolor}\begin{kframe}
\begin{alltt}
\hlstd{R> }\hlkwd{cv}\hlstd{(weights,} \hlkwc{type} \hlstd{=} \hlkwd{c}\hlstd{(}\hlstr{"bootstrap"}\hlstd{,} \hlstr{"kfold"}\hlstd{,} \hlstr{"subsampling"}\hlstd{),}
\hlstd{+ }  \hlkwc{B} \hlstd{=} \hlkwd{ifelse}\hlstd{(type} \hlopt{==} \hlstr{"kfold"}\hlstd{,} \hlnum{10}\hlstd{,} \hlnum{25}\hlstd{))}
\end{alltt}
\end{kframe}
\end{knitrout}

The argument \code{weights} is used to specify the weights of the original model, which can be extracted using \code{model.weights(object)}. Usually all model weights are one. 
Via argument \code{type} the resampling scheme is defined: \code{"bootstrap"} for non-parametric bootstrap, \code{"kfold"} for cross-validation and \code{"subsampling"} for resampling half of all observations for each fold. The number of folds is defined by~\code{B}. Per default, 10 folds are used for cross-validation and 25 folds for bootstrap as well as for subsampling.   

The function \code{cvLong()} is especially suited for functional response and has the additional argument \code{id}, which is used to specify which observations belong to the same response curve. For scalar response, \code{id = 1:N}.

\subsubsection*{Case study (ctd.): Fossil fuel data}
To tune the scalar-on-function regression model (\ref{eq.modSOF}), we search the optimal stopping iteration by 10-fold bootstrapping. 
First, the bootstrap folds are created using the function \code{cv()}. Second, for each bootstrap fold, the out-of-bag risk is computed for models with 1 to 1000 boosting iterations using the \code{cvrisk} function. The choice of the grid is independent of the number of boosting iterations of the fitted model object.  

\begin{knitrout}
\definecolor{shadecolor}{rgb}{1, 1, 1}\color{fgcolor}\begin{kframe}
\begin{alltt}
\hlstd{R> }\hlkwd{set.seed}\hlstd{(}\hlnum{123}\hlstd{)}
\hlstd{R> }\hlstd{folds_sof} \hlkwb{<-} \hlkwd{cv}\hlstd{(}\hlkwc{weights} \hlstd{=} \hlkwd{model.weights}\hlstd{(sof),} \hlkwc{type} \hlstd{=} \hlstr{"bootstrap"}\hlstd{,} \hlkwc{B} \hlstd{=} \hlnum{10}\hlstd{)}
\hlstd{R> }\hlstd{cvm_sof} \hlkwb{<-} \hlkwd{cvrisk}\hlstd{(sof,} \hlkwc{folds} \hlstd{= folds_sof,} \hlkwc{grid} \hlstd{=} \hlnum{1}\hlopt{:}\hlnum{1000}\hlstd{)}
\end{alltt}
\end{kframe}
\end{knitrout}

The object \code{cvm\_sof} contains the out-of-bag risk of each fold for all $1000$ iterations. 
\hfill $\blacklozenge$

\subsubsection{Methods to extract and visualize results from the resampling object}
\label{subsec.sof_results}
%
For a \code{cvrisk}-object as created by \code{cvrisk()}, the method \code{mstop()} extracts the estimated optimal number of boosting iterations, which corresponds to the number of boosting iterations yielding the minimal mean out-of-bag risk. 
\code{plot()} generates a plot of the estimated out-of-bag risk per stopping iteration in each fold. In addition, the mean out-of-bag risk per stopping iteration is displayed. The estimated optimal stopping iteration is marked by a dashed vertical line. 
In such a plot, the convergence behavior can be graphically examined. 

\subsubsection*{Case study (ctd.): Fossil fuel data} 
We generate a plot that displays for each fold the estimated out-of-bag risk per stopping iteration for each fold; see Figure~\ref{fig:fuel_plot_cvm}. 

\begin{knitrout}
\definecolor{shadecolor}{rgb}{1, 1, 1}\color{fgcolor}\begin{kframe}
\begin{alltt}
\hlstd{R> }\hlkwd{plot}\hlstd{(cvm_sof,} \hlkwc{ylim} \hlstd{=} \hlkwd{c}\hlstd{(}\hlnum{2}\hlstd{,} \hlnum{15}\hlstd{))}
\end{alltt}
\end{kframe}\begin{figure}[ht]

{\centering \includegraphics[width=4in]{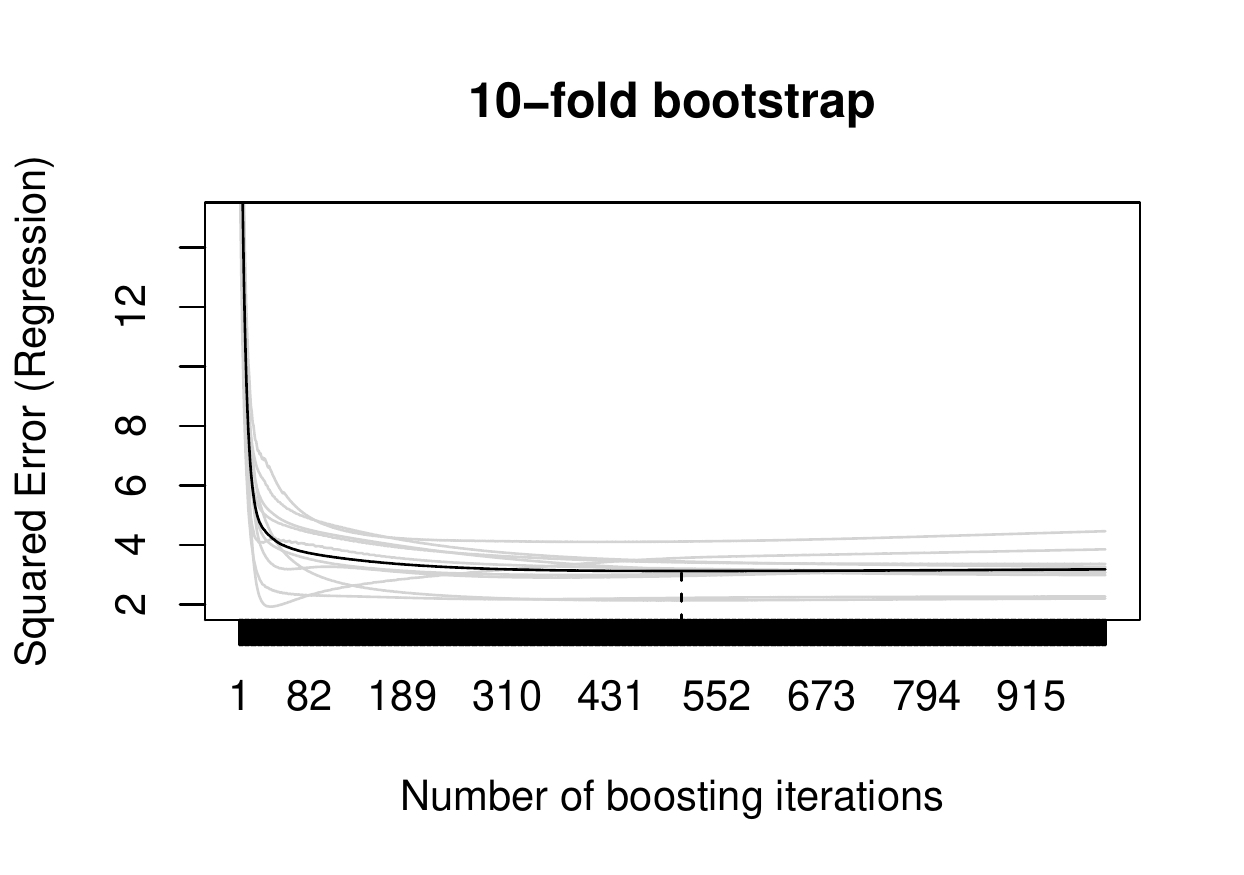} 

}

\caption{Bootstrapped out-of-bag risk for the model of the fossil fuels. For each fold, the out-of-bag risk is displayed as a gray line. The mean out-of-bag risk is visualized by a black line. The optimal number of boosting iterations is marked by a dashed vertical line.}\label{fig:fuel_plot_cvm}
\end{figure}

\end{knitrout}

For small numbers of boosting iterations, the out-of-bag risk declines sharply with a growing number of boosting iterations. With more and more iterations the model gets more complex and the out-of-bag risk starts to slowly increase.  
The dashed vertical line marks the estimated optimal stopping iteration of 511, which can be accessed using the function \code{mstop()}: 

\begin{knitrout}
\definecolor{shadecolor}{rgb}{1, 1, 1}\color{fgcolor}\begin{kframe}
\begin{alltt}
\hlstd{R> }\hlkwd{mstop}\hlstd{(cvm_sof)}
\end{alltt}
\begin{verbatim}
[1] 511
\end{verbatim}
\end{kframe}
\end{knitrout}

\hfill $\blacklozenge$

\subsubsection{Methods to extract and display results from the model object} 
Fitted \code{FDboost} objects inherit methods from class \code{mboost}. Thus, all methods available for \code{mboost} objects can also be applied to models fitted by \code{FDboost()}. 
The design and penalty matrices that are constructed by the base-learners can be extracted using the \code{extract()} function. For example, \code{extract(object, which = 1)} returns the design matrix of the first base-learner and \code{extract(object, which = 1, what = "penalty")} the corresponding penalty matrix. 
The number of boosting iterations for an \code{FDboost} object can be changed afterwards using the subset operator; e.g., \code{object[50]} sets the number of boosting iterations for \code{object} to 50. Note that the subset operator directly changes \code{object}, and hence no assignment is necessary.  

One can access the estimated coefficients by the \code{coef()} function. \textcolor{myc}{The function takes a fitted \code{object} produced by \code{FDboost()} and returns estimated coefficient functions such as $\hat\beta(s)$, $\hat\beta(s,t)$, $\hat g(x)$ or other estimated effects.} For smooth effects, \code{coef()} returns the smooth estimated effects evaluated on a regular grid. \textcolor{myc}{The resolution of the grid can be specified by the arguments \code{n1, n2} and \code{n3} for $1$-, $2$- and $3$-dimensional smooth terms, respectively, which define the number of equidistantly spaced grid points over the range of the covariate. The resulting object is a list containing an element for the offset and a named list with one entry for each further model term. The value of the offset for each observation can be accessed with \code{coef(object)\$offset\$value}. List entries for model terms in \code{coef(object)\$smterms} are, in turn, lists with different entries, in particular, including \code{\$x} (\code{\$y, \$z}) representing unique grid-points used to evaluate the coefficient function and \code{\$value} representing a vector, matrix or list of matrices with the coefficient values.} The estimated spline-coefficients $\hat{\bm{\theta}}_j$ of smooth effects can be obtained by \code{object\$coef()}, \textcolor{myc}{which is equal to setting the argument \code{raw} to \code{TRUE} in the \code{coef} function.}

The estimated effects can be graphically displayed by the \code{plot()} function. The coefficient plots can be customized by various arguments. For example, coefficient surfaces can be displayed as image plots, setting \code{pers = FALSE}, or as perspective plots, setting \code{pers = TRUE}. To plot only some of the base-learners, the argument \code{which} can be used. For instance, \code{plot(object, which = c(1,3))} plots the estimated effects of the first and the third base-learner.  
The fitted values and predictions for new data can be obtained by the methods \code{fitted()} and \code{predict()}, respectively. 

\subsubsection*{Case study (ctd.): Fossil fuel data}  
\textcolor{myc}{To better understand the penalization used in the \code{sof} model, we can exemplarily extract the marginal penalty matrix for UVVIS as follows:}
\begin{knitrout}
\definecolor{shadecolor}{rgb}{1, 1, 1}\color{fgcolor}\begin{kframe}
\begin{alltt}
\hlstd{R> }\hlstd{marg_pen} \hlkwb{<-} \hlkwd{extract}\hlstd{(sof,} \hlstr{"penalty"}\hlstd{,} \hlkwc{which} \hlstd{=} \hlnum{2}\hlstd{)}
\hlstd{R> }\hlstd{marg_pen[[}\hlnum{1}\hlstd{]][}\hlnum{1}\hlopt{:}\hlnum{5}\hlstd{,}\hlnum{1}\hlopt{:}\hlnum{5}\hlstd{]}
\end{alltt}
\begin{verbatim}
     [,1] [,2] [,3] [,4] [,5]
[1,]    1   -1    0    0    0
[2,]   -1    2   -1    0    0
[3,]    0   -1    2   -1    0
[4,]    0    0   -1    2   -1
[5,]    0    0    0   -1    2
\end{verbatim}
\end{kframe}
\end{knitrout}

In order to continue working with the optimal model, we set the number of boosting iterations to the estimated optimal value.  

\begin{knitrout}
\definecolor{shadecolor}{rgb}{1, 1, 1}\color{fgcolor}\begin{kframe}
\begin{alltt}
\hlstd{R> }\hlstd{sof} \hlkwb{<-} \hlstd{sof[}\hlkwd{mstop}\hlstd{(cvm_sof)]}
\end{alltt}
\end{kframe}
\end{knitrout}

\textcolor{myc}{We can access estimated coefficients using \code{coef()}, e.g., by extracting the estimated coefficient function $\hat\beta_\NIR(s_\NIR)$ contained in \code{\$value} evaluated at grid points \code{\$x}}

\begin{knitrout}
\definecolor{shadecolor}{rgb}{1, 1, 1}\color{fgcolor}\begin{kframe}
\begin{alltt}
\hlstd{R> }\hlstd{coef_sof} \hlkwb{<-} \hlkwd{coef}\hlstd{(sof)}
\hlstd{R> }\hlkwd{str}\hlstd{(coef_sof}\hlopt{$}\hlstd{smterms}\hlopt{$}\hlstd{`bsignal(NIR)`)}
\end{alltt}
\end{kframe}
\end{knitrout}

\textcolor{myc}{To display the estimated effects, \code{plot()} can be called on the fitted FDboost object. }

Per default, \code{plot()} only displays effects of base-learners that were selected at least once. See Figure~\ref{fig:fuel_results} for the resulting plots. 

\begin{knitrout}
\definecolor{shadecolor}{rgb}{1, 1, 1}\color{fgcolor}\begin{kframe}
\begin{alltt}
\hlstd{R> }\hlkwd{par}\hlstd{(}\hlkwc{mfrow} \hlstd{=} \hlkwd{c}\hlstd{(}\hlnum{1}\hlstd{,}\hlnum{3}\hlstd{))}
\hlstd{R> }\hlkwd{plot}\hlstd{(sof,} \hlkwc{ask} \hlstd{=} \hlnum{FALSE}\hlstd{,} \hlkwc{ylab} \hlstd{=} \hlstr{""}\hlstd{)}
\end{alltt}
\end{kframe}\begin{figure}[ht]

{\centering \includegraphics[width=6in]{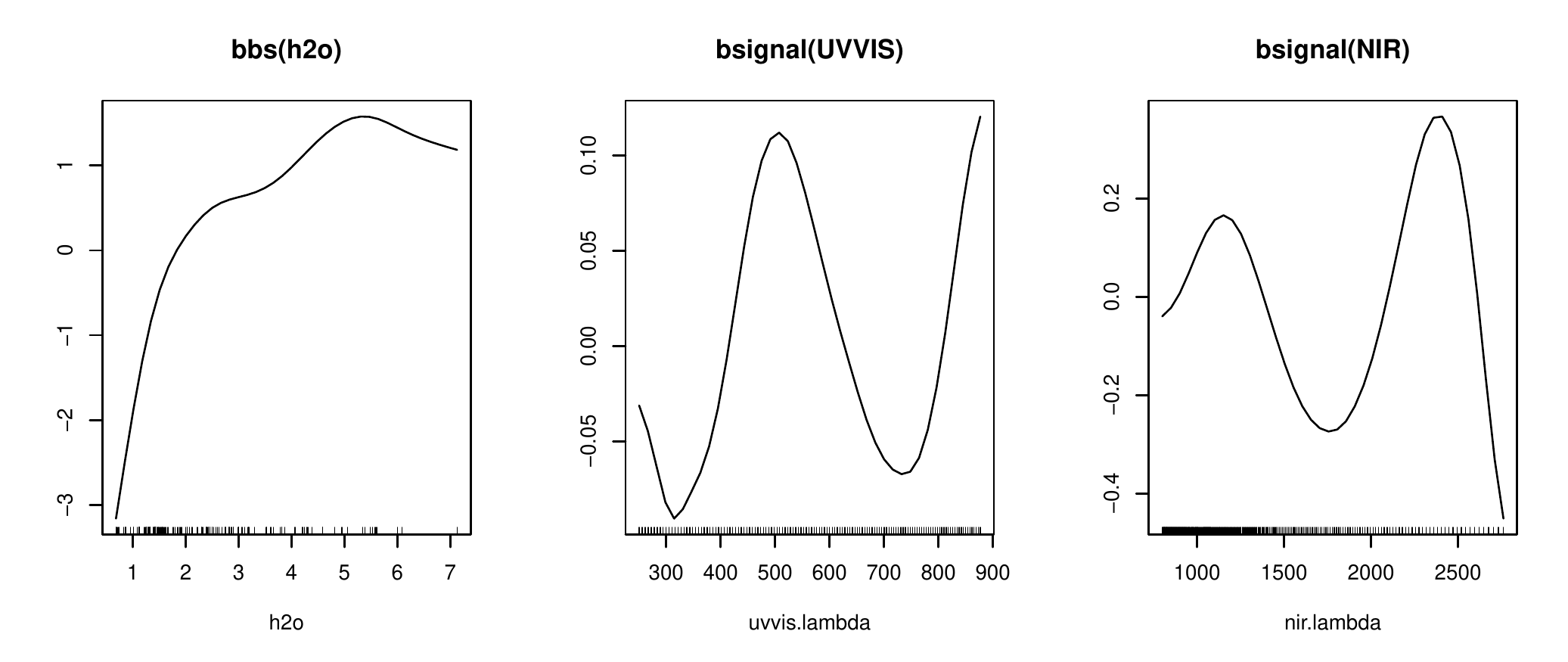} 

}

\caption{Coefficient estimates of the model for the heat value of the fossil fuels with optimal number of boosting iterations. The smooth effect of the water content (left), the linear effect of the UVVIS spectrum (center) and the NIR spectrum (right) are displayed.}\label{fig:fuel_results}
\end{figure}

\end{knitrout}

The mean heat value is estimated to be higher for higher water content and lower for lower water content (see Figure~\ref{fig:fuel_results} left). High values of the UVVIS spectrum at a wavelength of around 500 and 850 nm are associated with higher heat values. Higher values of the UVVIS spectrum at wavelength around 300 and 750 nm are associated with lower heat values (see Figure~\ref{fig:fuel_results} middle). The effect of the NIR spectrum can be interpreted analogously. 
\hfill $\blacklozenge$

\subsubsection{Bootstrapped coefficient estimates} 
In order to get a measure for the uncertainty associated with the estimated coefficient functions, one can employ nested bootstrap. The optimal number of boosting iterations in each bootstrap fold, in turn, is estimated by an inner resampling procedure. 
The bootstrapped coefficients are shrunken towards zero as boosting shrinks coefficients towards zero due to early stopping. Thus, the resulting bootstrap ``confidence'' interval is biased towards zero but still captures the variability of the coefficient estimates. \textcolor{myc}{While they do not have proper coverage properties due to shrinkage bias, these bootstrap intervals capture all the sources of uncertainty (induced by the resampling, the model selection as well as the actual uncertainty of coefficients). They may be used to check, e.g., for the existence of certain effects by examining whether the resulting intervals contain the value zero, which was found to work well in \citet{ruegamer2018}. Having no formal inference procedure clearly is a limitation of the model-based boosting framework in general and users who want to formally test pre-specified hypotheses are referred to alternative software packages such as \texttt{refund} \citep{refund_gen} for cases where these are applicable and the particular strengths of model-based boosting (high-dimensional data and models, model selection, general loss-functions) are not needed.} In \pkg{FDboost} the function \code{bootstrapCI()} can be used to conveniently compute bootstrapped coefficients: 

\begin{knitrout}
\definecolor{shadecolor}{rgb}{1, 1, 1}\color{fgcolor}\begin{kframe}
\begin{alltt}
\hlstd{R> }\hlkwd{bootstrapCI}\hlstd{(object,} \hlkwc{B_outer} \hlstd{=} \hlnum{100}\hlstd{,} \hlkwc{B_inner} \hlstd{=} \hlnum{25}\hlstd{, ...)}
\end{alltt}
\end{kframe}
\end{knitrout}

The argument \code{object} is the fitted model object. The maximal number of boosting iterations for each bootstrap fold is the number of boosting iterations of the model-object. 
Per default bootstrap is used with \code{B\_outer = 100} outer folds and \code{B\_inner = 25} inner folds. The dots argument, \code{...} can be used to pass further arguments to \code{applyFolds()}, which is used for the outer bootstrap. 
In particular, setting the argument \code{mc.cores} to an integer greater $1$ will run the outer bootstrap in parallel on the number of cores that are specified via \code{mc.cores} (this does not work under Windows, as the parallelization is based on the function \code{mclapply()}). \textcolor{myc}{As for the resampling scheme, which determines the number of iterations, the bootstrap which is done to quantify uncertainty of coefficient estimates should be conducted on the level of independent observations. This is particularly relevant for functional responses, where both resampling procedures should be done on the level of curves. Additional dependence in the data, such as observations sampled from clusters or in a longitudinal fashion, should also be taken into account for scalar-on-function models. To this end, observations should be sampled on the levels of clusters, subjects, or in nested designs, by a nested sampling for each of the levels. This yields a limitation of our method in cases, in which observations can not be separated into independent units (e.g., for spatially correlated observations with a strong dependence among all observations). However, costumized solutions such as a block-wise bootstrap \citep[cf.][]{brockhaus2016} for time-series data can be employed as in the scalar case.}

\subsubsection*{Case study (ctd.): Fossil fuel data}  
We recompute the model on 100 bootstrap samples to compute bootstrapped coefficient estimates. In each bootstrap fold the optimal number of boosting iterations is estimated by an inner bootstrap with 10 folds. 
In contrast to other methods and analytic inference concepts, employing bootstrap for coefficient uncertainty is much more time consuming but can be easily parallelized. See the help page of \code{bootstrapCI()} for example code. 
The resulting estimated coefficients can be seen in Figure~\ref{fig:sof_boot}. 

\begin{knitrout}
\definecolor{shadecolor}{rgb}{1, 1, 1}\color{fgcolor}\begin{kframe}
\begin{alltt}
\hlstd{R> }\hlkwd{set.seed}\hlstd{(}\hlnum{123}\hlstd{)}
\hlstd{R> }\hlstd{sof_bootstrapCI} \hlkwb{<-} \hlkwd{bootstrapCI}\hlstd{(sof[}\hlnum{1000}\hlstd{],} \hlkwc{B_outer} \hlstd{=} \hlnum{100}\hlstd{,} \hlkwc{B_inner} \hlstd{=} \hlnum{10}\hlstd{,}
\hlstd{+ }  \hlkwc{mc.cores} \hlstd{=} \hlnum{10}\hlstd{)}
\hlstd{R> }\hlkwd{par}\hlstd{(}\hlkwc{mfrow} \hlstd{=} \hlkwd{c}\hlstd{(}\hlnum{1}\hlstd{,}\hlnum{3}\hlstd{))}
\hlstd{R> }\hlkwd{plot}\hlstd{(sof_bootstrapCI,} \hlkwc{ask} \hlstd{=} \hlnum{FALSE}\hlstd{,} \hlkwc{commonRange} \hlstd{=} \hlnum{FALSE}\hlstd{,} \hlkwc{ylab} \hlstd{=} \hlstr{""}\hlstd{)}
\end{alltt}
\end{kframe}\begin{figure}[ht]

{\centering \includegraphics[width=\maxwidth]{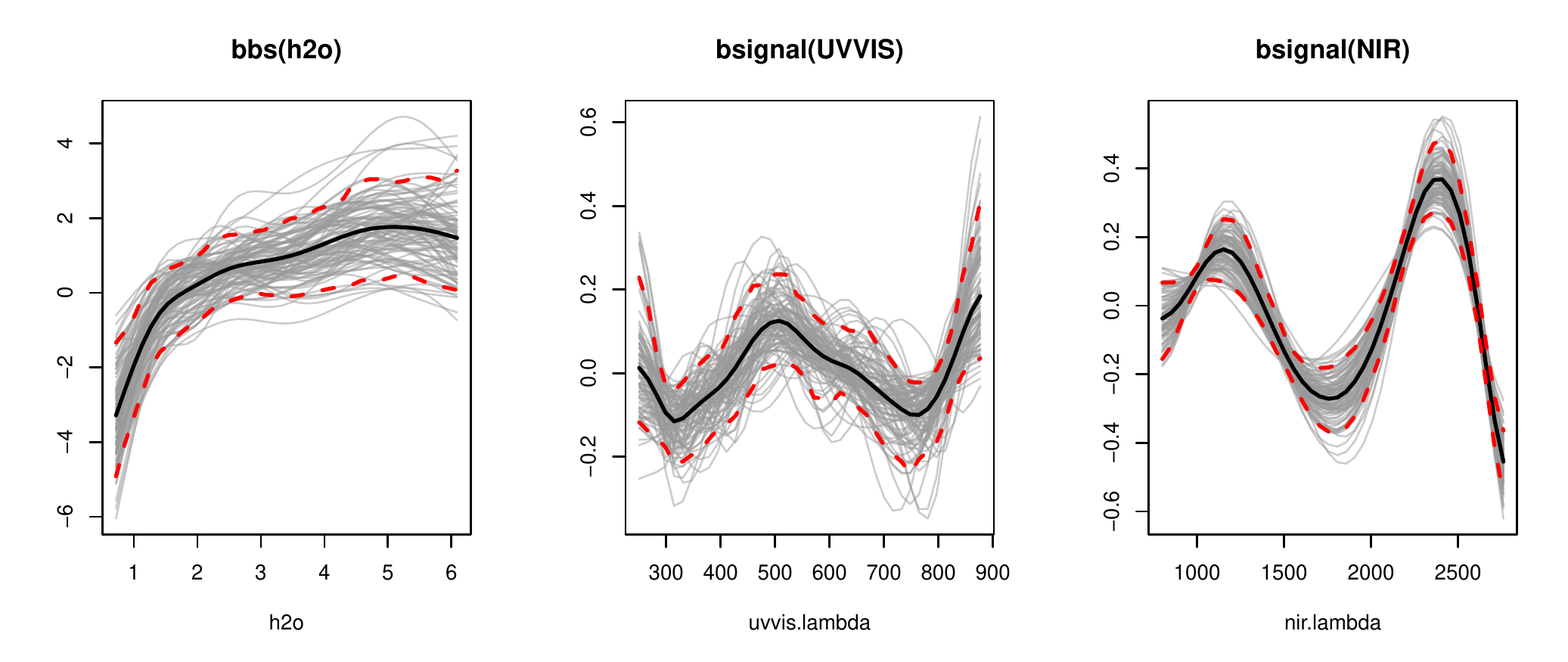} 

}

\caption[Bootstrapped coefficient estimates of the model for the heat value of the fossil fuels]{Bootstrapped coefficient estimates of the model for the heat value of the fossil fuels. The coefficient estimates in the bootstrap samples for the smooth effect of the water content (left), the linear effect of the UVVIS spectrum (middle) and the NIR spectrum (right) are displayed. The pointwise 5\% and the 95\%  quantiles are marked with dashed red lines. The pointwise 50\% quantile is marked by a black line.}\label{fig:sof_boot}
\end{figure}

\end{knitrout}



\hfill $\blacklozenge$


\subsection{Functional response} 
\label{sec.fdboost_fof}

In this subsection, we explain how to fit models with functional response like model (\ref{eq.fof}). Models with scalar and functional covariates are treated, thus covering function-on-scalar and function-on-function regression models. 


\subsubsection{Specification of functional response} 
\label{subsec.fof_response}
If a functional variable is observed on one common grid, its observations can be represented by a matrix. In \pkg{FDboost}, such functional variables have to be supplied as $<$number of curves$>$ by $<$number of evaluation points$>$ matrices. That is, a functional response $y_i(t_g)$, with $i=1,\ldots,N$ curves and $g=1,\ldots,G$ evaluation points, is stored in an $N \times G$ matrix with cases in rows and evaluation points in columns. This corresponds to a data representation in wide format. 
The $t$~variable must be given as vector $(t_1, \ldots, t_G)^\top$. 
%

For the functional response, curve-specific observation grids are possible, i.e., the $i$th response curve is observed at evaluation points $(t_{ig},\ldots,t_{iG_i})^\top$ specific for each curve~$i$.  
In this case, three pieces of information must be supplied: the values of the response, the evaluation points and the curve to which each of the observations belongs. 
The response is supplied as the vector $\left(y_1(t_{11}), \ldots, y_N(t_{N G_N}) \right)^\top$. This vector has length $n=\sum_{i=1}^N G_i$. The $t$~variable contains all evaluation points $(t_{11}, \ldots, t_{N G_N})^\top$. The argument \code{id} contains the information on which observation corresponds to which response curve. 
The argument \code{id} must be supplied as a right-sided formula \code{id = ~ idvariable}. 

\subsubsection*{Case study (ctd.): Emotion components data}
In the following, we give an example for a model fit with a functional response.  
In the first model fit, the response is stored in the matrix \code{EMG}, in the second in the vector \code{EMG\_long}. 
We fit an intercept model by defining the \code{formula} as \code{y ~ 1} and the \code{timeformula} as \code{~ bbs(t)}.   
\begin{knitrout}
\definecolor{shadecolor}{rgb}{1, 1, 1}\color{fgcolor}\begin{kframe}
\begin{alltt}
\hlstd{R> }\hlcom{# fit intercept model with response matrix}
\hlstd{R> }\hlstd{fos_intercept} \hlkwb{<-} \hlkwd{FDboost}\hlstd{(EMG} \hlopt{~} \hlnum{1}\hlstd{,}
\hlstd{+ }  \hlkwc{timeformula} \hlstd{=} \hlopt{~} \hlkwd{bbs}\hlstd{(t,} \hlkwc{df} \hlstd{=} \hlnum{3}\hlstd{),}
\hlstd{+ }  \hlkwc{data} \hlstd{= emotionHGL)}
\end{alltt}
\end{kframe}
\end{knitrout}

The corresponding mathematical formula is
\begin{equation*}
\label{eq.fos_intercept}
 \EV( Y_\EMG(t)) = \beta_0(t),
\end{equation*}
i.e., we simply estimate the mean curve $\beta_0(t)$ of the functional EMG signal.

To fit a model with response in long format, we first have to convert the data into the corresponding format. We therefore construct a dataset \code{data\_emotion\_long} that contains the response in long format. Usually, the long format specification is only necessary for responses that are observed on curve specific grids. We here provide this version for illustrative purposes, but in this example the following model specification is  equivalent to the previous model fit \code{fos\_intercept}.  
\begin{knitrout}
\definecolor{shadecolor}{rgb}{1, 1, 1}\color{fgcolor}\begin{kframe}
\begin{alltt}
\hlstd{R> }\hlstd{emotion_long} \hlkwb{<-} \hlstd{emotionHGL}
\hlstd{R> }\hlstd{emotion_long}\hlopt{$}\hlstd{EMG_long} \hlkwb{<-} \hlkwd{as.vector}\hlstd{(emotion_long}\hlopt{$}\hlstd{EMG)}
\hlstd{R> }\hlstd{emotion_long}\hlopt{$}\hlstd{time_long} \hlkwb{<-} \hlkwd{rep}\hlstd{(emotionHGL}\hlopt{$}\hlstd{t,} \hlkwc{each} \hlstd{=} \hlkwd{nrow}\hlstd{(emotionHGL}\hlopt{$}\hlstd{EMG))}
\hlstd{R> }\hlstd{emotion_long}\hlopt{$}\hlstd{curveid} \hlkwb{<-} \hlkwd{rep}\hlstd{(}\hlnum{1}\hlopt{:}\hlkwd{nrow}\hlstd{(emotionHGL}\hlopt{$}\hlstd{EMG),} \hlkwd{ncol}\hlstd{(emotionHGL}\hlopt{$}\hlstd{EMG))}
\end{alltt}
\end{kframe}
\end{knitrout}

\begin{knitrout}
\definecolor{shadecolor}{rgb}{1, 1, 1}\color{fgcolor}\begin{kframe}
\begin{alltt}
\hlstd{R> }\hlstd{fos_intercept_long} \hlkwb{<-} \hlkwd{FDboost}\hlstd{(EMG_long} \hlopt{~} \hlnum{1}\hlstd{,}
\hlstd{+ }  \hlkwc{timeformula} \hlstd{=} \hlopt{~} \hlkwd{bbs}\hlstd{(time_long,} \hlkwc{df} \hlstd{=} \hlnum{3}\hlstd{),}
\hlstd{+ }  \hlkwc{id} \hlstd{=} \hlopt{~} \hlstd{curveid,} \hlkwc{data} \hlstd{= emotion_long)}
\end{alltt}
\end{kframe}
\end{knitrout}
\hfill $\blacklozenge$

%
%


\subsubsection*{Effects in the formula that are combined with the timeformula}
Many covariate effects can be represented by the Kronecker product of two marginal bases as in Equation~\ref{eq.effect_o}. The response and the bases in covariate direction $\mb_j(x)$ are specified in \code{formula} as \code{Y ~ b\_1 + \ldots + b\_J}. The base-learner for the expansion along $t$ is specified in \code{timeformula} as \code{~ b\_Y}. 
Each base-learner in \code{formula} is combined with the base-leaner in \code{timeformula} using the operator \code{\%O\%}. This operator implements the Kronecker product of two basis vectors as in Equation~\ref{eq.effect_o}. 
Consider, for example, \code{formula = Y ~ b\_1 + b\_2}\textcolor{myc}{. If, \code{b\_1} is defined by \code{bols(z)} with covariate $z$ and a scalar response is given, using \code{timeformula = NULL} specifies a model with linear effect $z \beta$. In the case of a functional response, we usually want the effect $z \beta$ to vary for each time-point $t\in\mathcal{T}$ of the response, i.e., $z\beta(t)$. This can be done by defining} \code{timeformula = ~ b\_Y}\textcolor{myc}{, where the base-learner \code{b\_Y} defines the form of variation in $t$-direction. Assuming a linear effect in~$t$, \code{b\_Y} is set to \code{bols(t)}. The combination of \code{timeformula} and \code{formula} yields} \code{Y ~ b\_1 \%O\% b\_Y + b\_2  \%O\% b\_Y}\textcolor{myc}{. For the particular example, \code{b\_1 \%O\% b\_Y} is equal to \code{bols(z) \%O\% bols(t)} yielding $z\beta(t)$.}

\textcolor{myc}{If marginal base-learners are specified with a penalty, the Kronecker product of the two basis vectors is defined with an isotropic penalty matrix as in~\ref{eq.penalty_iso}. 
If the effect should only be penalized in $t$ direction, the operator \code{\%A0\%} can be used as it sets up the penalty as Equation~\ref{eq.penalty_0}.} 
If \code{formula} contains base-learners that are composed of two base-learners by \code{\%O\%} or \code{\%A0\%}, those effects are not expanded with \code{timeformula}, allowing for model specifications with different effects in $t$~direction. This can be used, for example, to model some effects linearly and others non-linearly in $t$ or to construct effects using \code{\%A0\%}.  
For further details on these operators and their use, we refer to Appendix~\ref{app.operatorsOX}. 

We start with base-learners for the \code{timeformula}. Theoretically, it is possible to use any base-learner which models the effect of a continuous variable. Usually, the effects are assumed to be smooth along~$t$. In this case, the base-learner \code{bbs()} can be used, which represents the smooth effect by P-splines \citep{schmid2008}.  Thus, \code{bbs()} uses a B-spline representation for the design matrix and a squared difference matrix as penalty matrix. 
Using the \code{bbs()} base-leaner in the \code{timeformula} corresponds to using a marginal basis $\mb_Y$ as described in Equation~\ref{eq.fof_example_fun_effect}. 

Base-learners that can be used in \code{formula} are listed in Table~\ref{tab.array_linpred}. In this table, a selection of additive predictors that can be represented within the array framework are listed in the left column. In the right column, the corresponding \code{formula} is given. The \code{timeformula} is set to \code{~ bbs(t)} to model all effects as smooth effects in~$t$. Thus, the specified effects in \code{formula} are combined with \code{timeformula} using the Kronecker product.  
%
%
\renewcommand\arraystretch{1.6}
\begin{table}[ht]\centering
\begin{small}
\begin{tabular}{p{.35\textwidth}| p{.6\textwidth} }
additive predictor\\ $h(\mx,t) = \sum_j h_j(\mx,t)$ & call  \\ 
\hline
 $\beta_0(t)$ & \code{y ~ 1} \\
 $\beta_0(t) + z_1 \beta_1(t)$ & \code{y ~ 1 + bolsc(z1)} \\
 $\beta_0(t) + f_1(z_1, t)$ & \code{y ~ 1 + bbsc(z1)} \\
 $\beta_0(t) + z_1 \beta_1(t) + z_2 \beta_2(t) + z_1 z_2 \beta_3(t)$ &  
 \code{y ~ 1 + bolsc(z1) + bolsc(z2) +} \\
 & \qquad \code{bols(z1) \%Xc\% bols(z2)} \\
 $\beta_0(t) + z_1 \beta_1(t) + f_2(z_2,t) + z_1 f_3(z_2,t)$ & \code{y ~ 1 + bolsc(z1) + bbsc(z2) + bols(z1) \%Xc\% bbs(z2)} \\
 $\beta_0(t) + f_1(z_1, t) + f_2(z_2, t) + f_3(z_1, z_2, t)$ & 
 \code{y ~ 1 + bbsc(z1) + bbsc(z2) + bbs(z1) \%Xc\% bbs(z2)} \\ 
\hline 
 $\beta_0(t) + \int_{\mathcal{S}} x(s) \beta_1(s,t)\,ds $ & 	\code{y ~ 1 + bsignal(x, s = s)} \\ 
 & \code{y ~ 1 + bfpc(x, s = s)} \\ 
 $\beta_0(t) + z \beta_1(t) + \int_{\mathcal{S}} x(s) \beta_2(s,t)\,ds$ &    \code{y ~ 1 + bolsc(z) + bsignal(x, s = s)} \\
 \qquad $ + z \int_{\mathcal{S}} x(s) \beta_3(s,t)\,ds $ & 	\qquad   \code{+ bsignal(x, s = s) \%X\% bolsc(z)}\\ 
\end{tabular}
\end{small}
\caption[Additive predictors that can be represented within the array framework.]{Additive predictors that can be represented within the array framework.}
\label{tab.array_linpred}
\end{table}
\renewcommand\arraystretch{1}


For \code{offset = NULL}, the model contains a smooth offset $\beta^\ast_0(t)$. The smooth offset is computed prior to the model fit as smoothed population minimizer of the loss. For mean regression, the smooth offset is the smoothed mean over~$t$. The specification \code{offset = "scalar"} yields a constant offset $\beta_0^\ast$. 
The resulting intercept in the final model is the sum of the offset and the smooth intercept $\tilde\beta_0(t)$ specified in the \code{formula} as \code{1}, i.e., $\beta_0(t) = \beta^\ast_0(t) + \tilde\beta_0(t)$.      

The upper part of Table~\ref{tab.array_linpred} gives examples for linear predictors with scalar covariates. 
A linear effect of a scalar covariate is specified using the base-learner \code{bolsc()}. This base-learner works for continuous and for factor variables. 
A smooth effect of a continuous covariate is obtained by using the base-learner \code{bbsc()}. 
The base-learners \code{bolsc()} and \code{bbsc()} are similar to the base-learners \code{bols()} and \code{bbs()} from the \pkg{mboost} package, but enforce pointwise sum-to-zero constraints to ensure identifiability for models with functional response (the suffix 'c' refers to 'constrained'). 
Since, for example, the effect $f_1(z_1, t)$ contains a smooth intercept as special case, the model would not be identifiable without constraints,  
see Appendix~\ref{app.constraints} for more details. We use the constraint $\sum_{i=1}^N h_j(\mx_i, t) = 0$ for all~$t$, which centers each effect for each point~$t$ \citep{scheipl2015}. This implies that effects varying over $t$ can be interpreted as deviations from the smooth intercept and that the intercept can be interpreted as global mean if all effects are centered in this way. 
It is possible to check whether all covariate effects sum to zero for all points $t$ by setting \code{check0 = TRUE} in the \code{FDboost()} call. 
To specify interaction effects of two scalar covariates, the base-learners for each of the covariates are combined using the operator \code{\%Xc\%} that applies the sum-to-zero constraint to the interaction effect.

The lower part of Table~\ref{tab.array_linpred} gives examples for linear predictors with functional covariates. 
In analogy to models with scalar response, the linear effect $\int_{\mathcal{S}} x(s)\beta(s,t)\, ds$ can be fitted by \code{bsignal()} or \code{bfpc()} and the interaction effect is formed using the operator \code{\%X\%} (see the explanations for Table~\ref{tab.sof_FDboost_effects}). 


\subsubsection*{Case study (ctd.): Emotion components data}
For the emotion components data with the EMG signal as functional response, $Y_\EMG(t)$, $t\in [0,1560]ms$, we fit models with scalar and functional covariate effects in the following. 

\subsubsection*{Function-on-scalar regression}
We specify a model for the conditional expectation of the EMG signal using a random intercept curve for each subject and a linear effect for the study setting \code{power}:   
\begin{equation}
\label{eq.fos_random_power}
 \EV( Y_\EMG(t)| \mx )  
 = \beta_0(t) + \sum_{k=1}^{23} I(x_{\text{subject}} = k) \beta_{\text{subject},k}(t) + x_{\text{power}} \beta_{\text{power}}(t),
\end{equation}
with \code{subject} having values 1 to 23 for the participants of the study, and $x_{\text{power}}$ taking values $\{-1,1\}$ for low and high power.  
Both covariate effects in the model are specified by using a centered base-learner. The linear effect of the factor variable \code{subject} and the effect of \code{power} are both specified using the \code{bolsc()} base-learner. Therefore, the effects sum up to zero for each time-point $t$ over all observations $i=1,\ldots,N=184$, i.e., $\sum_{i=1}^N \sum_{k=1}^{23} I(x_{\text{subject},i}=k) \beta_{\text{subject},k}(t) = 0$ for all~$t$.  
\begin{knitrout}
\definecolor{shadecolor}{rgb}{1, 1, 1}\color{fgcolor}\begin{kframe}
\begin{alltt}
\hlstd{R> }\hlstd{fos_random_power} \hlkwb{<-} \hlkwd{FDboost}\hlstd{(EMG} \hlopt{~} \hlnum{1} \hlopt{+} \hlkwd{bolsc}\hlstd{(subject,} \hlkwc{df} \hlstd{=} \hlnum{2}\hlstd{)}
\hlstd{+ }  \hlopt{+} \hlkwd{bolsc}\hlstd{(power,} \hlkwc{df} \hlstd{=} \hlnum{1}\hlstd{)} \hlopt{%A0%} \hlkwd{bbs}\hlstd{(t,} \hlkwc{df} \hlstd{=} \hlnum{6}\hlstd{),}
\hlstd{+ }  \hlkwc{timeformula} \hlstd{=} \hlopt{~} \hlkwd{bbs}\hlstd{(t,} \hlkwc{df} \hlstd{=} \hlnum{3}\hlstd{),}
\hlstd{+ }  \hlkwc{data} \hlstd{= emotion)}
\end{alltt}
\end{kframe}
\end{knitrout}

As described in Section~\ref{sec.boosting}, it is important that all base-learners have the same number of degrees of freedom. In this model the degrees of freedom for each base-learner are $2 * 3 = 6$. 
By specifying the \code{bolsc}-baselearner with \code{df = 2} for \code{subject}, the subject effect is estimated with a Ridge penalty similar to a random effect, whereas the \code{power} effect is estimated unpenalized due to the use of the \code{\%A0\%}-operator. 

Analogously, a model with response in long format as in \code{fos\_intercept\_long} could be specified by changing the formula to the formula of \code{fos\_random\_power}. 

\subsubsection*{Function-on-function regression}
For the data subset for one specific game condition, we use the effect of the EEG signal to model the EMG signal:  
\begin{equation}
\label{eq.fof_signal}
 \EV(Y_\EMG(t)| \mx) = \beta_0(t) + \int_{\mathcal{S}} x_\EEG(s)\beta_\EEG(s,t)\,ds. 
\end{equation} 
In this model each time-point of the covariate $x_\EEG(s)$ potentially influences each time-point of the response $Y_\EMG(t)$. We center the EEG signal per time point such that $\sum_{i=1}^N x_{\EEG,i}(s) = 0$ for each $s$ to center its effect per time-point.

\begin{knitrout}
\definecolor{shadecolor}{rgb}{1, 1, 1}\color{fgcolor}\begin{kframe}
\begin{alltt}
\hlstd{R> }\hlstd{emotionHGL}\hlopt{$}\hlstd{EEG} \hlkwb{<-} \hlkwd{scale}\hlstd{(emotionHGL}\hlopt{$}\hlstd{EEG,} \hlkwc{scale} \hlstd{=} \hlnum{FALSE}\hlstd{)}
\end{alltt}
\end{kframe}
\end{knitrout}

\begin{knitrout}
\definecolor{shadecolor}{rgb}{1, 1, 1}\color{fgcolor}\begin{kframe}
\begin{alltt}
\hlstd{R> }\hlstd{fof_signal} \hlkwb{<-} \hlkwd{FDboost}\hlstd{(EMG} \hlopt{~} \hlnum{1} \hlopt{+} \hlkwd{bsignal}\hlstd{(EEG,} \hlkwc{s} \hlstd{= s,} \hlkwc{df} \hlstd{=} \hlnum{2}\hlstd{),}
\hlstd{+ }  \hlkwc{timeformula} \hlstd{=} \hlopt{~} \hlkwd{bbs}\hlstd{(t,} \hlkwc{df} \hlstd{=} \hlnum{3}\hlstd{),}
\hlstd{+ }  \hlkwc{data} \hlstd{= emotionHGL)}
\end{alltt}
\end{kframe}
\end{knitrout}

We will show and interpret plots of the estimated coefficients later on. 
Assuming that the brain activity (measured via the EEG) triggers the muscle activity (measured via the EMG), it is reasonable to assume that EMG signals are only influenced by past EEG signals. Such a relationship can be represented using a historical effect 
$\int_{T_1}^{t} x(s) \beta(s,t)\,ds$, which will be discussed in the following paragraph. 
\hfill $\blacklozenge$


\subsubsection*{Effects in the formula comprising both the effect in covariate and $t$-direction} 
If the covariate varies with~$t$, the effect cannot be separated into a marginal basis depending on the covariate and a marginal basis depending only on~$t$. In this case the effects are represented as in Equation~\ref{eq.b_jY}. 
Examples for such effects are historical and concurrent functional effects, as discussed in \cite{brockhaus2015hist}. In Table~\ref{tab.nonarray_linpred} we give an overview of possible additive predictors containing such effects. 
\renewcommand\arraystretch{1.6}
\begin{table}[ht]\centering
\begin{small}
\begin{tabular}{p{.35\textwidth}| p{.6\textwidth} }
additive predictor $h(x,t) = \sum_j h_j(x,t)$ & call  \\ 
\hline 
 $\beta_0(t) + x(t) \beta(t)$ & \code{y ~ 1 + bconcurrent(x, s = s, time = t)} \\
 $\beta_0(t) + \int_{T_1}^{t} x(s) \beta(s,t)\,ds $ &  \code{y ~ 1 + bhist(x, s = s, time = t)} \\
 $\beta_0(t) + \int_{t-\delta}^{t} x(s) \beta(s,t)\,ds $ & \code{y ~ 1 + bhist(x, s = s, time = t,}\\
 & \qquad \code{limits = limitsLag)}$^\ast$ \\ 
 $\beta_0(t) + \int_{T_1}^{t-\delta} x(s) \beta(s,t)\,ds $ & \code{y ~ 1 + bhist(x, s = s, time = t,} \\
& \qquad \code{limits = limitsLead)}$^\ast$ \\ 
 $\int_{l(t)}^{u(t)} x(s) \beta(s,t)\,ds $ & \code{y ~ 1 + bhist(x, s = s, time = t, limits = mylimits)} \\ 
\hline 
 $\beta_0(t) + z \beta_1(t) + \int_{T_1}^{t} x(s) \beta_2(s,t)\,ds$ &  \code{y ~ 1 + bolsc(z) + bhist(x, s = s, time = t)}	\\
$\qquad +\; z \int_{T_1}^{t} x(s) \beta_3(s,t)\,ds $ & \qquad \code{+ bhistx(x) \%X\% bolsc(z)}\\ 
\end{tabular}
\end{small}
\caption[Additive predictors.]{Additive predictors that contain effects that cannot be separated into an effect in covariate direction and an effect in $t$ direction. These effects in \code{formula} are not expanded by the \code{timeformula}. 
We give examples for general limit functions \code{mylimits} in this section. 
In \code{bhistx()}, the variable \code{x} has to be of class \code{hmatrix}, please see the manual of \code{bhistx()} for details. }
\label{tab.nonarray_linpred}
\end{table}
\renewcommand\arraystretch{1}

The concurrent effect $\beta(t)x(t)$ is only meaningful if the functional response and the functional covariate are observed over the same domain. Models with concurrent effects can be seen as varying-coefficient models \citep{hastie1993}, where the effect varies over~$t$. The base-learner \code{bconcurrent()} expands the smooth concurrent effect $\beta(t)$ in P-splines. 
The historical effect $\int_{T_1}^{t} x(s) \beta(s,t)\,ds$ uses only covariate information up to the current observation point of the response. The base-learner \code{bhist()} expands the coefficient surface $\beta(s,t)$ in $s$ and in $t$ direction using P-splines to fit the historical effect. 
In Appendix~\ref{app.funbl}, details on the representation of functional effects are given. 

The interface of \code{bhist()} is:  
\begin{knitrout}
\definecolor{shadecolor}{rgb}{1, 1, 1}\color{fgcolor}\begin{kframe}
\begin{alltt}
\hlstd{R> }\hlkwd{bhist}\hlstd{(x, s, time,} \hlkwc{limits} \hlstd{=} \hlstr{"s<=t"}\hlstd{,} \hlkwc{knots} \hlstd{=} \hlnum{10}\hlstd{,} \hlkwc{degree} \hlstd{=} \hlnum{3}\hlstd{,} \hlkwc{differences} \hlstd{=} \hlnum{1}\hlstd{,}
\hlstd{+ }  \hlkwc{df} \hlstd{=} \hlnum{4}\hlstd{,} \hlkwc{lambda} \hlstd{=} \hlkwa{NULL}\hlstd{,} \hlkwc{check.ident} \hlstd{=} \hlnum{FALSE}\hlstd{)}
\end{alltt}
\end{kframe}
\end{knitrout}
Most arguments of \code{bhist()} are analogous to those of \code{bsignal()}.
\code{bhist()} has the additional argument \code{time} to specify the observation points of the response. Via the argument \code{limits} in \code{bhist()} the user can specify integration limits depending on~$t$. Per default a historical effect with limits $s \leq t$ is used. Other integration limits can be specified by using a function with arguments \code{s} and \code{t}, which returns \code{TRUE} for combinations of \code{s} and \code{t} that lie within the integration interval and \code{FALSE} otherwise. In the following, we give three examples for functions that can be used for \code{limits} resulting in a classical historical effect, a lag effect or a lead effect, respectively:   

\begin{knitrout}
\definecolor{shadecolor}{rgb}{1, 1, 1}\color{fgcolor}\begin{kframe}
\begin{alltt}
\hlstd{R> }\hlstd{limitsHist} \hlkwb{<-} \hlkwa{function}\hlstd{(}\hlkwc{s}\hlstd{,} \hlkwc{t}\hlstd{) \{}
\hlstd{+ }  \hlstd{s} \hlopt{<=} \hlstd{t}
\hlstd{+ }\hlstd{\}}
\hlstd{R> }\hlstd{limitsLag} \hlkwb{<-} \hlkwa{function}\hlstd{(}\hlkwc{s}\hlstd{,} \hlkwc{t}\hlstd{,} \hlkwc{delta} \hlstd{=} \hlnum{5}\hlstd{) \{}
\hlstd{+ }  \hlstd{s} \hlopt{>=} \hlstd{t} \hlopt{-} \hlstd{delta} \hlopt{&}  \hlstd{s} \hlopt{<=} \hlstd{t}
\hlstd{+ }\hlstd{\}}
\hlstd{R> }\hlstd{limitsLead} \hlkwb{<-} \hlkwa{function}\hlstd{(}\hlkwc{s}\hlstd{,} \hlkwc{t}\hlstd{,} \hlkwc{delta} \hlstd{=} \hlnum{5}\hlstd{) \{}
\hlstd{+ }  \hlstd{s} \hlopt{<=} \hlstd{t} \hlopt{-} \hlstd{delta}
\hlstd{+ }\hlstd{\}}
\end{alltt}
\end{kframe}
\end{knitrout}

The base-learner \code{bhistx()} is especially suited to form interaction effects such as factor-specific historical effects \citep{ruegamer2018}, as \code{bhist()} cannot be used in combination with the row-wise tensor product operator \code{\%X\%} to form interaction effects. 
\code{bhistx()} requires the data to be supplied as an object of type \code{hmatrix}; see the manual of \code{bhistx()} for its setup. 

\subsubsection*{Case study (ctd.): Emotion components data}
Again, we use the subset of the data for one specific game condition. 
We start with a simple function-on-function regression model by specifying a concurrent effect of the EEG signal on the EMG signal:  
\begin{equation*}
 \EV(Y_\EMG(t)| \mx)  = \beta_0(t) + x_\EEG(t)\beta(t). 
\end{equation*} 
A concurrent effect is obtained by the base-learner \code{bconcurrent()}, which is not expanded by the base-learner in \code{timeformula}. In this model, \code{timeformula} is only used to expand the smooth intercept. 

\begin{knitrout}
\definecolor{shadecolor}{rgb}{1, 1, 1}\color{fgcolor}\begin{kframe}
\begin{alltt}
\hlstd{R> }\hlstd{fof_concurrent} \hlkwb{<-} \hlkwd{FDboost}\hlstd{(EMG} \hlopt{~} \hlnum{1} \hlopt{+} \hlkwd{bconcurrent}\hlstd{(EEG,} \hlkwc{s} \hlstd{= s,} \hlkwc{time} \hlstd{= t,} \hlkwc{df} \hlstd{=} \hlnum{6}\hlstd{),}
\hlstd{+ }  \hlkwc{timeformula} \hlstd{=} \hlopt{~} \hlkwd{bbs}\hlstd{(t,} \hlkwc{df} \hlstd{=} \hlnum{3}\hlstd{),} \hlkwc{data} \hlstd{= emotionHGL,}
\hlstd{+ }  \hlkwc{control} \hlstd{=} \hlkwd{boost_control}\hlstd{(}\hlkwc{mstop} \hlstd{=} \hlnum{300}\hlstd{))}
\end{alltt}
\end{kframe}
\end{knitrout}

Assuming that the activity in the muscle can be completely traced back to previous activity in the brain, a more appropriate model seems to be a historical model including a historical effect
\begin{equation}
\label{eq.fof_historical}
 \EV(Y_\EMG(t)| \mx)  = \beta_0(t) + \int_{l(t)}^{u(t)} x_\EEG(s)\beta_\EEG(s,t)\,ds. 
\end{equation} 
From a neuro-anatomy perspective, the signal from the brain requires time to reach the muscle. We therefore set $l(t) = 0$ and $u(t) = t - 3$, which is in line with \citet{ruegamer2018}. 

\begin{knitrout}
\definecolor{shadecolor}{rgb}{1, 1, 1}\color{fgcolor}\begin{kframe}
\begin{alltt}
\hlstd{R> }\hlstd{fof_historical} \hlkwb{<-} \hlkwd{FDboost}\hlstd{(EMG} \hlopt{~} \hlnum{1} \hlopt{+} \hlkwd{bhist}\hlstd{(EEG,} \hlkwc{s} \hlstd{= s,} \hlkwc{time} \hlstd{= t,}
\hlstd{+ }  \hlkwc{limits} \hlstd{=} \hlkwa{function}\hlstd{(}\hlkwc{s}\hlstd{,} \hlkwc{t}\hlstd{) s} \hlopt{<=} \hlstd{t} \hlopt{-} \hlnum{3}\hlstd{,} \hlkwc{df} \hlstd{=} \hlnum{6}\hlstd{),}
\hlstd{+ }  \hlkwc{timeformula} \hlstd{=} \hlopt{~} \hlkwd{bbs}\hlstd{(t,} \hlkwc{df} \hlstd{=} \hlnum{3}\hlstd{),} \hlkwc{data} \hlstd{= emotionHGL,}
\hlstd{+ }  \hlkwc{control} \hlstd{=} \hlkwd{boost_control}\hlstd{(}\hlkwc{mstop} \hlstd{=} \hlnum{300}\hlstd{))}
\end{alltt}
\end{kframe}
\end{knitrout}
 
More complex historical models are discussed in \citet{ruegamer2018}. In particular, a model containing random effects for the participants, effects for the game conditions and game condition- as well as subject-specific historical effects of the EEG signal. 
\hfill $\blacklozenge$ 
\\\\
It is also possible to combine effects listed in Table~\ref{tab.array_linpred} and Table~\ref{tab.nonarray_linpred} to form more complex models. In particular, base-learners with and without array structure can be combined within one model. As in the component-wise boosting procedure each base-learner is evaluated separately, the array structure of the Kronecker product base-learners can still be exploited in such hybrid models.


\subsubsection{Model tuning and early stopping}
\label{subsec.fof_tune} 
For a fair selection of base-learner, additional care is needed for functional responses as only some of the base-learners in the \code{formula} are expanded by the base-learner in \code{timeformula}. In particular, all base-learners listed in Table~\ref{tab.array_linpred} are expanded by \code{timeformula}, whereas base-learners given in Table~\ref{tab.nonarray_linpred} are not expanded by the \code{timeformula}.   
For the row-wise tensor product and the Kronecker product of two base-learners, the degrees of freedom for the combined base-learner is computed as product of the two marginally specified degrees of freedom. 
For instance, \code{formula = y ~ bbsc(z, df = 3) + bhist(x, s = s, df = 12)} and \code{timeformula = ~ bbs(t, df = 4)} implies $ 3 \cdot 4 = 12$ degrees of freedom for the first combined base-learner and $12$ degrees of freedom for the second base-learner.   
The call \code{extract(object, "df")} displays the degrees of freedom for each base-learner in an \code{FDboost} object. For other tuning options such as the number of iterations and the specification of the step-length see Section~\ref{sec.fdboost_sof}.

To find the optimal number of boosting iterations for a model fit with functional response, {\pkg{FDboost}} provides two resampling functions. 
Depending on the specified model, some parameters are computed from the data prior to the model fit: per default a smooth functional offset $\beta_0^\ast(t)$ is computed (\code{offset = NULL} in \code{FDboost()}) and for linear and smooth effects of scalar variables, defined by \code{bolsc()} and \code{bbsc()}, transformation matrices for the sum-to-zero constraints are computed. 
The function \code{cvrisk.FDboost()} uses the smooth functional offset and the transformation matrices from the original model fit in all folds. Thus, these parameters are treated as fixed and the uncertainty induced by their estimation is not considered in the resampling. 
On the other hand, \code{applyFolds()} recomputes the whole model in each fold.   
The two resampling methods are equal if no smooth offset is used and if the model does not contain any base-learner with a sum-to-zero constraint (i.e., neither \code{bolsc()} nor \code{bbsc()}). In general, we recommend to use the function \code{applyFolds()} to determine the optimal number of boosting iterations for a model with functional response.  
The interface of \code{applyFolds()} is: 

\begin{knitrout}
\definecolor{shadecolor}{rgb}{1, 1, 1}\color{fgcolor}\begin{kframe}
\begin{alltt}
\hlstd{R> }\hlkwd{applyFolds}\hlstd{(object,}
\hlstd{+ }  \hlkwc{folds} \hlstd{=} \hlkwd{cv}\hlstd{(}\hlkwd{rep}\hlstd{(}\hlnum{1}\hlstd{,} \hlkwd{length}\hlstd{(}\hlkwd{unique}\hlstd{(object}\hlopt{$}\hlstd{id))),} \hlkwc{type} \hlstd{=} \hlstr{"bootstrap"}\hlstd{),}
\hlstd{+ }  \hlkwc{grid} \hlstd{=} \hlnum{1}\hlopt{:}\hlkwd{mstop}\hlstd{(object))}
\end{alltt}
\end{kframe}
\end{knitrout}

The interface is in analogy to the interface of \code{cvrisk()}. In the argument \code{object}, the fitted model object is specified. 
\code{grid} defines the grid on which the optimal stopping iteration is searched. 
Via the argument \code{folds} the resampling folds are defined by suitable weights. 
The function \code{applyFolds()} expects resampling weights that are defined on the level of curves, $i=1,\ldots,N$. That means that the folds must contain weights $w_i$, $i=1, \ldots, N$, which can be done easily using the function \code{cv()}. 

\subsubsection{Methods to extract and display results}
Methods to extract and visualize results are the same irrespective of scalar or functional response. Thus, we refer to the corresponding paragraphs at the end of Section~\ref{sec.fdboost_sof}. 

\subsubsection*{Case study (ctd.): Emotion components data}
\textcolor{myc}{Exemplarily, the penalty matrix for the historical effect can be extracted as follows:}
\begin{knitrout}
\definecolor{shadecolor}{rgb}{1, 1, 1}\color{fgcolor}\begin{kframe}
\begin{alltt}
\hlstd{R> }\hlstd{kron_pen} \hlkwb{<-} \hlkwd{extract}\hlstd{(fof_historical,} \hlstr{"penalty"}\hlstd{)}
\hlstd{R> }\hlkwd{as.matrix}\hlstd{(kron_pen[[}\hlnum{1}\hlstd{]][}\hlnum{1}\hlopt{:}\hlnum{5}\hlstd{,}\hlnum{1}\hlopt{:}\hlnum{5}\hlstd{])}
\end{alltt}
\end{kframe}
\end{knitrout}
\textcolor{myc}{This is equal to the kronecker sum of two marginal B-Spline penalties with isotropic penalization (as defined by Equation~\ref{eq.penalty_ani} with $\lambda_j = \lambda_Y$):}
\begin{knitrout}
\definecolor{shadecolor}{rgb}{1, 1, 1}\color{fgcolor}\begin{kframe}
\begin{alltt}
\hlstd{R> }\hlstd{margPen} \hlkwb{<-} \hlkwd{extract}\hlstd{(}\hlkwd{with}\hlstd{(emotionHGL,}
\hlstd{+ }  \hlkwd{bbs}\hlstd{(s,} \hlkwc{knots}\hlstd{=}\hlnum{10}\hlstd{,} \hlkwc{differences} \hlstd{=} \hlnum{1}\hlstd{)),} \hlstr{"penalty"}\hlstd{)}
\hlstd{R> }\hlstd{(}\hlkwd{kronecker}\hlstd{(margPen,} \hlkwd{diag}\hlstd{(}\hlkwd{ncol}\hlstd{(margPen)))} \hlopt{+}
\hlstd{+ }  \hlkwd{kronecker}\hlstd{(}\hlkwd{diag}\hlstd{(}\hlkwd{ncol}\hlstd{(margPen)), margPen))[}\hlnum{1}\hlopt{:}\hlnum{5}\hlstd{,}\hlnum{1}\hlopt{:}\hlnum{5}\hlstd{]}
\end{alltt}
\begin{verbatim}
     [,1] [,2] [,3] [,4] [,5]
[1,]    2   -1    0    0    0
[2,]   -1    3   -1    0    0
[3,]    0   -1    3   -1    0
[4,]    0    0   -1    3   -1
[5,]    0    0    0   -1    3
\end{verbatim}
\end{kframe}
\end{knitrout}

As for scalar response, the \code{plot}-function can be used to access the estimated effects in a function-on-function regression. In the following, we compare the three basic types of functional covariate effects, which can be used in conjunction with a functional response. 
We first determine the optimal number of stopping iterations for all three presented models. 

\begin{knitrout}
\definecolor{shadecolor}{rgb}{1, 1, 1}\color{fgcolor}\begin{kframe}
\begin{alltt}
\hlstd{R> }\hlkwd{set.seed}\hlstd{(}\hlnum{123}\hlstd{)}
\hlstd{R> }\hlstd{folds_bs} \hlkwb{<-} \hlkwd{cv}\hlstd{(}\hlkwc{weights} \hlstd{=} \hlkwd{rep}\hlstd{(}\hlnum{1}\hlstd{, fof_signal}\hlopt{$}\hlstd{ydim[}\hlnum{1}\hlstd{]),}
\hlstd{+ } \hlkwc{type} \hlstd{=} \hlstr{"kfold"}\hlstd{,} \hlkwc{B} \hlstd{=} \hlnum{5}\hlstd{)}
\end{alltt}
\end{kframe}
\end{knitrout}

\begin{knitrout}
\definecolor{shadecolor}{rgb}{1, 1, 1}\color{fgcolor}\begin{kframe}
\begin{alltt}
\hlstd{R> }\hlstd{cvm_concurrent} \hlkwb{<-} \hlkwd{applyFolds}\hlstd{(fof_concurrent,} \hlkwc{folds} \hlstd{= folds_bs,} \hlkwc{grid} \hlstd{=} \hlnum{1}\hlopt{:}\hlnum{300}\hlstd{)}
\hlstd{R> }\hlstd{ms_conc} \hlkwb{<-} \hlkwd{mstop}\hlstd{(cvm_concurrent)}
\hlstd{R> }\hlstd{fof_concurrent} \hlkwb{<-} \hlstd{fof_concurrent[ms_conc]}
\end{alltt}
\end{kframe}
\end{knitrout}

\begin{knitrout}
\definecolor{shadecolor}{rgb}{1, 1, 1}\color{fgcolor}\begin{kframe}
\begin{alltt}
\hlstd{R> }\hlstd{cvm_signal} \hlkwb{<-} \hlkwd{applyFolds}\hlstd{(fof_signal,} \hlkwc{folds} \hlstd{= folds_bs,} \hlkwc{grid} \hlstd{=} \hlnum{1}\hlopt{:}\hlnum{300}\hlstd{)}
\hlstd{R> }\hlstd{ms_signal} \hlkwb{<-} \hlkwd{mstop}\hlstd{(cvm_signal)}
\hlstd{R> }\hlstd{fof_signal} \hlkwb{<-} \hlstd{fof_signal[ms_signal]}
\end{alltt}
\end{kframe}
\end{knitrout}

\begin{knitrout}
\definecolor{shadecolor}{rgb}{1, 1, 1}\color{fgcolor}\begin{kframe}
\begin{alltt}
\hlstd{R> }\hlstd{cvm_historical} \hlkwb{<-} \hlkwd{applyFolds}\hlstd{(fof_historical,} \hlkwc{folds} \hlstd{= folds_bs,} \hlkwc{grid} \hlstd{=} \hlnum{1}\hlopt{:}\hlnum{300}\hlstd{)}
\hlstd{R> }\hlstd{ms_hist} \hlkwb{<-} \hlkwd{mstop}\hlstd{(cvm_historical)}
\hlstd{R> }\hlstd{fof_historical} \hlkwb{<-} \hlstd{fof_historical[ms_hist]}
\end{alltt}
\end{kframe}
\end{knitrout}

Then, we plot the estimated effects into one figure:

\begin{knitrout}
\definecolor{shadecolor}{rgb}{1, 1, 1}\color{fgcolor}\begin{kframe}
\begin{alltt}
\hlstd{R> }\hlkwd{par}\hlstd{(}\hlkwc{mfrow} \hlstd{=} \hlkwd{c}\hlstd{(}\hlnum{1}\hlstd{,}\hlnum{3}\hlstd{))}
\hlstd{R> }\hlkwd{plot}\hlstd{(fof_concurrent,} \hlkwc{which} \hlstd{=} \hlnum{2}\hlstd{,} \hlkwc{main} \hlstd{=} \hlstr{"Concurrent EEG effect"}\hlstd{)}
\hlstd{R> }\hlkwd{plot}\hlstd{(fof_signal,} \hlkwc{which} \hlstd{=} \hlnum{2}\hlstd{,} \hlkwc{main} \hlstd{=} \hlstr{"Signal EEG effect"}\hlstd{,}
\hlstd{+ }  \hlkwc{n1} \hlstd{=} \hlnum{80}\hlstd{,} \hlkwc{n2} \hlstd{=} \hlnum{80}\hlstd{,} \hlkwc{zlim} \hlstd{=} \hlkwd{c}\hlstd{(}\hlopt{-}\hlnum{0.02}\hlstd{,} \hlnum{0.025}\hlstd{),}
\hlstd{+ }  \hlkwc{col} \hlstd{=} \hlkwd{terrain.colors}\hlstd{(}\hlnum{20}\hlstd{))}
\hlstd{R> }\hlkwd{plot}\hlstd{(fof_historical,} \hlkwc{which} \hlstd{=} \hlnum{2}\hlstd{,} \hlkwc{main} \hlstd{=} \hlstr{"Historical EEG effect"}\hlstd{,}
\hlstd{+ }  \hlkwc{n1} \hlstd{=} \hlnum{80}\hlstd{,} \hlkwc{n2} \hlstd{=} \hlnum{80}\hlstd{,} \hlkwc{zlim} \hlstd{=} \hlkwd{c}\hlstd{(}\hlopt{-}\hlnum{0.02}\hlstd{,} \hlnum{0.025}\hlstd{),}
\hlstd{+ }  \hlkwc{col} \hlstd{=} \hlkwd{terrain.colors}\hlstd{(}\hlnum{20}\hlstd{))}
\end{alltt}
\end{kframe}\begin{figure}[ht]

{\centering \includegraphics[width=\maxwidth]{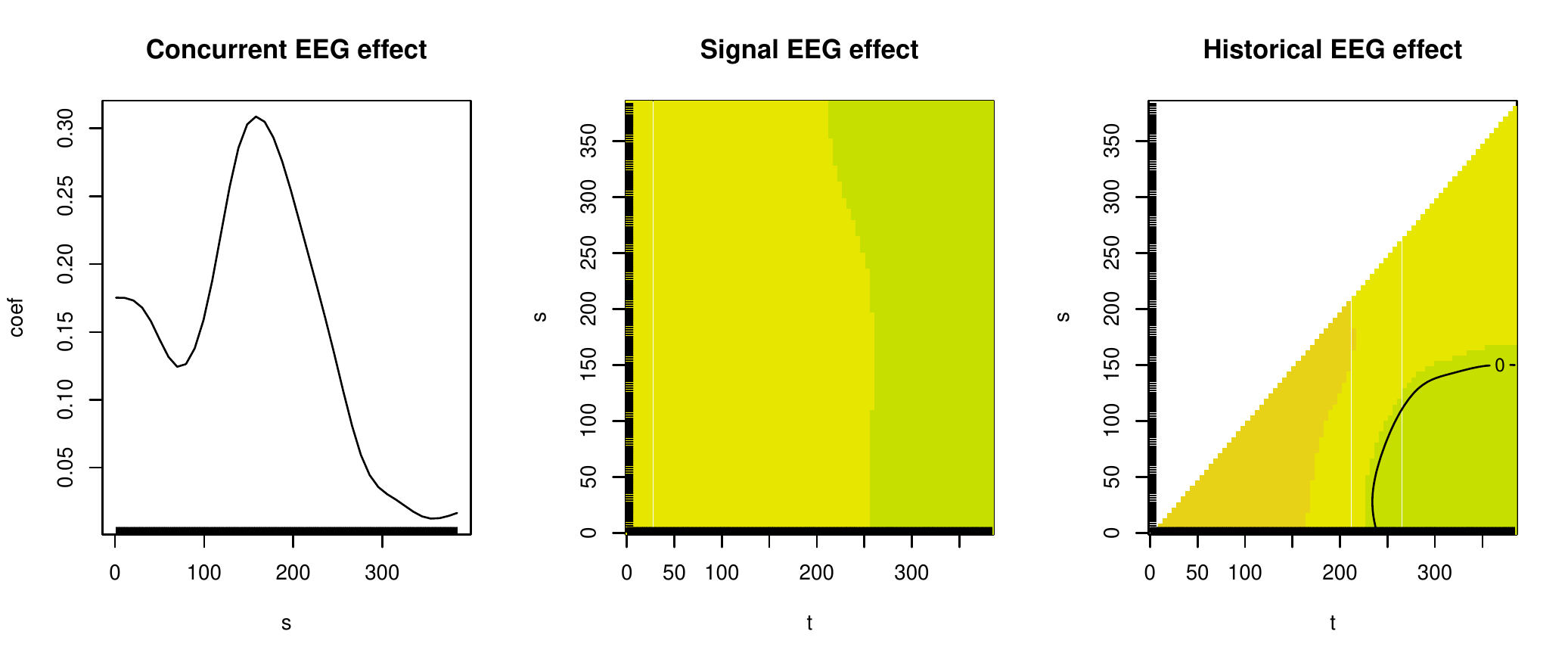} 

}

\caption{Visualization of estimated concurrent EEG effect (left panel), signal EEG effect (center panel) and historical EEG effect (right panel).}\label{fig:visualFOF}
\end{figure}

\end{knitrout}

The concurrent effect corresponds to the diagonal of the other two surfaces in Figure~\ref{fig:visualFOF} and assumes that off-diagonal time-points have no association. Due to the temporal lag between EEG and EMG discussed for model (\ref{eq.fof_historical}), there is no meaningful interpretation for this model and the effect is only shown for demonstrative purposes. 
The historical effect corresponds to the assumption that the upper triangle in the signal EEG effects should be zero, as future brain activity should not influence the present muscle activity. 
The results in Figure~\ref{fig:visualFOF} (right panel) can be interpreted in the same manner as results of a scalar-on-function regression when keeping a certain time point $t$ fixed. For the time point $t = 180$ of the EMG signal, for example, time points $s \approx 100$ to $s \approx 177$ of the EEG signal do not show an effect, but for $s < 100$ the estimated effect on the expected EMG signal is positive. For a detailed description of the interpretation of historical effect surfaces as shown in Figure~\ref{fig:visualFOF}, we refer to the online appendix of \citet{ruegamer2018}.

Careful interpretation has to take into account that this data set has a rather small signal-to-noise ratio due to the oscillating nature of both signals. In such cases, it is recommended to check the uncertainty of estimated effects via bootstrap, e.g., by using the \code{bootstrapCI()} function as exemplarily shown in Figure~\ref{fig:visualFOF2}. 

\begin{knitrout}
\definecolor{shadecolor}{rgb}{1, 1, 1}\color{fgcolor}\begin{kframe}
\begin{alltt}
\hlstd{R> }\hlstd{fof_historical_bci} \hlkwb{<-} \hlkwd{bootstrapCI}\hlstd{(fof_historical,} \hlkwc{mc.cores} \hlstd{=} \hlnum{2}\hlstd{,}
\hlstd{+ }  \hlkwc{B_inner} \hlstd{=} \hlnum{10}\hlstd{,} \hlkwc{type_inner} \hlstd{=} \hlstr{"kfold"}\hlstd{)}
\hlstd{R> }\hlkwd{par}\hlstd{(}\hlkwc{mfrow}\hlstd{=}\hlkwd{c}\hlstd{(}\hlnum{1}\hlstd{,}\hlnum{3}\hlstd{))}
\hlstd{R> }\hlkwd{plot}\hlstd{(fof_historical_bci,} \hlkwc{which} \hlstd{=} \hlnum{2}\hlstd{,} \hlkwc{ask} \hlstd{=} \hlnum{FALSE}\hlstd{,} \hlkwc{pers} \hlstd{=} \hlnum{FALSE}\hlstd{,}
\hlstd{+ }  \hlkwc{col} \hlstd{=} \hlkwd{terrain.colors}\hlstd{(}\hlnum{20}\hlstd{),} \hlkwc{probs} \hlstd{=} \hlkwd{c}\hlstd{(}\hlnum{0.05}\hlstd{,} \hlnum{0.5}\hlstd{,} \hlnum{0.95}\hlstd{))}
\end{alltt}
\end{kframe}\begin{figure}[ht]

{\centering \includegraphics[width=\maxwidth]{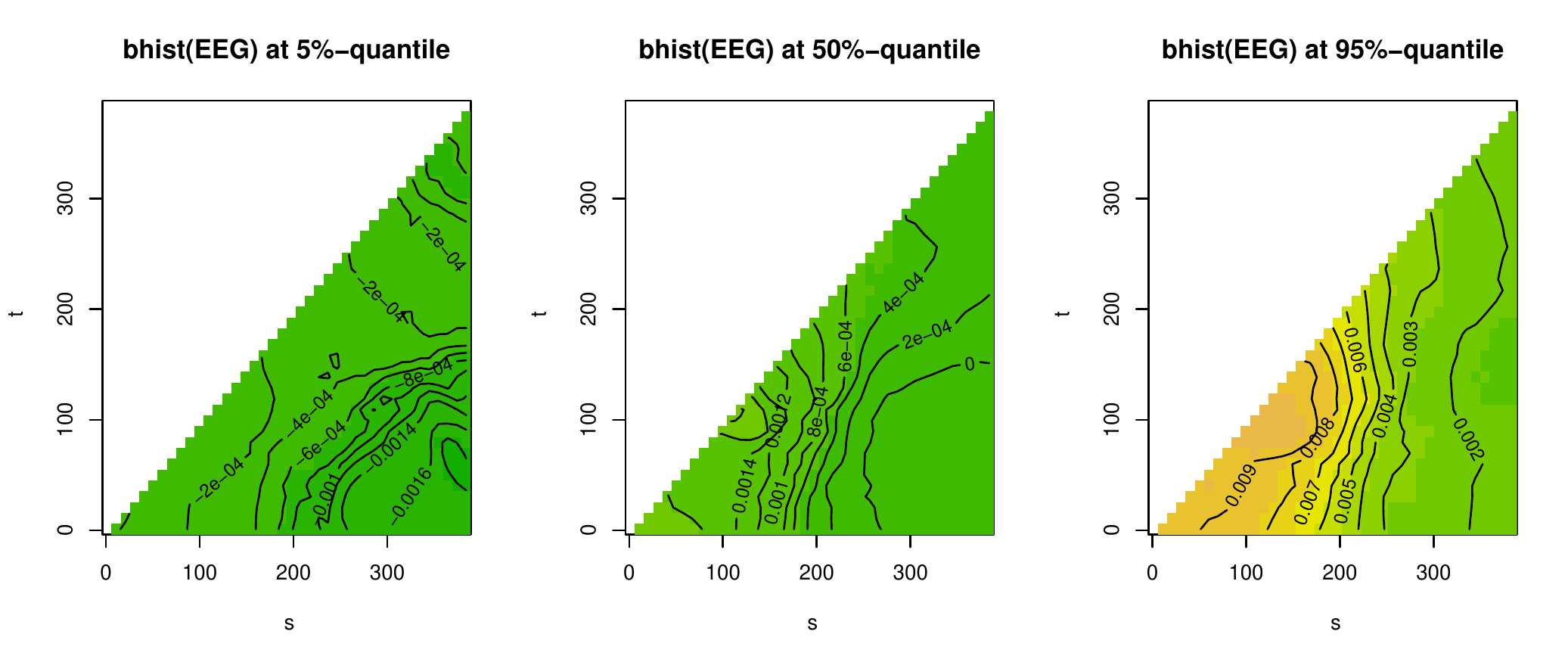} 

}

\caption{Visualization of three bootstrap quantiles for the historical EEG effect based on 100 bootstrap samples and a 10-fold cross-validation to optimize the stopping iteration for each bootstrap sample.}\label{fig:visualFOF2}
\end{figure}

\end{knitrout}

\hfill $\blacklozenge$

\subsection{Functional regression models beyond the mean}
\label{sec.fdboost_family}
Using boosting for model estimation it is possible to optimize other loss functions than the squared error loss. This allows to fit, e.g., generalized linear models (GLMs) and quantile regression models \citep{koenker2005}. 
It is also possible to fit models for several parameters of the conditional response distribution in the framework of generalized additive models for location, scale and shape \citep[GAMLSS,][]{rigby2005}.  

%

For the estimation of these more general models, a suitable loss function in accordance with the modeled characteristic of the response distribution is defined and optimized. The absolute error loss ($L_1$~loss), for instance, implies median regression, and minimizing the $L_2$-loss yields mean regression.  

In \code{FDboost()}, the regression type is specified by the \code{family} argument. The \code{family} argument expects an object of class \code{Family}, which implements the respective loss function with its corresponding negative gradient and link function.   
The default is \code{family = Gaussian()} which yields $L_2$-boosting \citep{buhlmann2003}. This means that the mean squared error loss is minimized, which is equivalent to maximizing the log-likelihood of the normal distribution. 
Table~\ref{tab.families} lists some loss functions currently implemented in \pkg{mboost}, which can be directly used in \pkg{FDboost} \citep[see][for more families]{hofner2014mboost}. 
\citet{hofner2014mboost} also give an example on how to implement new families via the function \code{Family()}. See also the help page \code{?Family} for more details on all families. 
\renewcommand\arraystretch{1.4}
\begin{table}[ht]\centering
\begin{small}
\begin{tabular}{l|lll}
		response type & regression type & loss & call \\ 
		\hline
		continuous response & mean regression & $L_2$ loss & \code{Gaussian()}   \\
		& median regression & $L_1$ loss & \code{Laplace()}  \\
		& quantile regression & check function & \code{QuantReg()}  \\
		& expectile regression & asymmetric $L_2$ & \code{ExpectReg()}\\
		& robust regression & Huber loss & \code{Huber()} \\
non-negative response & gamma regression & $-l_\text{gamma}$ & \code{GammaReg()}  \\ 
\hline
binary response & logistic regression & $-l_\text{Bernoulli}$ & \code{Binomial()} \\
		& AdaBoost classification & exponential loss & \code{AdaExp()} \\
\hline
count response & Poisson model & $-l_\text{Poisson}$ & \code{Poisson()} \\
		&  neg.~binomial model & $-l_\text{neg.~binomial}$ & \code{NBinomial()} \\
\hline
scalar ordinal response  & proportional odds model & $-l_\text{proportional odds model}$ & \code{ProppOdds()} \\
\hline
scalar categorical response & multinomial model & $-l_\text{multinomial}$ & \code{Multinomial()}\\
\hline
scalar survival time & Cox model & $-l_\text{cox}$ & \code{CoxPH()} \\
\end{tabular}
\end{small}
\caption[Families that are implemented in \pkg{mboost}.]{Overview of some families that are implemented in \pkg{mboost}. $-l_{F}$ denotes the negative log-likelihood of the distribution or model $F$.}
\label{tab.families}
\end{table}
\renewcommand\arraystretch{1}

For a continuous response, several model types are available \citep{buhlmann2007}: $L_2$-boosting yields mean regression; a more robust alternative is median regression, which optimizes the absolute error loss; the Huber loss is a combination of $L_1$ and $L_2$~loss \citep{huber1964}; quantile regression can be used to model a certain quantile of the conditional response distribution \citep{fenske2011}; and expectile regression for modeling an expectile \citep{newey1987,sobotka2012}. 
For a non-negative continuous response, models assuming the gamma distribution can be useful.   
A binary response can be modeled in a GLM framework with a logit model or by minimizing the exponential loss, which corresponds to the first boosting algorithm 'AdaBoost' \citep{friedman2001,buhlmann2007}. 
Count data can be modeled assuming a Poisson or negative binomial distribution \citep{schmid2010}. 

For functional response, we compute the loss point-wise and integrate over the domain of the response.  

The following models can only be applied for scalar and not for functional response. 
For ordinal response, a proportional odds model can be used \citep{schmid2011}. 
For categorical response, the multinomial logit model is available.  
For survival models, boosting Cox proportional hazard models and accelerated failure time models have been introduced by \citet{schmid2008failure}. 

\subsubsection*{Case study (ctd.): Emotion components data}
So far, we fitted a model for the conditional mean of the response. As a more robust alternative, we consider median regression by setting \code{family = QuantReg(tau = 0.5)}. We use the \code{update} function, to update the functional model with the new family.   

\begin{knitrout}
\definecolor{shadecolor}{rgb}{1, 1, 1}\color{fgcolor}\begin{kframe}
\begin{alltt}
\hlstd{R> }\hlstd{fof_signal_med} \hlkwb{<-} \hlkwd{update}\hlstd{(fof_signal,} \hlkwc{family} \hlstd{=} \hlkwd{QuantReg}\hlstd{(}\hlkwc{tau} \hlstd{=} \hlnum{0.5}\hlstd{))}
\end{alltt}
\end{kframe}
\end{knitrout}

For median regression, the smooth intercept is the estimated median at each time-point and the effects are deviations from the median. 

 \textcolor{myc}{Similarily, if a certain quantile of the functional response is of interest, for example the 90\% quantile, the model can be updated as follows}

\begin{knitrout}
\definecolor{shadecolor}{rgb}{1, 1, 1}\color{fgcolor}\begin{kframe}
\begin{alltt}
\hlstd{R> }\hlstd{fof_historical_q90} \hlkwb{<-} \hlkwd{update}\hlstd{(fof_historical,} \hlkwc{family} \hlstd{=} \hlkwd{QuantReg}\hlstd{(}\hlkwc{tau} \hlstd{=} \hlnum{0.9}\hlstd{))}
\end{alltt}
\end{kframe}
\end{knitrout}

\textcolor{myc}{which is equivalent to the following initial model specification:}

\begin{knitrout}
\definecolor{shadecolor}{rgb}{1, 1, 1}\color{fgcolor}\begin{kframe}
\begin{alltt}
\hlstd{R> }\hlstd{fof_historical_q90} \hlkwb{<-} \hlkwd{FDboost}\hlstd{(EMG} \hlopt{~} \hlnum{1} \hlopt{+} \hlkwd{bhist}\hlstd{(EEG,} \hlkwc{s} \hlstd{= s,} \hlkwc{time} \hlstd{= t,}
\hlstd{+ }  \hlkwc{limits} \hlstd{=} \hlkwa{function}\hlstd{(}\hlkwc{s}\hlstd{,} \hlkwc{t}\hlstd{) s} \hlopt{<=} \hlstd{t} \hlopt{-} \hlnum{3}\hlstd{,} \hlkwc{df} \hlstd{=} \hlnum{6}\hlstd{),}
\hlstd{+ }  \hlkwc{timeformula} \hlstd{=} \hlopt{~} \hlkwd{bbs}\hlstd{(t,} \hlkwc{df} \hlstd{=} \hlnum{3}\hlstd{),} \hlkwc{data} \hlstd{= emotionHGL,}
\hlstd{+ }  \hlkwc{control} \hlstd{=} \hlkwd{boost_control}\hlstd{(}\hlkwc{mstop} \hlstd{=} \hlnum{300}\hlstd{),}
\hlstd{+ }  \hlkwc{family} \hlstd{=} \hlkwd{QuantReg}\hlstd{(}\hlkwc{tau} \hlstd{=} \hlnum{0.9}\hlstd{))}
\end{alltt}
\end{kframe}
\end{knitrout}
 
\textcolor{myc}{To illustrate an example for scalar-on-function regression with binary response, consider the case, in which the goal is to predict the \code{game\_outcome} in the case study for the emotions component data using only the muscle activity measured via the EMG. Consider the model $$g(\mathbb{P}(Y_{i,j}|\bm{x}_{i,j})) = \beta_0 + \gamma_j + \int_\mathcal{S} x_{\EMG,i,j}(s) \beta_{\EMG}(s) ds + \int_\mathcal{S} x_{\EMG,i,j}(s) \gamma_{\EMG,j}(s) ds,$$ for observation $i = 1,\ldots,8$ of subject $j=1,\ldots,23$, where $g$ is the inverse of the logit function, $Y_{i,j} \in \{0,1\}$ determines the game outcome (\textit{gain} and \textit{loss}, respectively) for participant $j$ in game $i$, $\gamma_j$ is a subject effect and the EMG is modeled using a global EMG effect $\beta_{\EMG}$ as well as a subject-specific EMG effect $\gamma_{\EMG,j}$. We first center the EMG-signal as it is now used as covariate}

\begin{knitrout}
\definecolor{shadecolor}{rgb}{1, 1, 1}\color{fgcolor}\begin{kframe}
\begin{alltt}
\hlstd{R> }\hlstd{emotion}\hlopt{$}\hlstd{EMG} \hlkwb{<-} \hlkwd{scale}\hlstd{(emotion}\hlopt{$}\hlstd{EMG,} \hlkwc{center} \hlstd{=} \hlnum{TRUE}\hlstd{,} \hlkwc{scale} \hlstd{=} \hlnum{FALSE}\hlstd{)}
\end{alltt}
\end{kframe}
\end{knitrout}

\textcolor{myc}{and specifiy the model in \code{FDboost} as follows}

\begin{knitrout}
\definecolor{shadecolor}{rgb}{1, 1, 1}\color{fgcolor}\begin{kframe}
\begin{alltt}
\hlstd{R> }\hlstd{sof_binary} \hlkwb{<-} \hlkwd{FDboost}\hlstd{(}
\hlstd{+ }  \hlstd{game_outcome} \hlopt{~} \hlnum{1} \hlopt{+}
\hlstd{+ }  \hlkwd{brandom}\hlstd{(subject,} \hlkwc{df} \hlstd{=} \hlnum{4}\hlstd{)} \hlopt{+}
\hlstd{+ }  \hlkwd{bsignal}\hlstd{(EMG,} \hlkwc{s} \hlstd{= s,} \hlkwc{df} \hlstd{=} \hlnum{4}\hlstd{)} \hlopt{+}
\hlstd{+ }  \hlkwd{brandom}\hlstd{(subject,} \hlkwc{df} \hlstd{=} \hlnum{2}\hlstd{)} \hlopt{%X%} \hlkwd{bsignal}\hlstd{(EMG,} \hlkwc{s} \hlstd{= s,} \hlkwc{df} \hlstd{=} \hlnum{2}\hlstd{),}
\hlstd{+ }  \hlkwc{data} \hlstd{= emotion,}
\hlstd{+ }  \hlkwc{family} \hlstd{=} \hlkwd{Binomial}\hlstd{(),}
\hlstd{+ }  \hlkwc{control} \hlstd{=} \hlkwd{boost_control}\hlstd{(}\hlkwc{mstop} \hlstd{=} \hlnum{5000}\hlstd{),}
\hlstd{+ }  \hlkwc{timeformula} \hlstd{=} \hlkwa{NULL}\hlstd{)}
\end{alltt}
\end{kframe}
\end{knitrout}

\textcolor{myc}{Note that the row-wise tensor product operator \code{\%X\%} in this case is used to specify a subject specific functional effect of the EMG-signal and the resulting degrees of freedom of this base learner are determined as the product of the \code{df}s of both base learners. To get a measure of the performance of this model, we could, e.g., compute predictions and look at the confusion matrix when simply rounding the predictions:}

\begin{knitrout}
\definecolor{shadecolor}{rgb}{1, 1, 1}\color{fgcolor}\begin{kframe}
\begin{alltt}
\hlstd{R> }\hlstd{predictions} \hlkwb{<-} \hlkwd{predict}\hlstd{(sof_binary,} \hlkwc{type} \hlstd{=} \hlstr{"response"}\hlstd{)}
\hlstd{R> }\hlstd{round_preds} \hlkwb{<-} \hlkwd{round}\hlstd{(predictions)}
\hlstd{R> }\hlkwd{table}\hlstd{(round_preds,} \hlkwd{as.numeric}\hlstd{(emotion}\hlopt{$}\hlstd{game_outcome))}
\end{alltt}
\end{kframe}
\end{knitrout}

\begin{knitrout}
\definecolor{shadecolor}{rgb}{1, 1, 1}\color{fgcolor}\begin{kframe}
\begin{verbatim}
   0  1
0 77 12
1 15 80
\end{verbatim}
\end{kframe}
\end{knitrout}

\hfill $\blacklozenge$
\\\\
The combination of GAMLSS with functional variables is discussed in \cite{brockhaus2016} and \cite{stoecker2017}. 
For GAMLSS models, \pkg{FDboost} builds on the package \pkg{gamboostLSS} \citep{gamboostLSS_gen}, in which families are implemented to fit GAMLSS. For details on the boosting algorithm to fit GAMLSS, see \cite{mayr2012} and \cite{thomas2016stability}.  
The families in \pkg{gamboostLSS} need to model at least two distribution parameters. 
For an overview of currently implemented response distributions for GAMLSS, we refer to \cite{hofner2015gamboostlss}. 
In \pkg{FDboost}, the function \code{FDboostLSS()} implements GAMLSS with functional data. 
The interface of \code{FDboostLSS()} is: 

\begin{knitrout}
\definecolor{shadecolor}{rgb}{1, 1, 1}\color{fgcolor}\begin{kframe}
\begin{alltt}
\hlstd{R> }\hlkwd{FDboostLSS}\hlstd{(formula, timeformula,} \hlkwc{data} \hlstd{=} \hlkwd{list}\hlstd{(),} \hlkwc{families} \hlstd{=} \hlkwd{GaussianLSS}\hlstd{(), ...)}
\end{alltt}
\end{kframe}
\end{knitrout}

In \code{formula} a named list of formulas is supplied. Each list entry in the \code{formula} specifies the potential covariate effects for one of the distribution parameters. The names of the list are the names of the distribution parameters.  
The argument \code{families} is used to specify the assumed response distribution with its modeled distribution parameters. The default \code{families = GaussianLSS()} yields a Gaussian location scale model.  
In the dots-argument further arguments passed to \code{FDboost()} can be supplied. 
The model object which is fitted by \code{FDboostLSS()} is a list of \code{FDboost} model objects. 
It is not possible to automatically fit a smooth offset within \code{FDboostLSS()}. Per default, a scalar offset value is used for each distribution parameter. 
For functional response, it can thus be useful to center the response prior to the model fit. 
All integration weights for the loss function are set to one, corresponding to the negative log-likelihood of the observation points. 

For model objects fitted by \code{FDboostLSS()}, methods to estimate the optimal stopping iterations, as well as methods for plotting and prediction exist. 
For more details on boosting GAMLSS models, we refer to \cite{hofner2015gamboostlss}, which is a tutorial for the package \pkg{gamboostLSS}.  

%

\subsubsection*{Case study (ctd.): Fossil fuel data}
We fit a Gaussian location scale model for the heat value. Such a model is obtained by setting \code{families = GaussianLSS()}, where the expectation is modeled using the identity link and the standard deviation by a log-link. 
Mean and standard deviation of the heat value are modeled by different covariates:   
\begin{align*}
Y_i | \mx_i  &\sim N( \mu_i, \sigma_i^2 ), \\
\mu_i &= \beta_0 + f(z_{\ho,i}) + \int_{\mathcal{S}_\NIR} x_{\NIR,i} (s_\NIR)  \beta_\NIR(s_\NIR) \,ds_\NIR + \int_{\mathcal{S}_\UVVIS} x_{\UVVIS,i} (s_\UVVIS)  \beta_\UVVIS(s_\UVVIS) \,ds_\UVVIS  \\
\log \sigma_i& = \alpha_0 + \alpha_1 z_{\ho,i}.  
\end{align*} 
The mean is modeled depending on the water content as well as depending on the NIR and the UVVIS spectrum. The standard deviation is modeled using a log-link and a linear predictor based on the water content.  
The \code{formula} has to be specified as a list of two formulas with names \code{mu} and \code{sigma} for mean and standard deviation of the normal distribution. We use the noncyclic fitting method that is introduced by \citet{thomas2016stability}. 

\begin{knitrout}
\definecolor{shadecolor}{rgb}{1, 1, 1}\color{fgcolor}\begin{kframe}
\begin{alltt}
\hlstd{R> }\hlstd{fuelSubset}\hlopt{$}\hlstd{h2o_center} \hlkwb{<-} \hlstd{fuelSubset}\hlopt{$}\hlstd{h2o} \hlopt{-} \hlkwd{mean}\hlstd{(fuelSubset}\hlopt{$}\hlstd{h2o)}
\hlstd{R> }\hlkwd{library}\hlstd{(}\hlstr{"gamboostLSS"}\hlstd{)}
\hlstd{R> }\hlstd{sof_ls} \hlkwb{<-} \hlkwd{FDboostLSS}\hlstd{(}\hlkwd{list}\hlstd{(}\hlkwc{mu} \hlstd{= heatan} \hlopt{~} \hlkwd{bbs}\hlstd{(h2o,} \hlkwc{df} \hlstd{=} \hlnum{4}\hlstd{)}
\hlstd{+ }  \hlopt{+} \hlkwd{bsignal}\hlstd{(UVVIS, uvvis.lambda,} \hlkwc{knots} \hlstd{=} \hlnum{40}\hlstd{,} \hlkwc{df} \hlstd{=} \hlnum{4}\hlstd{)}
\hlstd{+ }  \hlopt{+} \hlkwd{bsignal}\hlstd{(NIR, nir.lambda,} \hlkwc{knots} \hlstd{=} \hlnum{40}\hlstd{,} \hlkwc{df} \hlstd{=} \hlnum{4}\hlstd{),}
\hlstd{+ }  \hlkwc{sigma} \hlstd{= heatan} \hlopt{~} \hlnum{1} \hlopt{+} \hlkwd{bols}\hlstd{(h2o_center,} \hlkwc{df} \hlstd{=} \hlnum{2}\hlstd{)),}
\hlstd{+ }  \hlkwc{timeformula} \hlstd{=} \hlkwa{NULL}\hlstd{,} \hlkwc{data} \hlstd{= fuelSubset,}
\hlstd{+ }  \hlkwc{families} \hlstd{=} \hlkwd{GaussianLSS}\hlstd{(),} \hlkwc{method} \hlstd{=} \hlstr{"noncyclic"}\hlstd{)}
\hlstd{R> }\hlkwd{names}\hlstd{(sof_ls)}
\end{alltt}
\begin{verbatim}
[1] "mu"    "sigma"
\end{verbatim}
\end{kframe}
\end{knitrout}


The optimal number of boosting iterations is searched on a grid of 1 to 2000 boosting iterations. The algorithm updates in each boosting iteration the base-learner that best fits the negative gradient. Thus, in each iteration the additive predictor for only one of the distribution parameters is updated. 

\begin{knitrout}
\definecolor{shadecolor}{rgb}{1, 1, 1}\color{fgcolor}\begin{kframe}
\begin{alltt}
\hlstd{R> }\hlkwd{set.seed}\hlstd{(}\hlnum{123}\hlstd{)}
\hlstd{R> }\hlstd{cvm_sof_ls} \hlkwb{<-} \hlkwd{cvrisk}\hlstd{(sof_ls,} \hlkwc{folds} \hlstd{=} \hlkwd{cv}\hlstd{(}\hlkwd{model.weights}\hlstd{(sof_ls[[}\hlnum{1}\hlstd{]]),} \hlkwc{B} \hlstd{=} \hlnum{5}\hlstd{),}
\hlstd{+ }  \hlkwc{grid} \hlstd{=} \hlnum{1}\hlopt{:}\hlnum{2000}\hlstd{,} \hlkwc{trace} \hlstd{=} \hlnum{FALSE}\hlstd{)}
\end{alltt}
\end{kframe}
\end{knitrout}

The estimated coefficients for the expectation are similar to the effects resulting from the pure mean model. The water content has a negative effect on the standard deviation, with higher water content being associated with lower variability. 

\subsection{Variable selection by stability selection}
\label{sec.fdboost_stabsel}
Variable selection can be refined using stability selection \citep{meinshausen2010,shah2013}. Stability selection is a procedure to select influential variables while controlling false discovery rates and maximal model complexity. For component-wise gradient boosting, it is implemented in \pkg{mboost} in the function \code{stabsel()} \citep{hofner2014}, which can also be used for model objects fitted by \code{FDboost()}. 
\cite{brockhaus2015hist} compute function-on-function regression models with more functional covariates than observations and perform variable selection by stability selection. 
\cite{thomas2016stability} discuss stability selection for GAMLSS estimated by boosting.


\textcolor{myc}{\subsection{Computational Characteristics and Costs}}
\label{sec.computational_cost}

\textcolor{myc}{In order to give rough estimates on how \texttt{FDboost} 
scales up with increasing number of obervations $N$, observation points per response curve 
$G$, number of base-learners $J$ as well as other 
data and run-time related setups, this section provides some further 
insights into the algorithm and bottlenecks to bear in mind.}\\

\textcolor{myc}{
Estimating the run-time of \texttt{FDboost} is not straightforward as 
it depends on the number of boosting iterations, the size of the 
data set, the number and complexity of base-learners, 
as well as 
the type and parallelization of resampling. 
Different loss-functions, i.e., different types 
of regression should not change the run-time directly, 
but may require a smaller step-length as explained before which in turn 
induces a higher number of boosting iterations. 
In the following simulation study, we use the default value 
$\nu = 0.1$. 
\texttt{FDboost}
scales linearly in the number of iterations, which is why we use a fixed 
number $m_{\text{stop}} = 50$ in the following. 
However, 
note that the initialization of the model can get computationally very 
expensive, if very complex base-learners are defined \citep[see, e.g.][]{ruegamer2018}.
This is due to a singular-value decomposion of the design matrix 
of each base-learner, which is needed to 
compute the smoothing parameter corresponding to the pre-defined degrees 
of freedom and which has cubic run-time in the number of columns of the 
design matrix. 
For smooth effects, the number of columns of the design matrix of a base-learner is defined 
by the number of knots. For the simulation study, we use $20$ knots for 
a historical or unrestricted functional effect base-learner 
for function-on-function and scalar-on-function models, respectively. This 
corresponds to the number of knots used in the \texttt{fuelSubset} data and 
yields rather flexible estimates of functions. 
For applications where less flexibility is needed, this simulation study 
can be seen as a worst-case scenario estimate of run-times.}

\textcolor{myc}{Furthermore, we define the number of observations to be 
$N \in \{10, 100, 1000\}$, the number of time-points to be 
$G \in \{1, 10, 100, 1000\}$ and the number of base-learners to be 
$J \in \{5, 10, 20\}$. For $G=1$ scalar-on-function regression is performed, 
the other settings correspond to function-on-function regression. 
Due to computational burden, we exclude 
settings, in which $N = 1000$ and $G = 1000$ at the same time. The simulation 
was conducted on a Linux server with \textit{Intel(R) Xeon(R) CPU E5-4620 0} with 
\textit{2.20GHz}, \textit{64 cores} and \textit{512 GB RAM}.} 

\textcolor{myc}{We do not consider resampling or validation here as resampling on 
$k$ folds should approximately yield a $k$-multiple of the original 
run-time if not parallelized, i.e. run-times scale linearly in the number 
of folds. With parallelization the run-time can be reduced to the run-time 
of a single model fit.}

\textcolor{myc}{The results of the simulation study are visualized in the following, 
indicating a roughly linear increase in run-time and total allocation of memory 
by the number of observations 
(note that both are plotted against $\log_{10}(N)$), a linear increase by the number of 
observed time points per curve $G$ as well as by the number of base-learners $J$. 
The $m_\text{stop} = 50$ iterations play a comparatively minor role in time and memory consumption 
after the model has been initialized. 
Note that the total amount of allocated memory can only be interpreted in relative 
terms and does not correspond to the maximum amount of consumed memory at one 
time-point, which is considerably smaller.}

\begin{knitrout}
\definecolor{shadecolor}{rgb}{1, 1, 1}\color{fgcolor}\begin{figure}[ht]

{\centering \includegraphics[width=\maxwidth]{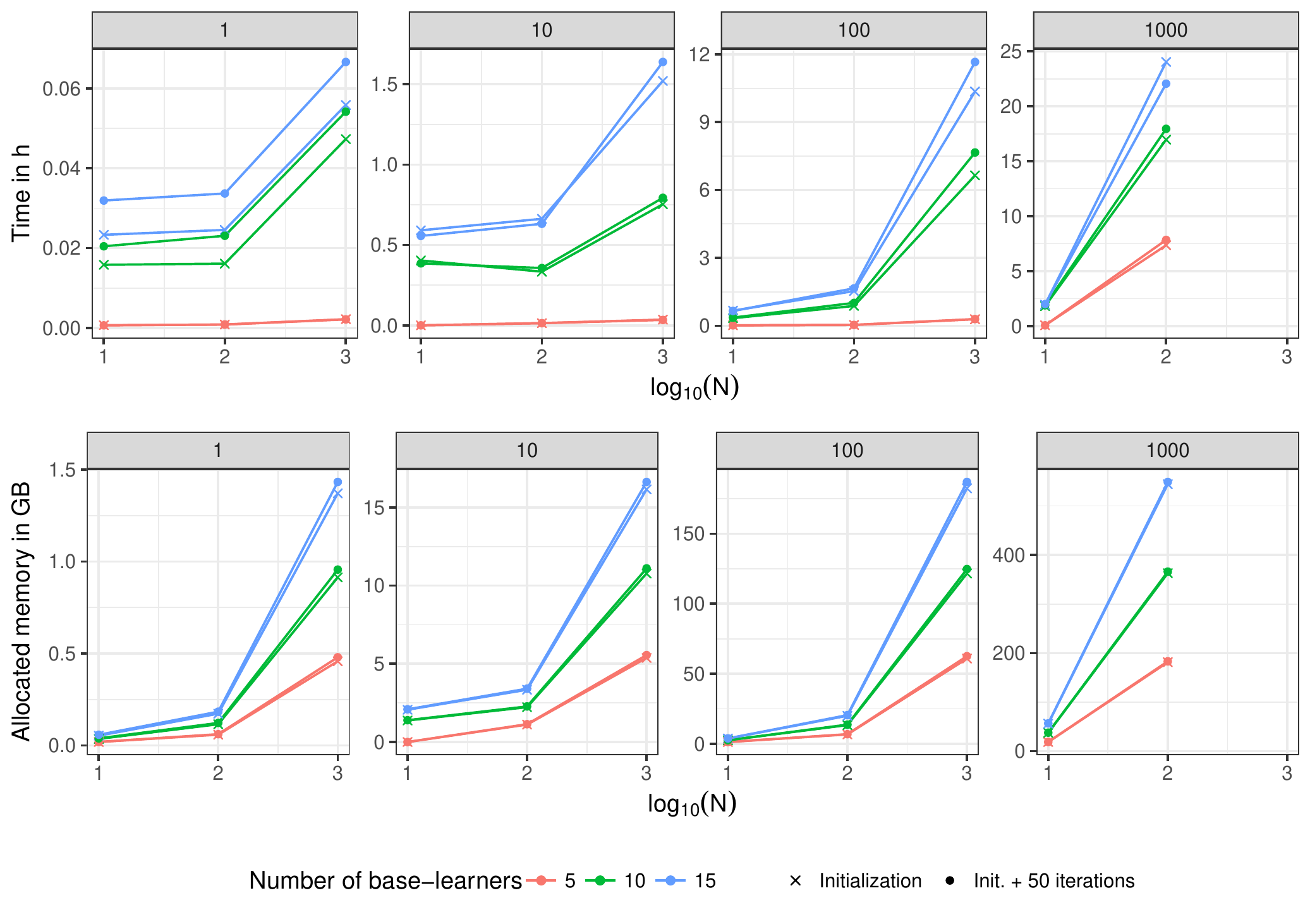} 

}

\caption{Estimated computational costs of FDboost in the simulation study. Different columns correspond to different numbers of observed time points per curve (G) and the number of base learners (J) is visualized by different colors.}\label{fig:compCosts}
\end{figure}

\end{knitrout}

 
\section{Discussion}
\label{sec.discussion} 

The \proglang{R}~add-on package \pkg{FDboost} provides a comprehensive implementation to fit functional regression models by gradient boosting. The implementation allows to fit regression models with scalar or functional response depending on many covariate effects. 
The framework includes mean, mean with link function, median and quantile regression models as well as GAMLSS.  
Various covariate effects are implemented including linear and smooth effects of scalar covariates, linear effects of functional covariates and interaction effects, also between scalar and functional covariates \citep{ruegamer2018}. The linear functional effects can have flexible integration limits, for example, to form historical or lag effects \citep{brockhaus2015hist}. 
Whenever possible, the effects are represented in the structure of linear array models \citep{currie2006} to increase computational efficiency \citep{brockhaus2015}.  
Component-wise gradient boosting allows to fit models in high-dimensional data situations and performs data-driven variable selection.  
\pkg{FDboost} builds on the well tested and modular implementation of \pkg{mboost} \citep{mboost_gen}. This facilitates the implementation of further base-learners in order to fit new covariate effects and that of families modeling other characteristics of the conditional response distribution. 


\appendix


\section{Constraints for effects of scalar covariates}
\label{app.constraints}

Consider a model for functional response with smooth intercept and an effect that contains a smooth intercept as special case, $\EV( Y_{i}(t) ) = \beta_0(t) + h_j(\mx_i,t)$, and define the mean effect at each point $t$ as $\bar h_j(\mx,t) = \mathbb{E}_X(h_j(\mX,t))$. 
This model can be parametrized in different ways, e.g., as   
\begin{align*}
\EV( Y_{i}(t) ) &= \beta_0(t) + h_j(\mx_i,t) \\
&= \left[ \beta_0(t) + \bar{h}_j(\mx,t) \right] + \left[ h_j(\mx_i,t) - \bar{h}_j(\mx,t) \right] \\ 
&= \tilde\beta_0(t) + \tilde h_j(\mx,t).    
\end{align*}
The problem arises as $\bar h_j(\mx,t)$ (or any other smooth function in $t$) can be shifted between the intercept and the covariate effect.  
At the level of the design matrices of these effects, this can be explained by the fact that the columns of the design matrix $\mB_{jY}$ and the columns of the design matrix of the functional intercept are linearly dependent. 
To obtain identifiable effects, \cite{scheipl2015} propose to center such effects $ h_j(\mx,t)$ at each point~$t$. The centering is achieved by setting the point-wise expectation over the covariate effects to zero on $\mathcal{T}$, i.e., $\EV_X(h_j(\mX, t)) = 0$ for all $t$, approximated by the sum-to-zero constraint $\sum_{i=1}^N h_j(\mx_i,t) = 0$ for all~$t$. 
How to enforce such constraints is described in Appendix~A of \cite{brockhaus2015}. 
Other constraints to obtain identifiable models are possible. However, this sum-to-zero constraint for each point~$t$ yields an intuitive interpretation: the intercept can be interpreted as global mean and the covariate effects can be interpreted as deviations from the smooth intercept. 

The constraint is enforced by a basis transformation of the design and penalty matrix. As shown in \citet{brockhaus2015}, it is sufficient to apply the constraint on the covariate-part of the design and the penalty matrix. Thus, it is not necessary to transform the basis in $t$~direction.  


\section{Base-learners for functional covariates}
\label{app.funbl}

The base-learner \code{bsignal()} sets up a linear effect of a functional variable $\int_{\mathcal{S}} x_j(s) \beta_j(s) \,ds \approx \mb_j (\mx)^\top \mtheta_j$ using P-splines. 
We approximate the integral numerically as a weighted sum using integration weights $\Delta(s)$ \citep{wood2011}, see Equation~\ref{eq.sof_example_fun_effect}: 
\begin{equation*}
\begin{aligned}
 \mb_j(\mx_i)^\top 
&= \left[\sum_{r=1}^R \Delta(s_r) x_i(s_r) \ba_1(s_r)\ \cdots\ \sum_{r=1}^R \Delta(s_r) x_i(s_r) \ba_{K_j}(s_r)\right] \\
&\approx \left[ \int_{\mathcal{S}}  x_i(s) \ba_1(s)\, ds \ \cdots\ \int_{\mathcal{S}} x_i(s) \ba_{K_j}(s)\, ds \right],
\end{aligned}
\end{equation*}
where $\ba_k(s_r)$, $k=1,\ldots, K_j$ are B-splines evaluated at~$s_r$. The corresponding penalty matrix $\mP_j$ is a squared difference matrix and thus, the smooth effect $\beta_j(s)$ in~$s$ is represented by P-splines. 

Using the base-learner \code{bfpc()} the linear functional effect $\int_{\mathcal{S}} x_j(s) \beta_j(s) \,ds$ is specified using an FPC basis. The functional covariate $x_j(s)$ and the coefficient $\beta_j(s)$ are both represented in the basis that is spanned by the functional principal components \citep[FPCs, see, e.g.,][Chap.~8 and 9]{ramsay2005} of $x_j(s)$. 
%
Let $X_j(s)$ be a zero-mean stochastic process in the space of all square-integrable functions $L^2(\mathcal{S})$. Let $x_{ij}(s)$ be the observations of the copies $X_{ij}(s)$ of this process. We denote the eigenvalues of the auto-covariance of $X_j(s)$ as $\zeta_1 \geq \zeta_2 \geq \cdots \geq 0$ and the corresponding eigenfunctions as $e_k(s)$,  $k \in \mathbb{N}$. The eigenfunctions $\{e_k(s), k \in \mathbb{N} \}$ form an orthonormal basis for the $L^2(\mathcal{S})$. Using the Karhunen-Lo\`eve theorem, the functional covariate can be represented as weighted sum  
\[
X_{ij}(s) = \sum_{k=1}^{\infty} Z_{ik} e_k(s),
\]
where $Z_{ik}$ are uncorrelated mean zero random variables with variance~$\zeta_k$ and realizations $z_{ik}$. In practice, the infinite sum is truncated at a certain value $K_j$. 
Representing the functional covariate and the coefficient function by this truncated basis with weights $\theta_l$ and $z_{ik}$, respectively, the effect simplifies to 
\begin{equation*}
\int_{\mathcal{S}} x_{ij}(s) \beta_j(s) \,ds
\approx \sum_{k,l=1}^{K_j} \int_{\mathcal{S}} z_{ik} e_k(s) e_l(s) \theta_l \,ds
= \sum_{k=1}^{K_j} z_{ik} \theta_k,  
\end{equation*}
as the eigenfunctions $e_k(s)$ are orthonormal. 
Thus, this approach is equivalent to using the (estimated) first $K_j$ FPC scores~$z_{ik}$ as linear covariates. The number of eigenfunctions is usually chosen such that the truncated basis explains a fixed proportion of the total variability of the covariate, for example 99\% \citep[cf.,][]{morris2015}. This truncation achieves regularized effects, as the effect can only lie in the space spanned by the first $K_j$ eigenfunctions. For the penalty matrix $\mP_j$ the identity matrix is used in \code{bfpc()}.     
%

For scalar response, the base-learners \code{bsignal()} and \code{bfpc()} yield the effect $\int_{\mathcal{S}} x_j(s) \beta_j(s) \,ds$. Combining them with a smooth effect in $t$ using \code{bbs()}, they can be used to fit effects for function-on-function regression $\int_{\mathcal{S}} x_j(s) \beta_j(s,t) \,ds$.   
\\\\
The base-learner \code{bhist()} allows to specify functional linear effects with integration limits depending on~$t$, $\int_{l(t)}^{u(t)} x(s) \beta(s,t)\,ds$. Per default, a historical effects with limits $[l(t), u(t)] = [T_1, t]$ is fitted. 
The integral is approximated by a numerical integration scheme \citep{scheipl2015}. 
We transform the observations of the functional covariate $x_j(s_r)$ such that they contain the integration limits and the weights for numerical integration. 
We define $ \tilde{x}_j(s_r,t) = I \left( l(t) \le s_r \le u(t) \right) \Delta(s_r) x_j(s_r),$ with indicator function~$I(\cdot)$ and integration weights $\Delta(s_r)$. 
The marginal basis over the covariates $\mx$, which in this case also depends on $t$, is:  
\begin{equation*}
\begin{aligned}
 \mb_{jY}(\mx_i, t)^\top 
&= \left[ \sum_{r=1}^R \tilde{x}_j(s_r,t) \ba_1(s_r)\ \cdots\ \sum_{r=1}^R \tilde{x}_j(s_r,t) \ba_{K_j}(s_r) \right]
 \otimes
\Big[ \ba_1(t_{g}) \; \cdots \; \ba_{K_Y}(t_{g}) \Big] \\ 
&\approx \left[ \int_{l(t)}^{u(t)}  x_i(s) \ba_1(s)\, ds \; \cdots\; \int_{l(t)}^{u(t)} x_i(s) \ba_{K_j}(s)\, ds \right]  
\otimes
\big[ \ba_1(t_{g}) \; \cdots \; \ba_{K_Y}(t_{g}) \big].
\end{aligned}
\end{equation*}
The isotropic penalty in Equation~\ref{eq.penalty_iso} is used with squared difference matrices as marginal penalties to form P-splines bases for the $s$ and $t$ direction of $\beta(s,t)$. 

For a concurrent effect $x(t)\beta(t)$, the base-learner \code{bconcurrent()} can be used. The smooth effect $\beta(t)$ in~$t$ is expanded by P-splines.


\section{Row tensor product and Kronecker product bases}
\label{app.operatorsOX}

In the \proglang{R}~package \pkg{mboost} \citep{mboost_gen}, 
the Kronecker product of two base-learners 
is implemented as \code{\%O\%}. 
The row-wise tensor product of two base-learners 
is implemented in the operator \code{\%X\%}. 
The row-wise tensor product of two marginal design matrices, $\mB_j \in \real^{n \times K_j}$ and $\mB_Y \in \real^{n \times K_Y}$, is defined as $n \times K_j K_Y$ matrix 
\begin{equation*}
\mB_j \odot \mB_Y = (\mB_j \otimes \bm{1}^{\top}_{K_Y} ) \cdot (\bm{1}^{\top}_{K_j} \otimes \mB_Y ),
\end{equation*}
where $\cdot$~denotes entry-wise multiplication and $\bm{1}_{K}$ is the $K$-dimensional vector of ones. 
The operators \code{\%X\%} and \code{\%O\%} use the Kronecker product or the row-wise tensor product to compute the design matrix. The penalty is computed according to Equation~\ref{eq.penalty_ani}. 
When \code{\%X\%} or \code{\%O\%} is called with specified argument \code{df} in both marginal base-learners, the degrees of freedom of the composed effect are computed as the product of the two specified degrees of freedom. Then, only one smoothing parameter is computed for an isotropic penalty like in Equation~\ref{eq.penalty_iso}. 
Consider, for example, the composed base-learner \code{bols(z1, df = df1) \%O\% bbs(t, df = df2)}. The base-learner \code{bols()} specifies a linear effect. The base-learner \code{bbs()} specifies a smooth effect represented by P-splines. Thus, the composed base-learner yields the effect $z_1 \beta_j(t)$, which is linear in~$z_1$ and smooth in~$t$.  
The global degrees of freedom for the composed base-learner are computed as df$_j = $ \code{df1 * df2}. The corresponding smoothing parameter $\lambda_j$ is computed by Demmler-Reinsch orthogonalization \citep[Appendix B.1.1]{ruppert2003}.  

For array models, \code{FDboost()} connects the effects of \code{formula} and \code{timeformula} by the operator \code{\%O\%}, yielding \code{b\_1 \%O\% b\_Y + \ldots + b\_J \%O\% b\_Y}. The operator \code{\%O\%} uses the array framework of \cite{currie2006} to efficiently implement such effects in boosting \citep{hothorn2013}. 
If it is not possible to use the array framework, e.g., if the response is observed on curve-specific grids or for historical effects, the design matrix is computed as row-wise tensor product basis, i.e., using the operator \code{\%X\%}. Within the function \code{FDboost()} the appropriate operator is used automatically. 
When the marginal base-learners are supplied with specified degrees of freedom (argument \code{df}), \code{\%O\%} and \code{\%X\%} use the isotropic penalty (\ref{eq.penalty_iso}).  

The anisotropic penalty (\ref{eq.penalty_ani}) is obtained if the smoothing parameter is specified in both marginal base-learners; for instance, as \code{bols(z1, lambda = lambda1) \%O\% bbs(t, lambda = lambda2)}.  
However, it is hard to control the degrees of freedom in this case such that each base-learner in the model has the same number of degrees of freedom. Thus, specifying the smoothing parameter $\lambda$ in both marginal base-learners is hardly applicable in practice.    

In some cases, one only wants to penalize the basis in $t$ direction. In this case, the penalty in Equation~\ref{eq.penalty_0} can be used. Such a penalty is obtained using the operators \code{\%A0\%} or \code{\%Xa0\%}, for the Kronecker and the row-wise tensor product basis, respectively. When \code{\%A0\%} or \code{\%Xa0\%} are used to form an effect with penalty (\ref{eq.penalty_0}), the number of degrees of freedom in the first base-learner has to be equal to the number of its columns. 
Consider, \code{bols(z1, df = 1, intercept = FALSE) \%A0\% bbs(t, df = df2)}, with a metric variable \code{z1}. This specification implies $\mb_j(\mx_i)=z_{i1}$ and $\mP_j = \boldsymbol{0}$ for the \code{bols()} base-learner. The \code{bbs()} base-learner sets up a design matrix of B-spline evaluations in $t$ and a squared difference matrix as penalty matrix.  

Linking \code{formula} and \code{timeformula} in \code{FDboost()} to representation (\ref{eq.effect_o}), the $J$ base-learners in \code{formula} correspond to the $J$ marginal bases $\mb_j$ and the base-learners in \code{timeformula} corresponds to the marginal basis $\mb_Y$. If it is possible to represent the effects as Kronecker product, the base-learners are combined by \code{\%O\%}. Otherwise, the row-wise tensor product \code{\%X\%} is used to combine the marginal bases.  

Consider, for example, \code{formula = Y ~ b\_1 + b\_2 + \ldots + b\_J}, and the \code{timeformula = ~ b\_Y}. For an array model, this yields \code{Y ~ b\_1 \%O\% b\_Y + b\_2  \%O\% b\_Y + ... + b\_J \%O\% b\_Y}. 
If \code{formula} contains base-learners that are composed of two base-learners by \code{\%O\%} or \code{\%A0\%}, those effects are not expanded with \code{timeformula}, allowing for model specifications with different effects in $t$~direction. 
For example, \code{formula = Y ~ b\_1 + b\_2 \%A0\% b\_{Y0}}, and \code{timeformula = ~ b\_Y}, with non-linear base-learner \code{b\_Y} and linear base-learner \code{b\_{Y0}}, yield \code{Y ~ b\_1 \%O\% b\_Y + b\_2  \%A0\% b\_{Y0}}.


\section{Example code for resampling with repeated measurements}
\label{app.resampling}

In the following, we search the optimal stopping iteration for model (\ref{eq.fos_random_power}), which contains a linear effect for the game condition power and a person-specific effect. 

We search the optimal stopping iteration by a 5-fold cross-validation. The resampling is done on the level of curves, assuming that the observations per subject are independent conditional on the subject specific effects. 
We use the function \code{applyFolds()} for the resampling.  

\begin{knitrout}
\definecolor{shadecolor}{rgb}{1, 1, 1}\color{fgcolor}\begin{kframe}
\begin{alltt}
\hlstd{R> }\hlkwd{set.seed}\hlstd{(}\hlnum{123}\hlstd{)}
\hlstd{R> }\hlstd{folds_bs} \hlkwb{<-} \hlkwd{cv}\hlstd{(}\hlkwc{weights} \hlstd{=} \hlkwd{rep}\hlstd{(}\hlnum{1}\hlstd{, fos_random_power}\hlopt{$}\hlstd{ydim[}\hlnum{1}\hlstd{]),}
\hlstd{+ }  \hlkwc{type} \hlstd{=} \hlstr{"kfold"}\hlstd{,} \hlkwc{B} \hlstd{=} \hlnum{5}\hlstd{)}
\hlstd{R> }\hlstd{cvm} \hlkwb{<-} \hlkwd{applyFolds}\hlstd{(fos_random_power,} \hlkwc{folds} \hlstd{= folds_bs,} \hlkwc{grid} \hlstd{=} \hlnum{1}\hlopt{:}\hlnum{200}\hlstd{)}
\end{alltt}
\end{kframe}
\end{knitrout}

%
%
%

The optimal stopping iteration is estimated to be 200, which is the upper limit of the searched grid. Thus, the resampling has to be rerun with a higher maximal number of boosting iterations. 

To resample the observations on the level of independent observation units, the folds can be set up on the level of subjects. The corresponding folds for a leave-on-subject out cross-validation, which are then passed to \code{applyFolds()}, could be constructed as follows:
  
\begin{knitrout}
\definecolor{shadecolor}{rgb}{1, 1, 1}\color{fgcolor}\begin{kframe}
\begin{alltt}
\hlstd{R> }\hlkwd{set.seed}\hlstd{(}\hlnum{123}\hlstd{)}
\hlstd{R> }\hlstd{folds_bs_long_subject} \hlkwb{<-} \hlkwd{sapply}\hlstd{(}\hlkwd{levels}\hlstd{(emotion}\hlopt{$}\hlstd{subject),}
\hlstd{+ }  \hlkwa{function}\hlstd{(}\hlkwc{x}\hlstd{)} \hlkwd{as.numeric}\hlstd{(x} \hlopt{!=} \hlstd{emotion}\hlopt{$}\hlstd{subject))}
\end{alltt}
\end{kframe}
\end{knitrout}

\section{Fitting factor-specific historical models}

\textcolor{myc}{
In this section we provide code to fit a more complex and realistic model to the emotion component data. As the EMG signal might depend on all three study settings (\code{power}, \code{game\_outcome}, \code{control}) as well as their interactions, and the influence of the EEG signal might also be specific for each setting as well as for each subject, we assume the following model \citep[cf.][]{ruegamer2018}: 
\begin{equation}
\begin{split}
\mathbb{E}(Y_{\EMG,i,j}(t)|\bm{x}_{i,j}) = \beta_0(t) 
&+ \gamma_{\text{subject},j}(t)\\
&+ I(x_{\text{power},i,j} = 1) \beta_{\text{power}}(t)\\
&+ I(x_{\text{outcome},i,j} = 1) \beta_{\text{outcome}}(t)\\
&+ I(x_{\text{control},i,j} = 1) \beta_{\text{control}}(t)\\
&+ I(x_{\text{power},i,j} = 1, x_{\text{outcome},i,j} = 1) \beta_{\text{power,outcome}}(t)\\
&+ I(x_{\text{outcome},i,j} = 1, x_{\text{control},i,j} = 1) \beta_{\text{outcome,control}}(t)\\
&+ I(x_{\text{power},i,j} = 1, x_{\text{control},i,j} = 1) \beta_{\text{power,control}}(t)\\
&+ I(x_{\text{power},i,j} = 1, x_{\text{outcome},i,j} = 1, x_{\text{control},i,j} = 1) \, \cdot\\ 
&\quad \beta_{\text{power,outcome,control},i}(t)\\
&+ \int_0^{t-3} x_{\EMG,i,j}(s) \beta_{\EMG}(s,t) ds\\
&+ \int_0^{t-3} x_{\EMG,i,j}(s) \gamma_{\EMG,i}(s,t) ds\\
&+ \int_0^{t-3} x_{\EMG,i,j}(s) \zeta_{\EMG,j}(s,t) ds + \varepsilon_{i,j}(t)
\end{split}
\end{equation}
for observation $i = 1,\ldots,8$ corresponding to the 8 different game conditions of subject $j=1,\ldots,23$. The model was proposed in \citet{ruegamer2018}, which extended historical models by allowing for factor-specific historical effects. To our knowledge, \pkg{FDboost} so far is the only software capable of fitting such effects.}

\textcolor{myc}{To this end, we have to define the 3 two-way interactions \code{power.outcome}, \code{outcome.control}, \code{power.control}, 1 three-way interaction \code{gamecondition} and an \code{hmatrix}-object \code{X1h}. The object is needed for the function \code{bhistx}, which in turn allows to combine historical effects with factor variables using the row-wise tensor product operator \code{\%X\%}. To construct a \code{hmatrix}-object, the time and an identifier for each curve in long format must be supplied along with the original response. The corresponding model fit in \proglang{R} takes around 75 minutes to fit the model with $5000$ iterations and needs approximately a maximum of 15GB RAM at once. We further allow for an anisotropic penalty for all factor effects that are time-dependent, which is achieved by using the \code{\%A\%}-operator.}\\

\textcolor{myc}{This example also demonstrates how the degrees of freedom can be defined to be equal across all base-learners (in this case $\text{df}_j=20$), which is explained in Appendix~C.}\\

\begin{knitrout}
\definecolor{shadecolor}{rgb}{1, 1, 1}\color{fgcolor}\begin{kframe}
\begin{alltt}
\hlstd{R> }\hlstd{N} \hlkwb{<-} \hlkwd{nrow}\hlstd{(emotion}\hlopt{$}\hlstd{EEG)}
\hlstd{R> }\hlstd{G} \hlkwb{<-} \hlkwd{ncol}\hlstd{(emotion}\hlopt{$}\hlstd{EEG)}
\hlstd{R> }
\hlstd{R> }\hlstd{emotion}\hlopt{$}\hlstd{id_repeated} \hlkwb{=} \hlkwd{rep}\hlstd{(}\hlnum{1}\hlopt{:}\hlstd{N, G)}
\hlstd{R> }
\hlstd{R> }\hlstd{emotion}\hlopt{$}\hlstd{EEG} \hlkwb{<-} \hlkwd{scale}\hlstd{(emotion}\hlopt{$}\hlstd{EEG,} \hlkwc{center} \hlstd{=} \hlnum{TRUE}\hlstd{,} \hlkwc{scale} \hlstd{=} \hlnum{FALSE}\hlstd{)}
\hlstd{R> }
\hlstd{R> }\hlstd{X1h} \hlkwb{<-} \hlkwd{hmatrix}\hlstd{(}\hlkwc{time} \hlstd{=} \hlkwd{rep}\hlstd{(emotion}\hlopt{$}\hlstd{t,} \hlkwc{each} \hlstd{= N),}
\hlstd{+ }  \hlkwc{id} \hlstd{= emotion}\hlopt{$}\hlstd{id_repeated,}
\hlstd{+ }  \hlkwc{x} \hlstd{= emotion}\hlopt{$}\hlstd{EEG)}
\hlstd{R> }\hlstd{emotion}\hlopt{$}\hlstd{power.outcome} \hlkwb{<-} \hlkwd{interaction}\hlstd{(emotion}\hlopt{$}\hlstd{power, emotion}\hlopt{$}\hlstd{game_outcome)}
\hlstd{R> }\hlstd{emotion}\hlopt{$}\hlstd{outcome.control} \hlkwb{<-} \hlkwd{interaction}\hlstd{(emotion}\hlopt{$}\hlstd{game_outcome, emotion}\hlopt{$}\hlstd{control)}
\hlstd{R> }\hlstd{emotion}\hlopt{$}\hlstd{power.control} \hlkwb{<-} \hlkwd{interaction}\hlstd{(emotion}\hlopt{$}\hlstd{power, emotion}\hlopt{$}\hlstd{control)}
\hlstd{R> }\hlstd{emotion}\hlopt{$}\hlstd{gamecondition} \hlkwb{<-} \hlkwd{interaction}\hlstd{(emotion}\hlopt{$}\hlstd{power, emotion}\hlopt{$}\hlstd{game_outcome,}
\hlstd{+ }                                      \hlstd{emotion}\hlopt{$}\hlstd{control)}
\hlstd{R> }
\hlstd{R> }\hlstd{emotion}\hlopt{$}\hlstd{X1h} \hlkwb{<-} \hlkwd{I}\hlstd{(X1h)}
\hlstd{R> }
\hlstd{R> }\hlstd{mod} \hlkwb{<-} \hlkwd{FDboost}\hlstd{(}
\hlstd{+ }  \hlstd{EMG} \hlopt{~} \hlnum{1} \hlopt{+} \hlkwd{brandomc}\hlstd{(subject,} \hlkwc{df} \hlstd{=} \hlnum{5}\hlstd{)} \hlopt{%A%} \hlkwd{bbs}\hlstd{(t,} \hlkwc{df} \hlstd{=} \hlnum{4}\hlstd{)} \hlopt{+}
\hlstd{+ }  \hlkwd{bolsc}\hlstd{(power,} \hlkwc{df} \hlstd{=} \hlnum{2}\hlstd{,} \hlkwc{intercept} \hlstd{=} \hlnum{TRUE}\hlstd{)} \hlopt{%A%} \hlkwd{bbs}\hlstd{(t,} \hlkwc{df} \hlstd{=} \hlnum{10}\hlstd{)} \hlopt{+}
\hlstd{+ }  \hlkwd{bolsc}\hlstd{(game_outcome,} \hlkwc{df} \hlstd{=} \hlnum{2}\hlstd{,} \hlkwc{intercept} \hlstd{=} \hlnum{TRUE}\hlstd{)} \hlopt{%A%} \hlkwd{bbs}\hlstd{(t,} \hlkwc{df} \hlstd{=} \hlnum{10}\hlstd{)} \hlopt{+}
\hlstd{+ }  \hlkwd{bolsc}\hlstd{(control,} \hlkwc{df} \hlstd{=} \hlnum{2}\hlstd{,} \hlkwc{intercept} \hlstd{=} \hlnum{TRUE}\hlstd{)}  \hlopt{%A%} \hlkwd{bbs}\hlstd{(t,} \hlkwc{df} \hlstd{=} \hlnum{10}\hlstd{)}\hlopt{+}
\hlstd{+ }  \hlkwd{bolsc}\hlstd{(power.outcome,} \hlkwc{intercept} \hlstd{=} \hlnum{TRUE}\hlstd{,} \hlkwc{df} \hlstd{=} \hlnum{2}\hlstd{)} \hlopt{%A%} \hlkwd{bbs}\hlstd{(t,} \hlkwc{df} \hlstd{=} \hlnum{10}\hlstd{)} \hlopt{+}
\hlstd{+ }  \hlkwd{bolsc}\hlstd{(outcome.control,} \hlkwc{intercept} \hlstd{=} \hlnum{TRUE}\hlstd{,} \hlkwc{df} \hlstd{=} \hlnum{2}\hlstd{)} \hlopt{%A%} \hlkwd{bbs}\hlstd{(t,} \hlkwc{df} \hlstd{=} \hlnum{10}\hlstd{)} \hlopt{+}
\hlstd{+ }  \hlkwd{bolsc}\hlstd{(power.control,} \hlkwc{intercept} \hlstd{=} \hlnum{TRUE}\hlstd{,} \hlkwc{df} \hlstd{=} \hlnum{2}\hlstd{)} \hlopt{%A%} \hlkwd{bbs}\hlstd{(t,} \hlkwc{df} \hlstd{=} \hlnum{10}\hlstd{)} \hlopt{+}
\hlstd{+ }  \hlkwd{bolsc}\hlstd{(gamecondition,} \hlkwc{intercept} \hlstd{=} \hlnum{TRUE}\hlstd{,} \hlkwc{df} \hlstd{=} \hlnum{2}\hlstd{)} \hlopt{%A%} \hlkwd{bbs}\hlstd{(t,} \hlkwc{df} \hlstd{=} \hlnum{10}\hlstd{)} \hlopt{+}
\hlstd{+ }  \hlkwd{bhistx}\hlstd{(X1h,}
\hlstd{+ }    \hlkwc{limits} \hlstd{=} \hlkwa{function}\hlstd{(}\hlkwc{s}\hlstd{,}\hlkwc{t}\hlstd{)\{ s} \hlopt{<} \hlstd{t} \hlopt{-} \hlnum{3} \hlstd{\},}
\hlstd{+ }    \hlkwc{df} \hlstd{=} \hlnum{20}\hlstd{,} \hlkwc{knots} \hlstd{=} \hlnum{10}\hlstd{,}
\hlstd{+ }    \hlkwc{differences} \hlstd{=} \hlnum{2}\hlstd{,}
\hlstd{+ }    \hlkwc{standard} \hlstd{=} \hlstr{"length"}
\hlstd{+ }  \hlstd{)} \hlopt{+}
\hlstd{+ }  \hlkwd{bhistx}\hlstd{(X1h,}
\hlstd{+ }    \hlkwc{limits} \hlstd{=} \hlkwa{function}\hlstd{(}\hlkwc{s}\hlstd{,}\hlkwc{t}\hlstd{)\{ s} \hlopt{<} \hlstd{t} \hlopt{-} \hlnum{3} \hlstd{\},}
\hlstd{+ }    \hlkwc{df} \hlstd{=} \hlnum{5}\hlstd{,} \hlkwc{knots} \hlstd{=} \hlnum{10}\hlstd{,}
\hlstd{+ }    \hlkwc{differences} \hlstd{=} \hlnum{2}\hlstd{,}
\hlstd{+ }    \hlkwc{standard} \hlstd{=} \hlstr{"length"}\hlstd{)} \hlopt{%X%}
\hlstd{+ }  \hlkwd{bolsc}\hlstd{(gamecondition,} \hlkwc{df} \hlstd{=} \hlnum{4}\hlstd{,} \hlkwc{intercept} \hlstd{=} \hlnum{TRUE}\hlstd{,}
\hlstd{+ }    \hlkwc{index} \hlstd{= id_repeated)} \hlopt{+}
\hlstd{+ }  \hlkwd{bhistx}\hlstd{(X1h,}
\hlstd{+ }    \hlkwc{limits} \hlstd{=} \hlkwa{function}\hlstd{(}\hlkwc{s}\hlstd{,}\hlkwc{t}\hlstd{)\{ s} \hlopt{<} \hlstd{t} \hlopt{-} \hlnum{3} \hlstd{\},}
\hlstd{+ }    \hlkwc{df} \hlstd{=} \hlnum{5}\hlstd{,} \hlkwc{knots} \hlstd{=} \hlnum{10}\hlstd{,}
\hlstd{+ }    \hlkwc{differences} \hlstd{=} \hlnum{2}\hlstd{,}
\hlstd{+ }    \hlkwc{standard} \hlstd{=} \hlstr{"length"}\hlstd{)} \hlopt{%X%}
\hlstd{+ }  \hlkwd{brandomc}\hlstd{(subject,} \hlkwc{df} \hlstd{=} \hlnum{4}\hlstd{,} \hlkwc{index} \hlstd{= id_repeated),}
\hlstd{+ }  \hlkwc{control} \hlstd{=} \hlkwd{boost_control}\hlstd{(}\hlkwc{mstop} \hlstd{=} \hlnum{5000}\hlstd{,} \hlkwc{trace} \hlstd{=} \hlnum{TRUE}\hlstd{),}
\hlstd{+ }  \hlkwc{timeformula} \hlstd{=} \hlopt{~} \hlkwd{bbs}\hlstd{(t),}
\hlstd{+ }  \hlkwc{data} \hlstd{= emotion}
\hlstd{+ }\hlstd{)}
\end{alltt}
\end{kframe}
\end{knitrout}


\bibliography{literatur_entries}

\end{document}

%% file: packages.tex
\usepackage[inner=26mm,outer=21mm,top=28mm,bottom=30mm,includehead]{geometry}
\usepackage{setspace}
\usepackage{lipsum}
\usepackage{graphicx}

\usepackage{alltt}
\usepackage{amsmath}
\usepackage{amssymb}
\usepackage{natbib}
\usepackage{bm}

\usepackage{amsmath}  
\usepackage{amssymb}  
\usepackage{amsfonts} 
\usepackage{amsthm}

\usepackage{dsfont}   
\usepackage{stmaryrd}
\usepackage{latexsym}
\usepackage{bm}       

\usepackage{rotating}
\usepackage{array}

\usepackage[small, normal]{caption}
\usepackage[labelformat=simple, position=top]{subfig}

\usepackage{tabularx}
\usepackage{booktabs}
\usepackage{multicol}
\usepackage{multirow}
\usepackage{hhline}                    

\usepackage{pifont}        
\usepackage[round]{natbib} 

\usepackage{url} 

\usepackage{threeparttable}

%% file: header.tex
\newcommand{\bvec}{\left[\begin{array}{c}}
\newcommand{\evec}{\end{array}\right]}
\newcommand{\bmat}[1]{\left[\begin{array}{*{#1}{c}}}
\newcommand{\emat}{\end{array}\right]}

\newcommand{\EEG}{\textnormal{\tiny \textsc{EEG}}}
\newcommand{\EMG}{\textnormal{\tiny \textsc{EMG}}}

\newcommand{\NIR}{\textnormal{\tiny \textsc{NIR}}}
\newcommand{\UVVIS}{\textnormal{\tiny \textsc{UV}}}
\newcommand{\ho}{\textnormal{\tiny \textsc{h2o}}}

\newcommand{\real}{\mathds{R}}

\newcommand{\EV}{\mathds{E}}

\newcommand{\ba}{\phi}

\newcommand{\mb}{\bm{b}}
\newcommand{\mB}{\bm{B}}

\newcommand{\mI}{\bm{I}}

\newcommand{\mP}{\bm{P}}

\newcommand{\mx}{\bm{x}}
\newcommand{\mX}{\bm{X}}

\newcommand{\mtheta}{\bm{\theta}}